\documentclass[format=acmsmall, review=false]{acmart}
\usepackage{acm-ec-21}

\setcitestyle{authoryear,sort,semicolon}

\usepackage{Preamble}
\usepackage{Macros} 

\title[Colonel Blotto Games with Favoritism]{Colonel Blotto Games with Favoritism: Competitions with Pre-allocations and Asymmetric Effectiveness}

\begin{abstract}

We introduce the \emph{Colonel Blotto game with favoritism}, an extension of the famous Colonel Blotto game where the winner-determination rule is generalized to include pre-allocations and asymmetry of the players' resources effectiveness on each battlefield. Such \emph{favoritism} is found in many classical applications of the Colonel Blotto game. We focus on the Nash equilibrium. First, we consider the closely related model of all-pay auctions with favoritism and completely characterize its equilibrium. Based on this result, we prove the existence of a set of optimal univariate distributions---which serve as candidate marginals for an equilibrium---of the Colonel Blotto game with favoritism and show an explicit construction thereof. In several particular cases, this directly leads to an equilibrium of the Colonel Blotto game with favoritism. In other cases, we use these optimal univariate distributions to derive an approximate equilibrium with well-controlled approximation error. Finally, we propose an algorithm---based on the notion of winding number in parametric curves---to efficiently compute an approximation of the proposed optimal univariate distributions with arbitrarily small error.

\end{abstract}

\begin{document}

\author{Dong Quan Vu}
\email{dong-quan.vu@inria.fr}
\affiliation{%
	\institution{Univ. Grenoble Alpes, CNRS, Inria, Grenoble INP, LIG}
	  \country{France}
}

\author{Patrick Loiseau}
\email{patrick.loiseau@inria.fr}
\affiliation{%
	\institution{Univ. Grenoble Alpes, Inria, CNRS, Grenoble INP, LIG}
	  \country{France}
}
\renewcommand{\shortauthors}{Vu and Loiseau}


%

\maketitle

%
\section{Introduction}
\label{sec:Intro}

The \emph{Colonel Blotto game}, first introduced by \cite{borel1921}, is a famous resource allocation games. Two players A and B compete over $n$ battlefields by simultaneously distributing resources such that the sum of each player's allocations does not exceed her budget (the so-called \emph{budget constraint}). Each battlefield has a certain value. In each battlefield, the player who has the higher allocation wins and gains the whole battlefield's value while the other player gains zero; this is the \emph{winner-determination rule}. The total payoff of each player is the sum of gains from all the battlefields. 
	
The Colonel Blotto game captures a large range of practical situations. Its original application is \emph{military logistic} \cite{gross1950,grosswagner}, where resources correspond to soldiers, equipment or weapons; but it is now also used to model \emph{security problems} where battlefields are security targets and resources are security forces or effort \cite{chia2012,schwartz2014}, \emph{political competitions} where players are political parties who distribute their time or money resources to compete over voters or states \cite{kovenock2012,myerson1993incentives,roberson2006}, \emph{competitions in online advertising} \cite{masucci2014,masucci2015}, or \emph{radio-spectrum management systems} \cite{hajimirsaadeghi2017dynamic}.

In many of these applications, however, the winner-determination rule of the Colonel Blotto game is too restrictive to capture practical situations because a player might have an advantage over some battlefields; we refer to this as \emph{favoritism}. There can be two basic types of favoritism:
\begin{trivlist}
    \item[\emph{First},] players may have resources committed to battlefields before the game begins---we refer to them as \emph{pre-allocations}. These pre-allocations then add up to the allocations to determine the winner in each battlefield. In military logistics for instance, before the start of military operations, it is often the case that one side (or both) already installed military forces on some battlefields. Pre-allocations can also be found in R\&D contests, where companies can use the technologies they currently possess to gain advantage while competing to develop new products/technologies. In political contests, it is often the case that voters have an a priori position that may be interpreted as a pre-allocation of the corresponding party (e.g., Californian voters are in majority pro-Democrats). 
    \item[\emph{Second},] the \emph{resources effectiveness} may not be the same for both players, and may vary across battlefields. For example, in airport-surveillance, it often requires several agents to patrol a security target while a single terrorist may suffice for a successful attack. In military logistics, the effectiveness of resources (equipment, soldiers, etc.) may differ amongst players and vary according to the landscapes/features of the battlefields. In R\&D contests, one unit of resources (researchers/machines) of a company often has different strengths and weaknesses than that of other companies.
\end{trivlist}



In this work, we propose and analyze an extension of the Colonel Blotto game with a winner-determination rule capturing pre-allocations and asymmetric effectiveness of resources. Specifically, we consider the following rule: in battlefield $i \in \{1,\cdots, n\}$, if the allocations of Players A and B are $x^A$ and $x^B$ respectively, Player A wins if $x^A > q_i \cdot x^B - p_i$ and Player B wins otherwise (we will specify the tie breaking rule below). Here, $p_i \in \mathbb{R}$ and $q_i >0$ are given parameters known to both players that represent pre-allocations and asymmetric effectiveness of resources respectively. We call this game the \emph{Colonel Blotto game with favoritism} and denote it by \FCB throughout the paper. We focus on characterizing and computing Nash equilibria of the \FCB game. 

	
Completely characterizing and computing a Nash equilibrium of the Colonel Blotto game, even without favoritism, is a notoriously challenging problem (see related works below). A standard approach consists in first identifying candidate equilibrium marginal distributions for each battlefield's allocation---called the \emph{optimal univariate distributions}. This is often done by looking for an equivalence to the related problem of all-pay auctions---the game where two bidders secretly bid on a common item and the higher bidder wins the item and gains its value but both players pay their bids. Then, constructing an equilibrium based on these univariate distributions can be done exactly for some particular cases of parameters configurations (see related works below). In cases where this is not possible, an alternative solution is to look for \emph{approximate equilibria} with well-controlled approximation errors \cite{vu2019approximate}. Several works also consider a relaxation of the game with  budget constraints in expectation only---which is called the \emph{General Lotto game}---as a relevant model for certain applications \cite{myerson1993incentives,kovenock2020generalizations}. 

In this paper, we analyze the Colonel Blotto game with favoritism by following a similar pattern and make four main contributions as follows: 
\begin{enumerate}
    \item We first consider the model of \emph{all-pay auction with favoritism} ($\FAPA$), where the rule determining the winning bidder is shifted with an additive and a multiplicative parameter. We completely characterize the equilibria in general parameters configurations (with asymmetric items evaluation and no restriction on which bidder has which kind of favoritism). While the \FAPA game was studied in prior works, this result fills a gap in the literature. 
    
    %
    \item We prove the existence of a set of optimal univariate distributions of the \FCB game and give a construction thereof. The main challenge is that it is equivalent to finding a fixed point, but for a complex two-dimensional function for which standard existence results fail to apply. We overcome this obstacle by drawing tools from topology and carefully tailoring them to our particular problem. 
    Based on this core result, we deduce the equilibrium of the \FCB game for particular cases for which it is known how to construct joint distributions from the optimal univariate distributions. For other cases we show that, by applying the rescaling technique of \cite{vu2019approximate}, we can obtain an approximate equilibrium of the $\FCB$ game with negligible approximation error when the number of the battlefields is~large. Finally, for any parameter configuration, we immediately obtain the equilibrium of the relaxed \emph{General Lotto game with favoritism} in which one can simply sample independently on each battlefield.
    %
    \item 
We propose an algorithm that efficiently finds an approximation of the proposed optimal univariate distributions with arbitrarily small error. This improves the scalability of our results upon the naive solution for exact computation (which is exponential in the number of battlefields). Our algorithm is based on approximately solving the two-dimensional fixed-point problem by a dichotomy procedure using a generalization of the intermediate value theorem with the notion of winding number of parametric curves. 
    
    %
    \item We conduct a number of numerical experiments to analyze and illustrate the effect of favoritism in the players' payoffs at equilibrium of the \FCB~game (and of the \FGL game).
\end{enumerate}
\paragraph{\textbf{Related Work}}
%
There is a rich literature on characterizing equilibria of the (classical) Colonel Blotto game. The common approach is to look for a set of \emph{optimal univariate distributions} of the game, and then construct $n$-variate joint distributions \emph{whose realizations satisfy the budget constraints} (in other words, their supports are subsets of the (mixed) strategy sets). These joint distributions are equilibria of the game. Constructing such joint distributions, however, is challenging and equilibria are only successfully characterized in several restricted instances: Colonel Blotto games where players have symmetric budgets \cite{borel1938,gross1950,grosswagner,laslier2002,thomas2017,Boix-Adsera20a}, Colonel Blotto games with asymmetric budgets and two battlefields \cite{macdonell2015} or with any number of battlefields but under assumptions on the homogeneity of battlefields' values \cite{roberson2006,schwartz2014}. The Colonel Blotto game still lacks a complete characterization of equilibrium in its generalized parameters configuration, i.e., with asymmetric budgets and heterogeneous battlefields (see \cite{kovenock2012conflicts} for a survey). An extension of the Colonel Blotto game is studied in \cite{kovenock2020generalizations}, where the two players can have different evaluations of the battlefields. The authors find a set of optimal univariate distributions based on a solution of a fixed-point equation, but they can construct the n-variate equilibrium distribution only in restricted settings. Our work follows a similar pattern in spirit, but the fixed-point equation supporting the optimal univariate distributions is different and harder to solve because it is two-dimensional.

While studying the Colonel Blotto game, many works also consider the corresponding General Lotto game \cite{kovenock2020generalizations,myerson1993incentives}, in which budget constraints are relaxed to hold in expectation. There, an equilibrium can be directly obtained from a set of optimal univariate distributions by independently drawing on each battlefield. In recent work, \cite{vu2019approximate} propose an alternative approach to find solutions of the Colonel Blotto game: it consists of independently drawing on each battlefield and then rescaling to meet the budget constraint with probability one. The authors show that this rescaling strategy (termed independently uniform ($\IU$) strategy) yields an approximate equilibrium with error decreasing with the number of battlefields. 

The problem of constructing sets of optimal univariate distributions in the Colonel Blotto game can be converted into the problem of searching for an equilibrium of an all-pay auction. The state-of-the-art in characterizing equilibria of all-pay auctions is as follows: equilibria of the (classical) all-pay auctions were completely characterized by \cite{baye1994,hillman1989} in games with any number of~bidders. The all-pay auction with favoritism (also referred to as all-pay auctions with head-starts and handicaps and all-pay auctions with incumbency advantages) was studied by \cite{konrad2002investment} but its equilibria were explicitly characterized only in cases where players assess the item with the same value and where both kinds of favoritism are in favor of one player. Therefore, it still lacks an explicit analysis of equilibria with a general configuration of parameters. Note also that the literature on the \FAPA game (and all-pay auctions) goes beyond the study on their equilibria, see e.g., \cite{fu2006theory,li2012contests,pastine2012incumbency,siegel2009all,siegel2014asymmetric} and surveys by \cite{corchon2007theory,fu2019contests,konrad2009}.

Our work is the first to introduce the Colonel Blotto game with pre-allocations and asymmetric effectiveness of players' resources. The only exception is the recent work by \cite{chandan2020showing}, where partial results are obtained with pre-allocations only but from a very different perspective: the authors study a three-stage Colonel Blotto game that allows players to pre-allocate their resources; several conditions where pre-allocating is advantageous are indicated and this result is extended to three-player Colonel Blotto games. Note finally that there is also a growing literature on the discrete Colonel Blotto game, where players' allocations must be integers, see \eg \cite{Ahmadinejad16a,Behnezhad19a,Behnezhad17a,behnezhad2018battlefields,hart2008,hortala2012,vu18a,vu2020}; but this literature did not consider favoritism and these results are based on a very different set of techniques in comparison to that of the game models considered in this work.

\paragraph{\textbf{Notation}}
Throughout the paper, we use bold symbols (e.g., $\boldsymbol{x}$) to denote vectors and subscript indices to denote its elements (e.g., $\boldsymbol{x} = (x_1, x_2, \ldots , x_n)$). We also use the notation $\R^{n}_{>0}:= \braces*{ \boldsymbol{x}^{n} \in \R^{n}: x_i >0, \forall i}$, \mbox{$\R^{n}_{\ge0}:= \braces*{ \boldsymbol{x}^{n} \in \R^{n}: x_i \ge 0, \forall i}$} and $[n] =\{1,2, \ldots , n\}$, for any $n \in \mathbb{N} \backslash \{0\}$. We use the notation $\play$ to denote a player and $-\play$ to indicate her opponent. We also use the standard asymptotic notation $\bigoh$ and its variant $\tilde{\bigoh}$ which ignores logarithmic terms.
Finally, we use $\prob(E)$ to denote the probability of an event $E$ and $\Ex [X] $ to denote the expectation of a random variable~$X$.
	
\paragraph {\textbf{Outline of the Paper}}
The remainder of this paper is organized as follows. \cref{sec:Form} introduces the formulations of the {$\FCB$}, \FGL and \FAPA games. We present our complete equilibria characterization of the \FAPA game in \cref{sec:EquiAPA}. Using this result, in \cref{sec:OptUni_GRCBC}, we prove the existence and show the construction of a set of optimal univariate distributions of the $\FCB$ game. In \cref{sec:Corollary_results}, we  derive several corollary results, concerning the equilibria and approximate equilibria of the \FCB and \FGL games, from these optimal univariate distributions. In \cref{sec:heuristic}, we then present an algorithm that efficiently finds an approximation of the proposed optimal univariate distributions with arbitrarily small errors. In \cref{sec:NumExp}, we conduct several numerical experiments illustrating the effect of two types of favoritism in the \FCB and \FGL~games. Finally, we give a concluding discussion in \cref{sec:conclu}.

%
%
\section{Games Formulations}
\label{sec:Form}

In this section, we define the Colonel Blotto game with favoritism ($\FCB$), and two related models: the General Lotto game with favoritism ($\FGL$) and the all-pay auction with favoritism ($\FAPA$).%

\subsection{Colonel Blotto and General Lotto Games with Favoritism}
\label{sec:FormCB}
The Colonel Blotto game with favoritism (\FCB game) is a one-shot complete-information game between two Players A and B. Each player has a fixed \emph{budget} of resources, denoted $X^A$ and $X^B$ respectively; without loss of generality, we assume that \mbox{$0< X^A \le X^B$}. There are $n$ battlefields ($n \ge 3$). Each battlefield $i \in [n]$ has a value $w_i>0$ and is embedded with two additional parameters: $p_i \in \mathbb{R}$ and $q_i>0$. Knowing these parameters, players compete over the $n$ battlefield by simultaneously allocating their resources. The summation of resources that a player allocates to the battlefields cannot exceed her budget; i.e., the {pure strategy} set of Player ${\play} \in \left\{ A, B \right\}$ is $S^{\play} = \left\{ \boldsymbol{x}^{\play} \in \mathbb{R}_ {\ge 0} ^n: \sum\nolimits_{i = 1}^n {x_i^{\play} \le {X^{\play}}} \right\}$. If Players A and B respectively allocate $x^A_i$ and $x^B_i$ to battlefield $i$, the winner in this battlefield is determined by the following rule: if $x^A_i > q_i x^B_i -p_i$, Player A wins and gains the value $w_i$; reversely, if $x^A_i < q_i x^B_i -p_i$, Player B wins and gains $w_i$; finally, if a tie occurs, i.e., $x^A_i = q_i x^B_i -p_i$, Player A gains $\alpha w_i$ and Player B gains $(1-\alpha) w_i$, where $\alpha \in [0,1]$ is a given parameter.\footnote{We call $\alpha$ the tie-breaking parameter. It can be understood as if we randomly break the tie such that Player A wins battlefield $i$ with probability $\alpha$ while Player B wins it with probability $(1-\alpha)$. This includes all the tie-breaking rules considered in classical CB games found in the literature.} The payoff of each player in the game is the summation of gains they obtain from all the battlefields. Formally, we have the following definition.

\begin{definition}[The \FCB game]
\label{def:FCB}
The Colonel Blotto game with favoritism (with $n$ battlefields), denoted $\FCBn$, is the game with the description above; in particular,  when players A and B play the pure strategies \mbox{$\boldsymbol{x}^A = \parens*{x^A_i}_{i \in [n]}\! \in \! S^A$} and \mbox{$\boldsymbol{x}^B  = \parens*{x^B_i}_{i \in [n]} \in S^B $} respectively, their payoffs are defined as:
\begin{equation*}
    \mathrm{\Pi}_{\FCBn}^A \left( \boldsymbol{x}^A, \boldsymbol{x}^B \right) = \sum \limits_{i \in [n]} {w_i   \be \left( x^A_i, q_i x^B_i \!- \!p_i \right)  } \textrm{ and } \mathrm{\Pi}_{\FCBn}^B \left( \boldsymbol{x}^A, \boldsymbol{x}^B  \right) =\sum \limits_{i \in [n]} {w_i   \left[ 1\!-\! \be \left( x^A_i, q_i x^B_i\! - \!p_i \right) \right]  }.
\end{equation*}
Here, $\be: \mathbb{R}^2_{\ge 0} \rightarrow [0,1]$, termed as the Blotto function, is defined as follows: $\be\left( {x,y} \right)=1$ if $x>y$, $\be\left( {x,y} \right)=\alpha$ if $x=y$ and $\be\left( {x,y} \right)=0$ if $x<y$. 
  %
\end{definition}

A \emph{mixed strategy} of a player, say $\play \in \braces*{A,B}$, in $\FCBn$ is an $n$-variate distribution such that any pure strategy drawn from it is an $n$-tuple satisfying the corresponding budget constraint of player~$\play$. We reuse the notations $\mathrm{\Pi}_{\FCBn}^A \left(\sigma_A, \sigma_B \right)$ and $\mathrm{\Pi}_{\FCBn}^B \left(\sigma_A, \sigma_B \right)$ to denote the payoffs of Players A and B when they play the mixed strategies $\sigma_A$ and~$\sigma_B$ respectively. Note that to lighten the notation $\FCBn$, we include only the subscript $n$ (the number of battlefields) and omit other parameters involved in the definition of the game (including $X^A, X^B, \alpha, w_i, p_i, q_i, \forall i \in [n]$). 

The game $\FCBn$ extends the classical Colonel Blotto game by including the favoritism that a player may have in battlefield $i$ through the parameters $p_i$ and $q_i$; which can be interpreted as~follows: 
\begin{trivlist}
	\item[$(i)$] $p_i$ represents the difference between pre-allocations that players have at battlefield $i$ before the game begins (note that pre-allocations are not included in the players' budget $X^A$ and $X^B$). If $p_i >0$, Player A's pre-allocation at battlefield $i$ is larger; if $p_i<0$, Player B has a larger pre-allocation.
	\item[$(ii)$] $q_i$ represents the asymmetry in the effectiveness of players' resources (\emph{not} including the pre-allocations). Specifically, in battlefield $i$, each unit of Player~B's resource is worth $q_i$ units of Player A's resource. If $0< q_i < 1$, Player A's resource is more effective than that of Player B; reversely, if $  q_i > 1$, Player B's resource is more effective. %
\end{trivlist}
%
Note that if $p_i =0$ and $q_i = 1, \forall i \in [n]$, the game $\FCBn$ coincides with the classical CB game. Unlike many works in the literature on classical CB game, in the $\FCBn$ game defined above, we do not make assumptions on the symmetry in players' budgets or on the homogeneity of the battlefields'~values.

For the sake of conciseness, in the remainder, we consider \FCB under the following assumptions:

\begin{assumption} \label{assum:1}
    $\sum_{i \in [n]} {\parens*{q_i X^B - p_i}} \ge X^A$ and $\sum_{i \in [n]} { \parens*{X^A + p_i} /q_i } \ge X^B $. 
\end{assumption}
\begin{assumption}\label{assum:2}
    For any $i \in [n]$, $\parens*{q_i X^B - p_i} \ge 0$ and $\parens*{X^A + p_i} /q_i \ge 0$.
\end{assumption}
These assumptions are used simply to exclude trivial cases where one player has too strong favoritism in one (or all) battlefields. Indeed, if \cref{assum:1} is violated, there exists trivial pure equilibria.\footnote{If $\sum_{i \in [n]} {\parens*{q_i X^B - p_i}} < X^A$, by allocating $q_i X^B - p_i + \varepsilon $ to battlefield $i$ ($\varepsilon$ is an arbitrarily small number), Player A guarantees to win all battlefields regardless of Player B' allocations. If $\sum_{i \in [n]} { \parens*{X^A + p_i} /q_i } < X^B $, by allocating $\parens*{X^A + p_i} /q_i + \varepsilon$ to battlefield $i$, Player B guarantees to win all battlefields.} On the other hand, if in battlefield $i^* \in [n]$, $\parens*{q_{i^*} X^B - p_{i^*}} < 0$ (resp. $\parens*{X^A + p_{i^*}} /q_{i^*} < 0$), then by allocating 0, Player A (resp. Player B) guarantees to win this battlefield regardless of her opponent's allocation. Therefore, if $\FCBn$ has an battlefield $i^*$ violating \cref{assum:2}, an equilibrium of $\FCBn$ is simply the strategies where both players allocate 0 to battlefield $i^*$ and play an equilibrium of the game $\mathcal{CB}^F_{n-1}$ having the same setting as $\FCBn$ but excluding battlefield $i^*$. Note that analogous assumptions (when $p_i = 0$ and $q_i = 1, \forall i$) are found in other works considering the classical CB game (see \eg \cite{roberson2006} and Figure 1 in \cite{kovenock2020generalizations}).

Next, similar to the definition of the General Lotto game obtained from relaxing the classical CB game, for each instance of the \FCB game, we define an instance of the General Lotto with favoritism ($\FGL$) where the budget constraint is requested to hold only in expectation. Formally:
\begin{definition}[The \FGL game]
    \label{def:FGL}
    The General Lotto game with favoritism (with $n$ battlefields), denoted $\FGLn$, is the game with the same setting and parameters as the $\FCBn$ game, but where a mixed strategy of Player $\play \in \{A,B\}$ in $\FGLn$ is an $n$-variate distribution with marginal distributions $(F^{\play}_i)_{i \in [n]}$ such that \mbox{$\sum_{i \in [n]} {\Ex_{x_i \sim F^{\play}_i }[ x_i ] } \le X^{\play}$}. 
\end{definition}

We finally define the notion of Optimal Univariate Distributions in the $\FCB$ game. This notion is of great importance in studying equilibria of the \FCB game since intuitively, they are the candidates for the marginals of the equilibria. Formally:
\begin{definition}[Optimal Univariate Distributions (OUDs)] \label{def:OUD}
    $\left\{F^A_i, F^B_i: i \in [n] \right\}$ is a set of \emph{OUDs} of the game $\FCBn$ if the following conditions are satisfied:
    \begin{enumerate}[label=(C.\arabic*),ref=\textit{(C.\arabic*)}]
        \item  the supports of $F^A_i, F^B_i$ are subset of $\mathbb{R}_{\ge 0}$, \label{condi:OUD1}
        \item \mbox{$\sum_{i \in [n]} {\Ex_{x_i \sim F^\play_i }[ x_i ] } \le X^{\play}$}, $\play \in \{A,B\}$, \label{condi:OUD2}
        \item if Player $\play$ draws her allocation to battlefield $i$ from $F^\play_i$, $\forall i \in [n]$, Player $-\play$ has no pure strategy inducing a better payoff than when she draws her allocation to battlefield $i$ from $F^{-\play}_i$, $\forall i \in [n]$.\label{condi:OUD3}
    \end{enumerate}
\end{definition}

\subsection{All-pay Auctions with Favoritism}
\label{sec:FormAPA}
All-pay auctions with favoritism ($\FAPA$) have been studied in the literature under different sets of assumptions (see \cref{sec:Intro}). For the sake of coherence, in this section, we re-define the formulation of the \FAPA game using our notation as follows:

In the \FAPA game, two players, A and B, compete for a common item that is evaluated by each player with a value, denoted $u^A$ and $u^B$ respectively ($u^A, u^B > 0$).\footnote{The case where either $u^A =0$ or $u^B =0$ is trivial (there exist trivial pure equilibria) and thus, is omitted.} The item is embedded with two additional parameters: $p \in \mathbb{R}$ and $q>0$. Players simultaneously submit their \emph{bids} $x^A, x^B \ge 0$ (unlike in the $\FCB$ game, players can bid as large as they want in $\FAPA$). If $x^A > q x^B - p$, Player A wins the item and gains the value $u^A$; if $x^A < q x^B - p$, Player B wins and gains the value $u^B$; and in case of a tie, i.e., $x^A = q x^B -p$, Player A gains $\alpha u^A$ and Player B gains $(1-\alpha) u^B$ ($\alpha \in [0,1]$). Finally, \emph{both players} pay their bids. 
\begin{definition}[The \FAPA game] \label{def:APA-F}
	\emph{\FAPA} is the game with the above description; in particular, when the players A and B bid $x^A$ and $x^B$ respectively, their payoffs are \mbox{$\mathrm{\Pi}^A_{\FAPA} \left( x^A, x^B \right) \!= \!u^A \be \left( x^A, q x^B \!-\! p \right) \!-\! x^A$} and \mbox{$\mathrm{\Pi}^B_{\FAPA} \left( x^A, x^B \right) \!= \!u^B [1-\be \left( x^A, q x^B \!-\! p \right)] \!-\! x^B$}. Here, the function $\be$ is defined in \cref{def:FCB}.
\end{definition}
The formulation of the \FAPA game presented above differs from classical all-pay auctions by the parameters $p$ and $q$. If $p>0$, Player A has an \emph{additive advantage} in competing to win the item and if $p<0$, Player B has this favoritism; likewise, when $0<q<1$, Player A has a \emph{multiplicative favoritism} to compete for the item and when $q>1$, it is in favor of Player B. Our formulation of \FAPA is more general than the models (with two players) considered by previous works in the literature. If $p=0$, $q=1$ and $\alpha=1/2$, the $\FAPA$ game coincides to the classic two-bidder (first-price) all-pay auction (e.g., in \cite{baye1994,hillman1989}). If $u^A = u^B$, $\alpha = 1/2$, $p>0$ and $0<q \le 1$ (i.e., Player A has both advantages), $\FAPA$ coincides with the framework of all-pay contests with incumbency advantages considered in \cite{konrad2002investment}. Moreover, we also define \FAPA with a generalization of the tie-breaking rule (with the parameter $\alpha$ involving in the function $\be$) covering other tie-breaking rules considered in previous works. Finally, our definition of $\FAPA$ and its equilibria characterization (see \cref{sec:EquiAPA}) can also be extended to cases involving more than two players/bidders; in this work, we only analyze the two-player $\FAPA$ since it relates directly to the \FCB game, which is our main~focus.

%
%
\section{Equilibria of All-pay Auctions with Favoritism}
\label{sec:EquiAPA}

In this section, we characterize the equilibrium of the $\FAPA$ game. The closed-form expression of the equilibrium depends on the relation between the parameters $u^A, u^B,p$ and $q$. We present two groups of results corresponding to two cases: $p \ge 0$ (\cref{sec:APAPos}) and $p<0$ (\cref{sec:APANeg}).

%
%
\subsection{Equilibria of $\FAPA$ with $p \ge 0$.}
\label{sec:APAPos}
We first focus on the case where $p \ge 0$; in other words, Player A has an additive advantage. In equilibrium, players choose their bids according to uniform-type distributions which depend on the relation between $u^A, u^B, p$ and $q$. Particularly, we obtain the following theorem:
\begin{theorem}
\label{theo:positive}
In the $\FAPA$ game where $p \ge 0$, we have the following results:
	
\begin{itemize}
    \item[$(i)$] If $q u^B - p \le 0$, there exists a unique pure equilibrium where players' bids are $x^A = x^B = 0$ and their equilibrium payoffs are $\mathrm{\Pi}^A_{\FAPA} = u^A$ and $\mathrm{\Pi}^B_{\FAPA} = 0$ respectively.
	\item[($ii$)] If $0 < q u^B - p \le u^A$, there exists no pure equilibrium; there is a unique mixed equilibrium where Player A (resp. Player B) draws her bid from the distribution $\Aiip$ (resp. $\Biip$) defined as follows.
	\begin{align}
	& \Aiip(x) =   \left\{ \begin{array}{l}
	\frac{p}{q u^B} + \frac{x}{q u^B}, \forall x \in \left[ 0, q u^B \!-\! p \right], \\
	1                  \qquad \qquad, \forall x > q u^B \!-\!p, 
	\end{array} \right.
	\textrm{and} 
	& \Biip(x) =   \left\{ \begin{array}{l}
	1 - \frac{q u^B}{u^A} + \frac{p}{u^A}, \forall x \in \left[ 0, \frac{p}{q} \right) \\
	1 - \frac{q u^B}{u^A} + \frac{q \cdot x}{u^A}, \forall x \in \left[\frac{p}{q}, u^B \right], \\
	1 \qquad \qquad \qquad, \forall x > u^B.
	\end{array} \right. \label{eq:A+Def}
	\end{align}
	In this mixed equilibrium, players' payoffs are $\mathrm{\Pi}^A_{\FAPA} = u^A - q u^B + p$ and $\mathrm{\Pi}^A_{\FAPA} =0$.
	\item[$(iii)$]  If $q u^B - p > u^A$, there exists no pure equilibrium; there is a unique mixed equilibrium where Player A (resp. Player B) draws her bid from the distribution $\Aiiip$ (resp. $\Biiip$) defined as follows.
	\begin{align}
	& \Aiiip(x) =   \left\{ \begin{array}{l}
	1- \frac{u^A}{q u^B} + \frac{x}{q u^B}, \forall x \in \left[ 0, u^A \right], \\
	1          \qquad  \qquad \qquad, \forall x > u^A, 
	\end{array} \right.
	\textrm{ and } 
	& \Biiip(x) =   \left\{ \begin{array}{l}
	0 \qquad \qquad, \forall x \in \left[ 0, \frac{p}{q} \right) \\
	- \frac{p}{u^A} + \frac{q \cdot x}{u^A}, \forall x \in \left[\frac{p}{q}, \frac{u^A + p}{q} \right], \\
	1 \qquad \qquad, \forall x > \frac{u^A + p}{q}.
	\end{array} \right. \label{eq:bFA+Def}
	\end{align}
	In this mixed equilibrium, players' payoffs are $\mathrm{\Pi}^A_{\FAPA}= 0$ and $\mathrm{\Pi}^B_{\FAPA} = u^B - (u^A+p)/q$.
			
	\end{itemize}
\end{theorem}
	
A formal proof of \cref{theo:positive} can be found in \ref{appen:proofAPAPos}; here we discuss an intuitive interpretation of the result. First, note that no player has an incentive to bid more than the value at which she assesses the item, otherwise she is guaranteed a negative payoff. Then, the condition in Result~$(i)$ of \cref{theo:positive} indicates that Player A has too large advantages such that she always wins regardless of her own bid and Player B's bid, hence it is optimal for both players to bid zero (see the proof for the case $q u^B - p =0$). The condition in Result~$(ii)$ of \cref{theo:positive} gives Player A a favorable position: she can guarantee to win with a non-negative payoff by bidding $u^A$ knowing that Player B will not bid more than $u^B$; reversely, the condition in Result~$(iii)$ implies that Player B has a favorable position: by bidding $u^B$, she guarantees to win with a non-negative payoff since Player A will not bid more than $u^A$. Importantly, in Result~$(ii)$ of this theorem, as long as the condition $0 < q u^B - p \le u^A$ is satisfied, when $p$ increases (and/or $q$ decreases), the equilibrium payoff of Player A increases. This is in coherence with the intuition that when Player A has larger advantages, she can gain more. However, if $p$ is too large (and/or $q$ is too small) such that the condition in Result~$(i)$ of \cref{theo:positive} is satisfied, Player B gives up totally and Player A gains a fixed payoff ($u^A$) even if $p$ keeps increasing (and/or $q$ keeps decreasing). A similar intuition can be deduced for Player B and~Result~$(iii)$.

\begin{figure*}[htb!]
\centering
\subfloat[$\FAPA$ instance with $u^A = 4$, $u^B = 2$, \mbox{$p = 1.5$}, $q =1.5$ (i.e., \mbox{$0 \le q u^B-p < u^A$}). ]{
	\begin{tikzpicture}
	\node (img){\includegraphics[height=0.20\textwidth]{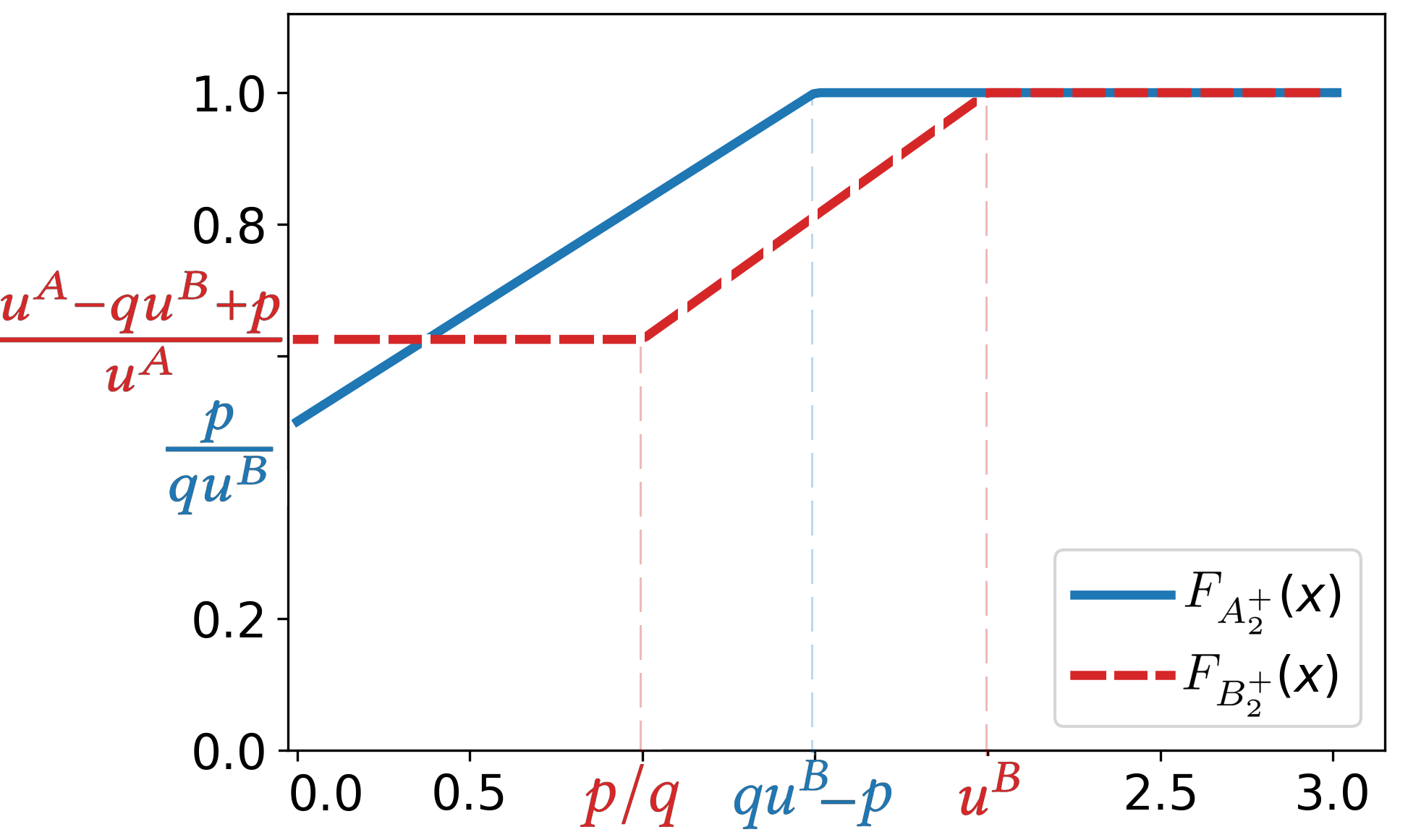}};
	\node[below=of img, node distance=0cm, yshift=1.1cm, xshift = 0.5cm] {\scriptsize$x$};
	\end{tikzpicture}
}   
\qquad \qquad
\subfloat[$\FAPA$ instance with $u^A = 2$, \mbox{$u^B = 4$}, \mbox{$p = 1$}, $q =1$ (i.e., \mbox{$ q u^B-p > u^A$}). ]{
	\begin{tikzpicture}
	\node (img){\includegraphics[height=0.205\textwidth]{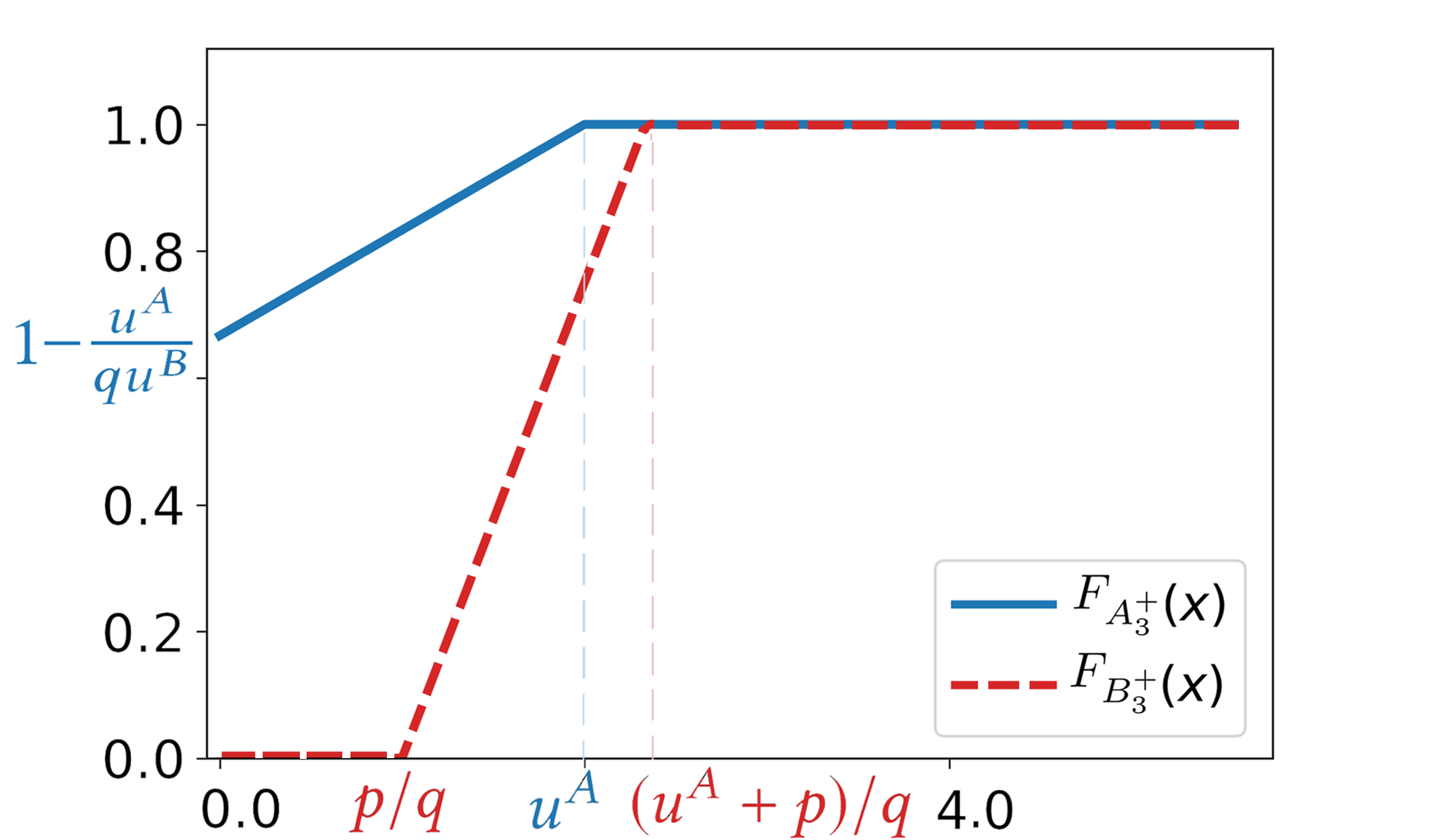}};
	\node[below=of img, node distance=0cm, yshift=1.1cm,xshift = 0.5cm] {\scriptsize$x$};
	\end{tikzpicture}
}%
\caption{The mixed equilibrium of the $\FAPA$ with $p\ge 0$.}
\label{fig:APA_pos}
\end{figure*}

We now turn our focus to the distributions $\Aiip, \Biip,\Aiiip$ and $\Biiip$ in Results~$(ii)$ and~$(iii)$ of \cref{theo:positive}. First, note that the superscript ${}^+$ in the notations of these distributions simply refers to the condition $p \ge 0$ being considered (to distinguish it with the case where $p < 0$ presented below) while the subscript index ($2$ or $3$) indicates that these distributions correspond to Results~$(ii)$ or~$(iii)$. These distributions all relate to uniform distributions: $\Aiip$ is the distribution placing a non-negative probability mass at zero, and then uniformly distributing the remaining mass on the range $\left(0, q u^B-p \right]$ while $\Biip$ places a non-negative mass at zero, then uniformly distributes the remaining mass on $\left[p/q,u^B \right]$; similarly, $\Aiiip$ places a mass at zero and uniformly distributes the remaining mass on $\left( 0, u^A \right]$ while $\Biiip$ is the uniform distribution on $\left[p/q, (u^A+p)/q \right]$; see  an illustration in \cref{fig:APA_pos}. Note finally that \cref{theo:positive} is consistent with results in the restricted cases of \FAPA presented in \cite{konrad2002investment} (where $u^A = u^B$, $p \ge 0$, $0 < p <1$ and $\alpha =1/2$) and with results for the classical APA from \cite{baye1996all,hillman1989} (i.e., when $p =0$, $q = 1$, $\alpha = 1/2$); see \ref{appen:preli} where we reproduce these results to ease comparison. 
%
%
%


\subsection{The $\FAPA$ game with $p < 0$}
\label{sec:APANeg}
We now consider the $\FAPA$ game in the case $p < 0$. We first define $p^{\prime} = -p/q$ and $q^{\prime} = 1/q$. Since $p<0$, we have $p^{\prime} >0$. Moreover, \mbox{$\be \left(x^A, q x^B - p \right) = \be \left( (x^A +p)/q, x^B\right) = \be \left(  q^{\prime} x^A - p^{\prime}, x^B \right)$} for any $x^A, x^B$. Therefore, the $\FAPA$ game with $p < 0$ (and $q>0$) is equivalent to an \FAPA with $p^{\prime} >0 $ (and $q^{\prime} >0$) in which the roles of players are exchanged. Applying \cref{theo:positive} with $p^{\prime}>0$ (and $q^{\prime} >0$), we can easily deduce the equilibrium of the \FAPA with $p<0$. Due to the limited space, we only present this result in \ref{sec:appen:APAneg}.
	

%
%
\section{Optimal Univariate Distributions of the Colonel Blotto Game with Favoritism}
\label{sec:OptUni_GRCBC}

The notion of optimal univariate distributions plays a key role in the equilibrium characterization of the Colonel Blotto game and its variants. In this section, we prove the existence of, and construct a set of optimal univariate distributions of the $\FCB$ game. This is the core result of our work.

As discussed in \cref{sec:Intro}, a classical approach in the Blotto games literature is to reduce the problem of constructing optimal univariate distributions (OUDs) to the problem of finding the equilibria of a set of relevant all-pay auction instances---each corresponding to players' allocations in one battlefield. The main question then becomes: \emph{which set of all-pay auction instances should we consider in order to find OUDs of the $\FCB$ game?} Naturally, from their formulations in \cref{sec:FormCB} and \cref{sec:FormAPA}, a candidate is the set of $\FAPA$ games in which the additive and multiplicative advantages of the bidders correspond to the parameters representing the pre-allocations and asymmetric effectiveness of players in each battlefield of the $\FCB$ game. The (uniquely defined) equilibrium distributions of these $\FAPA$ games satisfy Condition~\ref{condi:OUD3} in \cref{def:OUD} (\ie they are the marginal best-response against one another in the corresponding battlefield of the $\FCB$ game). Now, we only need to define the items' values in these $\FAPA$ games in such a way that their corresponding equilibrium distributions also hold Condition~\ref{condi:OUD2} in \cref{def:OUD} (\ie they satisfy the budget constraints in expectation). To do this, we first parameterize the items' values assessed by the bidders in the involved $\FAPA$ games, then we match these parameters with equations defining Condition~\ref{condi:OUD2}. Summarizing the above discussion, we define a particular set of distributions (on $\Rpos$) as~follows:
\begin{definition}	\label{def:OptDis_GRCBC}
		Given a game $\FCBn$ and a pair of positive real numbers $\kappa = (\gA, \gB)\! \in\! \mathbb{R}^{2}_{> 0}$, for each $i\! \in\! [n]$, we define $\Ai$ and $\Bi$ to be the pair of distributions that forms the equilibrium of the $\FAPA$ game with $p:= p_i$, $q:= q_i$, $u^A:= w_i \cdot \gA$ and $u^B:= w_i \cdot \gB$. The explicit formulas of $\Ai$ and $\Bi$ are given in \cref{tab:Opt_Uni_GRCBC} for each configuration of $w_i, p_i, q_i, \gA$ and $\gB$.	
	\end{definition}
	\begingroup
	\renewcommand{\arraystretch}{1.2} 
	\begin{table}[tb!]
	\footnotesize
		\centering
		\caption[]{$\Ai$ and $\Bi$ corresponding to $\kappa  =  (\gA, \gB)$ and a $\FCBn$ game. The notation $I^+_j(\gA,\gB)$ and $I^-_j(\gA,\gB)$ for $j=1,2,3$ denote the set of indices of battlefields satisfying the corresponding conditions; for example, $I^+_1(\gA, \gB)=\{i  \in  [n]:p_i \ge 0, q_i w_i \gB - p_i \le 0\}$ and $\forall i \in I^+_1 (\gA, \gB)$, $\Ai(x)  = 1$, $\Bi(x)  = 1, \forall x~\ge~0$.} \label{tab:Opt_Uni_GRCBC}
		\begin{tabular}{ |l|l|l| } 
			\hline
			\rowcolor{Gray}
			Indices Sets & Conditions & Definition  \\
			\hline
			$I^+_1(\gA, \gB)$ &  {$i  \in  [n]:p_i \ge 0$, $q_i w_i \gB - p_i \le 0$}      & $\Ai(x)  = 1, \forall x \ge 0$ and $\Bi(x)  = 1, \forall x \ge 0$.   \\ 
			\hline
			%
			$I^+_2(\gA, \gB)$ & \multirow{2}{3cm}{$i  \in  [n]: p_i \ge 0$, \mbox{$0 < q_i w_i \gB  -  p_i  \le  w_i \gA$}}  
			& $\Ai(x)  =  \left\{ \begin{array}{l}
			\frac{p_i}{q_i w_i \gB} +  \frac{x}{q_i w_i \gB}, \forall x  \in  \left[ 0, q_i w_i \gB  -  p_i \right], \\
			1              \qquad  \quad  \qquad \qquad, \forall x > q_i w_i \gB -p_i, 
			\end{array} \right.$     \\      
			& & $\Bi(x)  = \left\{ \begin{array}{l}
			1 - \frac{q_i \gB}{\gA} +  \frac{p_i}{w_i \gA}, \forall x  \in  \left[ 0, \frac{p_i}{q_i} \right), \\
			1 - \frac{q_i \gB}{\gA} +  \frac{q_i \cdot x}{w_i \gA}, \forall x  \in  \left[\frac{p_i}{q_i}, w_i \gB \right], \\
			1 \qquad \quad \qquad \qquad, \forall x > w_i \gB.
			\end{array} \right.$   \\ 
			\hline
			\ %
			$I^+_3(\gA, \gB)$ & \multirow{2}{3cm}{$i  \in  [n]: p_i \ge 0$, \mbox{$q_i w_i \gB - p_i > w_i \gA$}}  
			& $\Ai(x)  =    \left\{ \begin{array}{l}
			1- \frac{\gA}{q_i \gB} +  \frac{x}{q_i w_i \gB}, \forall x  \in  \left[ 0, w_i \gA \right], \\
			1         \qquad \qquad  \qquad \qquad, \forall x > w_i \gA, \end{array} \right.$\\
			& & $\Bi(x)  =    \left\{ \begin{array}{l}
			0 \qquad \qquad \qquad, \forall x  \in  \left[ 0, \frac{p_i}{q_i} \right), \\
			- \frac{p_i}{w_i \gA} +  \frac{q_i \cdot x}{w_i \gA}, \forall x  \in  \left[\frac{p_i}{q_i}, \frac{w_i \gA +  p_i}{q_i} \right], \\
			1 \qquad \qquad \qquad, \forall x > \frac{w_i \gA +  p_i}{q_i}.
			\end{array} \right.  $ \\ 
			\hline
			$I^-_1(\gA, \gB)$  & {$i  \in  [n]: p_i <0$, ${w_i \gA}  \le - p_i$}      & $\Ai(x)  = 1, \forall x \ge 0$ and $\Bi(x)  = 1, \forall x \ge 0$. \\ 
			\hline
			%
			$I^-_2(\gA, \gB)$ & \multirow{2}{3cm}{$i  \in  [n]: p_i <0$, \mbox{$-p_i< {w_i \gA} \le q_i w_i \gB  -  p_i$}}  
			& $\Ai(x)  = \left\{ \begin{array}{l}
			1-\frac{\gA}{q_i \gB} - \frac{p_i}{q_i w_i \gB}, \forall x  \in  \left[ 0, - p_i \right), \\
			1-\frac{\gA}{q_i \gB} +  \frac{x}{q_i w_i \gB}, \forall x  \in  \left[ -p_i, w_i \gA \right],\\
			1             \qquad  \quad   \qquad \qquad, \forall x > w_i \gA, 
			\end{array} \right.$         \\ 
			&  & $\Bi(x) : \left\{ \begin{array}{l}
			- \frac{p_i}{w_i \gA} +  \frac{q_i \cdot x}{w_i \gA}, \forall x  \in  \left[ 0, \frac{w_i \gA +  p_i}{q_i} \right], \\
			1 \qquad \qquad \qquad, \forall x > \frac{w_i \gA +  p_i}{q_i}.
			\end{array} \right.$   \\ 
			\hline
			\ %
			$I^-_3(\gA, \gB)$ & \multirow{2}{3cm}{$i  \in  [n]: p_i <0$, \mbox{${w_i \gA } > q_i w_i \gB  -  p_i$}}  
			& $\Ai(x)  =   \left\{ \begin{array}{l}
			0 \qquad \quad \qquad \qquad, \forall x  \in  \left[ 0, -p_i \right), \\
			\frac{p_i}{q_i w_i \gB} +  \frac{x}{q_i w_i \gB}, \forall x  \in  [  - p_i, q_i w_i \gB  -  p_i],\\
			1 \qquad \quad \qquad , \forall x > q_i w_i \gB -p_i,
			\end{array} \right.$ \\ 
			& & $\Bi(x)  =   \left\{ \begin{array}{l}
			1- \frac{q_i \gB}{\gA} +  \frac{q_i \cdot x}{w_i \gA}, \forall x  \in  \left[0, w_i \gB \right], \\
			1 \qquad \quad \qquad \qquad, \forall x > w_i \gB.
			\end{array} \right. $\\
			\hline
		\end{tabular}
	\end{table}
\endgroup
We consider the following system of equations (with variables~$\gA, \gB$):
\begin{align}
	\left\{ \begin{array}{l}
	\sum_{i \in [n]} \Ex_{x \sim \Ai} \left[ x  \right] = X^A, \\
	\sum_{i \in [n]} \Ex_{x \sim \Bi} \left[ x  \right] = X^B.
	\end{array} \right.
	\label{eq:system_Ex}
\end{align}
By defining the sets $\Iplus:= \braces*{j: p_j \ge 0, {\gB > \frac{p_j}{q_j w_j }} }$, \mbox{$\Iminus := \braces*{j: p_j < 0, {\gA > \frac{-p_j}{w_j }} }$} and the term \mbox{$\hi := \min \{ q_i w_i \gB, w_i \gA + p_i \}$}, and computing the expected values of $\Ai$ and $\Bi$ for $i \in [n]$ (see the details in \ref{sec:appen_proof_OUD}), we can rewrite System~\eqref{eq:system_Ex} as:
\begin{align}
	\left\{ \begin{array}{l}
	g^A (\gA, \gB) = 0, \\
	g^B (\gA,\gB) = 0, 
	\end{array} \right.
	\label{eq:system_f}
\end{align}
where $g^A, g^B: \mathbb{R}^2 \rightarrow \mathbb{R}$ are the following functions (for each given instance of the $\FCBn$~game):
\begin{subequations}
    \begin{align}
    	g^A (\gA,\gB )   &\!=\!   \sum \nolimits_{i  \in  \Iplus} \frac{ \bracks*{\hi}^2 \!-\!  {p_i}^2 }{2 q_i w_i } + \sum\nolimits_{i  \in  \Iminus} \frac{ \bracks*{\hi}^2 }{2 q_i w_i} - X^B \gA, \label{eq:fA} \\
    	g^B (\gA,\gB)   &\!=\!   \sum\nolimits_{i  \in  \Iplus} \frac{ \left[\hi \!-\! p_i\right]^2}{2 q_i w_i } \!+\!  \sum\nolimits_{i  \in  \Iminus} \frac{ \left[ \hi \!-\!p_i \right]^2 \!-\! p_i^2 }{2 q_i w_i}  \!-\! X^A \gB. \label{eq:fB}
    \end{align}
\end{subequations}
With these definitions, we can state our main result as the following theorem:
\begin{theorem} \label{theo:OUDs}
    For any game $\FCBn$,
        \begin{itemize}
            \item[$(i)$] There exists a positive solution $\kappa = (\gA, \gB) \in \mathbb{R}^2_{>0}$ of System~\eqref{eq:system_f}.
            \item[$(ii)$] For any positive solution \mbox{$\kappa = (\gA, \gB)\! \in \! \mathbb{R}^2_{>0}$} of System~\eqref{eq:system_f}, the corresponding set of distributions \mbox{$\left\{\Ai,\Bi, i \in [n] \right\}$} from \cref{def:OptDis_GRCBC} is a set of optimal univariate distributions of~$\FCBn$.
        \end{itemize}
\end{theorem}
\cref{theo:OUDs} serves as a core result for other analyses in this paper; it is interesting and important in several aspects. First, it shows that in any instance of the $\FCB$ game, there always exists a set of OUDs with the form given in \cref{def:OptDis_GRCBC}. Second, by comparing these OUDs of the $\FCB$ game with that of the classical Colonel Blotto game (see \eg results from \cite{kovenock2020generalizations}), we can see how the pre-allocations and the asymmetric effectiveness affect players' allocations at equilibrium; we will return to this point in \cref{sec:NumExp} with more discussions. Moreover, as candidates for marginals of the equilibrium of the $\FCB$ game (in cases where it exists), the construction of such OUDs allows us to deduce a variety of corollary results concerning equilibria and approximate equilibria of $\FCB$ and $\FGL$ games (we present and discuss them in \cref{sec:Corollary_results} and~\cref{sec:heuristic}). 

We give a detailed proof of \cref{theo:OUDs} in \ref{sec:appen_proof_OUD} and only discuss its main intuition here. First, we can prove Result~$(ii)$ of \cref{theo:OUDs} by simply checking the three conditions of \cref{def:OUD} defining the OUDs of the $\FCB$ game: it is trivial that for $\kappa \in \mathbb{R}^2_{>0}$, the supports of $\Ai, \Bi, \forall i \in [n]$ are subsets of $\Rpos$ (thus, they satisfy Condition~\ref{condi:OUD1}); moreover, if \mbox{$\kappa=(\gA,\gB) \in \R^{2}_{>0}$} is a solution of System~\eqref{eq:system_f}, then it is a solution of System~\eqref{eq:system_Ex} and trivially, $\Ai, \Bi, \forall i \in [n]$ satisfy Condition~\ref{condi:OUD2};\footnote{Note that due to the ``use-it-or-lose-it" rule of the $\FCB$ game, among the existing equilibria (if any), there exists at least one equilibrium in which players use all their resources, thus we only need to consider the equality case of Condition~\ref{condi:OUD2}.} finally, we can check that for each configuration of $w_i, p_i, q_i, \gA, \gB$ (given that $\gA, \gB >0$), the distributions $\Ai, \Bi$ form the equilibrium of the corresponding $\FAPA$ game; thus \mbox{$\Ai,\Bi, i \in [n]$} satisfy Condition~\ref{condi:OUD3}.

On the other hand, proving Result~$(i)$ of \cref{theo:OUDs} is a challenging problem in itself: $g^A$ and $g^B$ are not simply quadratic functions of $\gA$ and $\gB$ since these variables also appear in the conditions of the involved summations. Note that the particular instance of System~\eqref{eq:system_f} where $p_i=0, q_i =1, \forall i \in [n]$ coincides with a system of equations considered in \cite{kovenock2020generalizations} for the case of the classical Colonel Blotto game (without favoritism). Proving the existence of positive solutions of this system \emph{in this particular case} can be reduced to showing the existence of positive solutions of a real-valued 1-dimensional function (with a single variable $\lambda = \gA/\gB$) which can be done by using the intermediate value theorem (see \cite{kovenock2020generalizations}). \emph{In the general case of the $\FCB$ game and System~\eqref{eq:system_f}}, this approach \emph{cannot} be applied due to the involvement of arbitrary parameters $p_i, q_i, i \in [n]$. Alternatively, one can see our problem as proving the existence of a fixed-point in $\R^2_{>0}$ of the function \mbox{$F:\mathbb{R}^2 \rightarrow \mathbb{R}^2$} such that $F \parens*{\gA,\gB} = \parens*{g^A \parens*{\gA,\gB}/X^B + \gA, g^B \parens*{\gA,\gB}/X^A + \gB}$. This direction is also challenging since the particular formulations of $g^A$ and $g^B$ (thus, of $F$) does not allow us to use well-known tools such as Brouwer's fixed-point theorem \cite{brouwer1911abbildung} and/or Poincaré-Miranda theorem \cite{kulpa1997poincare}. 

\begin{figure*}[htb!]%
\scriptsize 
		\centering
		\subfloat[][\scriptsize Output plane as a heatmap to be used as reference]{{\includegraphics[height=0.16\textwidth]{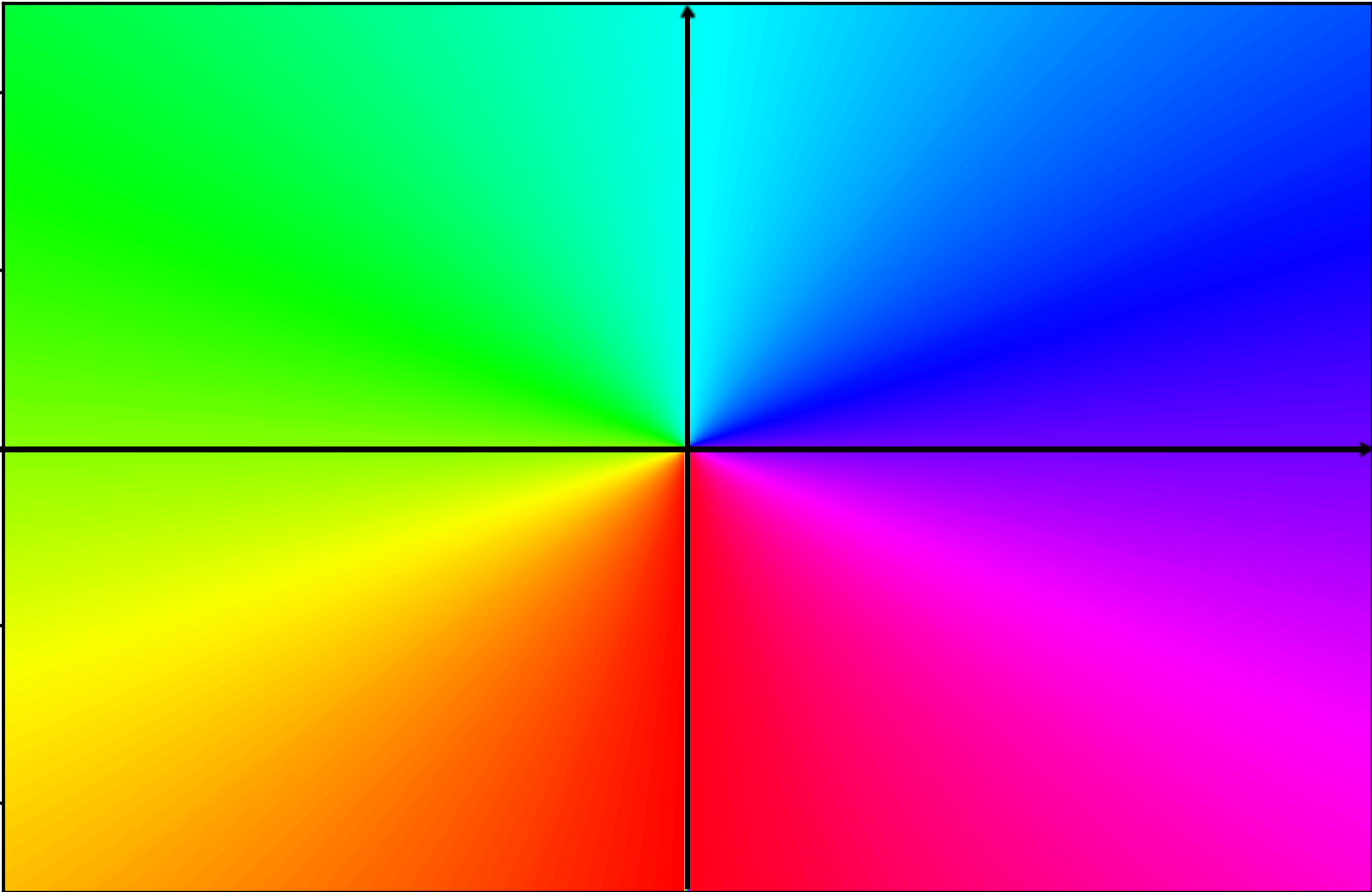} }}%
		\qquad \qquad
		\subfloat[][ \scriptsize Input plane with the colors of the values of function $G$]{
			\begin{tikzpicture}
			\node (img){\includegraphics[height=0.14\textwidth]{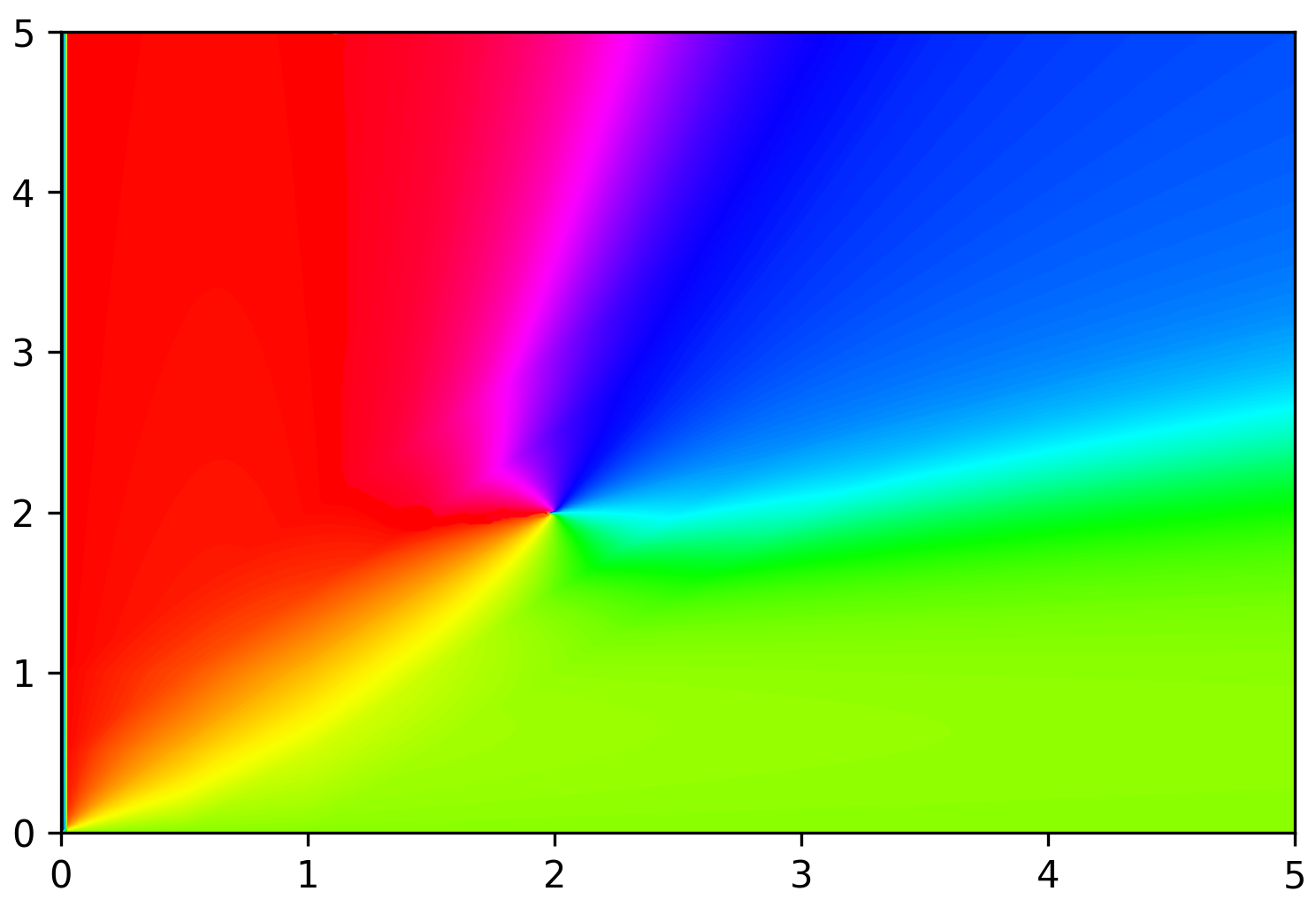}};
			\node[below=of img, node distance=0cm, yshift=1.2cm, xshift = 0.20cm] {\scriptsize$\gA$};
			\node[left=of img, node distance=-0.2cm, anchor=center,yshift=-0cm,xshift = 0.9cm] {\scriptsize$\gB$};
			\end{tikzpicture}	
		}%
		\qquad \qquad
		\subfloat[][\scriptsize The image via function $G$ of the rectangle~$D$]{{\includegraphics[height=0.16\textwidth]{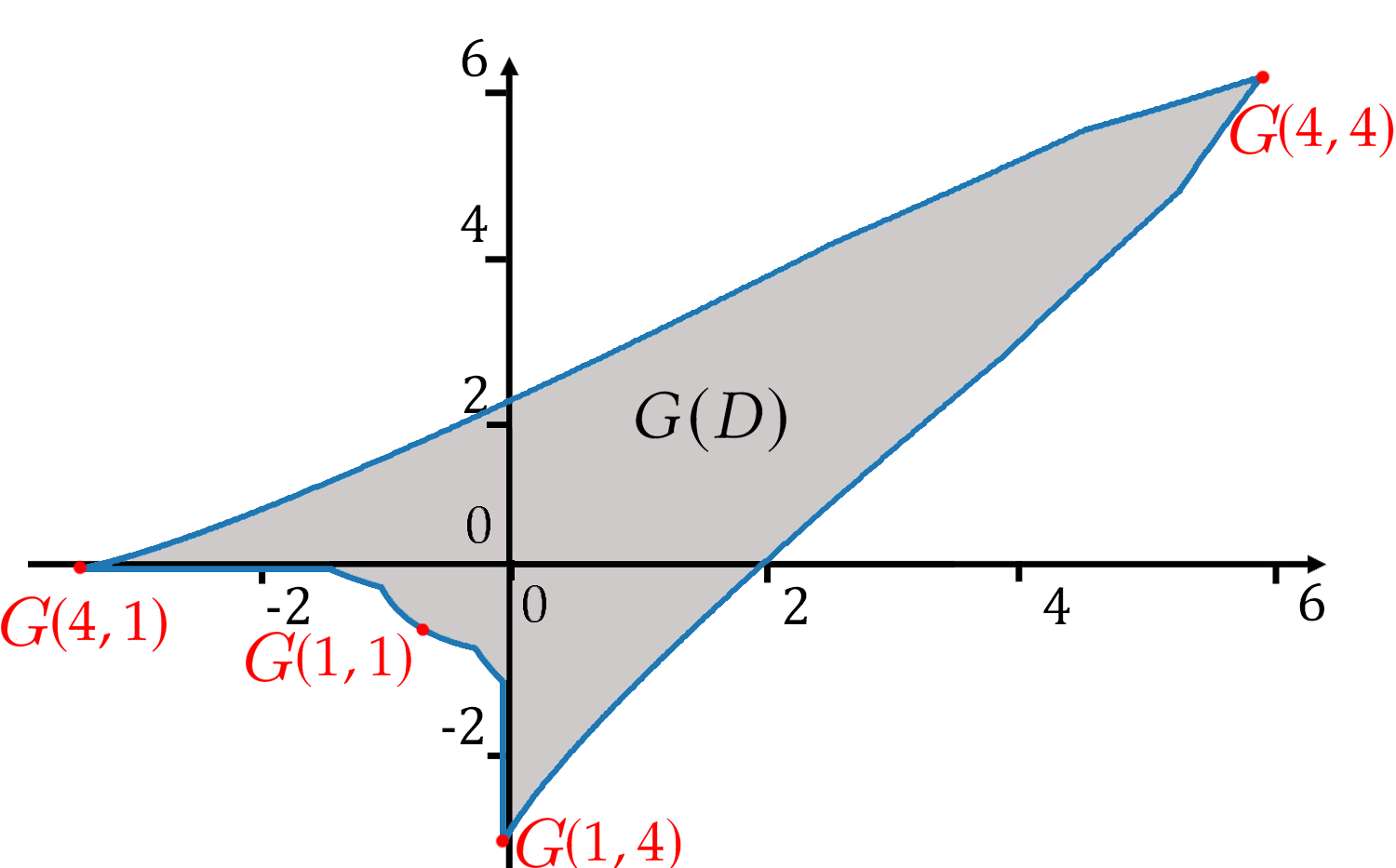} }}%
		\caption[]{Illustration of the function $G$ on an instance of $\FCBn$ (\cref{ex:example_1234}).}%
		\label{fig:example_1234}%
\end{figure*}

Instead of the approaches discussed above, in this work, we prove Result~${(i)}$ of \cref{theo:OUDs} via the following equivalent formulation: \emph{proving the existence of a {positive zero}, \ie the existence of a point $(a,b) \in \R^2$ such that $G(a,b)=(0,0)$, of the $G:\R^2 \rightarrow\R^2$} defined as follows:
\begin{align}
    &G\parens{\gA,\gB} = \parens*{ g^A \parens{\gA,\gB} , g^B \parens{\gA,\gB}} \in \R^2, \forall (\gA, \gB) \in \R^2. \label{eq:G_func}
\end{align}
Note also that although $(0,0)$ is a trivial solution of System~\eqref{eq:system_f} (\ie it is a zero of $G$), we can only construct $\Ai, \Bi$ (as in \cref{def:OptDis_GRCBC}) from solutions whose coordinates are \emph{strictly} positive (\ie only from positive zeros of $G$). In this proof, we work with the notion of \emph{winding numbers} which is intuitively defined as follows: the winding number, denoted $\W \parens*{\curve,y}$, of a parametric (2-dimensional) closed curve $\curve$ around a point $y \in \R^2$ is the number of times that $\curve$ travels counterclockwise around $y$ (see formal definitions of parametric curves, winding numbers and other related notions in \ref{sec:appen_winding}). This notion allows an important result as follows:\footnote{See \cite{chinn1966first} for a more general statement of \cref{lem:wind}. It is also considered in the literature as a variant of the \emph{main theorem of connectedness} in topology (see e.g., Theorem 12.N in \cite{viro2008elementary}).} %
\begin{lemma} \label{lem:wind}
    If $G$ is a continuous mapping, for any set $D \subset \R^2$ which is topologically equivalent to a disk such that $\W(\curve, (0,0)) \neq 0$ where $\curve$ is the $G$-image of the boundary of $D$, then $(0,0) \in G(D)$.
\end{lemma}
Our proof proceeds by crafting a tailored set $D \subset \R^2{>0}$ such that the function $G$ from \eqref{eq:G_func} satisfies all sufficient conditions of \cref{lem:wind};\footnote{It is trivial that $g^A( \cdot , \gB), g^A( \gA , \cdot) $ and $g^B( \cdot , \gB), g^B( \gA , \cdot) $ are all continuous and monotone functions in $\mathbb{R}_{>0}$; therefore, from Proposition 1 of \cite{kruse1969joint}, $g^A(\gA, \gB)$ and $g^B(\gA,\gB) $ are continuous functions in~$R^2_{>0}$.} then, we conclude that $G$ has a zero in $D$ and Result~$(ii)$ of \cref{theo:OUDs} follows. 
%
Note that finding such a set $D$ and quantifying the involved winding number are non-trivial due to the complexity in the expressions of $g^A$ and $g^B$. We illustrate \cref{lem:wind} and how the proof of Result~$(ii)$ of \cref{theo:OUDs} proceeds in a particular instance of $\FCB$ in \cref{ex:example_1234}.
%
\begin{example}\label{ex:example_1234}
	\emph{Consider a game $\FCBn$ with $n = 4$, $X^A=4, X^B=4$, \mbox{$w_1=w_3 =1$}, $w_2 = w_4 =2$, $p_1 = p_2 =1$, \mbox{$p_3 = p_4 =-1$},  $q_i =1, \forall i$. We illustrate in \cref{fig:example_1234} the values of the function $G:\mathbb{R}^2 \rightarrow \mathbb{R}^2$ corresponding to this game. \cref{fig:example_1234}(a) represents the output plane where each point is mapped with a color; e.g., if a point has the color blue, we know that both coordinates of this point are positive. \cref{fig:example_1234}(b) presents the input plane. Function $G$ maps each point in this input plane with a point in the output plane. Then, in \cref{fig:example_1234}(b), we colorize each point in the input plane with the corresponding color of its output (colors are chosen according to \cref{fig:example_1234}(a)). 
	 Solving \eqref{eq:system_f} in this case, we see that $(2,2)$ is the unique zero of $G$. We observe that in \cref{fig:example_1234}(b), when one choose a disk containing $(2,2)$, its boundary passes through all colors, which indicates that the $G$-image of its boundary goes around the origin $(0,0)$ of the output plane. This is confirmed by \cref{fig:example_1234}(c) showing the $G$-image of a rectangle $D$ having the vertices $(1,1)$, $(1, 4)$, $(4, 4)$, $(4, 1)$ (thus, it contains $(2,2)$); we observe that $G(\partial D)$, which is the blue curve, travels 1 time around $(0,0)$, thus $\W(G(D),(0,0)) \neq 0$.}
\end{example}

To complete this section, we compute the players' payoffs in the $\FCB$ game in the case where their allocations follow the proposed OUDs. Recall the notation of the indices sets defined in \cref{tab:Opt_Uni_GRCBC}, we have the following proposition (its proof is given in \ref{sec:appen_proof_payoffEQ}):
\begin{proposition}\label{propo:payoffEQ}
Given a game $\FCBn$ and $\kappa = \parens*{\gA, \gB} \in \R^2_{>0}$, if Players A and B play strategies such that the marginal distributions corresponding to battlefield $i \in [n]$ are $\Ai$ and $\Bi$ respectively, then their payoffs are:
\begin{subequations}
    \begin{align}
    	\Pi^A_{\FCBn} & = 	\sum \limits_{i \in I^+_1(\gA,\gB)} {\left[w_i \mathbb{I}_{\{p_i > 0\}} + \alpha w_i \mathbb{I}_{\{p_i =0\}} \right]} \! + \! \sum \limits_{i \in I^+_2(\gA,\gB)} {\left[ w_i \left(1 \!-\! \frac{q_i \gB}{\gA} \!+\! \frac{p_i}{w_i \gA} \right)  \!+\! \frac{(q_i w_i \gB \!-\! p_i )^2}{2 w_i \gA q_i \gB} \right]} \nonumber \\
    	& \hspace{4mm} +\!  \sum \limits_{i \in I^+_3(\gA,\gB)} \left[ \frac{w_i \gA}{2 q_i \gB} \right] \!+\! \sum \limits_{i \in I^-_2(\gA,\gB)} \left[ \frac{w_i \gA}{2 q_i \gB} \!-\! \frac{p_i ^2}{2 w_i \gA q_i \gB} \right] \!+\!  \sum \limits_{i \in I^-_3 (\gA,\gB)} \left[w_i \!-\! \frac{q_i \gB w_i }{2\gA} \right] \label{eq:payoff_A_GL},\\
        \Pi^B_{\FCBn} & = \sum_{i \in [n]} w_i -  	\Pi^A_{\FCBn}. \label{eq:payoff_B_GL}		
    \end{align}
\end{subequations}
\end{proposition}

If there exists an equilibrium of the game $\FCBn$ whose marginals are $\Ai, \Bi, i \in [n]$, then \eqref{eq:payoff_A_GL} and \eqref{eq:payoff_B_GL} are formulations of the equilibrium payoffs in this game. Observe, however, that as $p_i$ and $q_i$ (and $w_i$) change, System~\eqref{eq:system_f} also changes; thus, its solutions also vary and the configuration of the corresponding indices sets change. Therefore, the relationship between the favoritism parameters and the payoffs induced by the corresponding OUDs is \emph{not} easily deducible from \eqref{eq:payoff_A_GL}-\eqref{eq:payoff_B_GL}. We delay our discussion on this to \cref{sec:NumExp} where we present results from numerical~experiments.

%
%

%
%
\section{Equilibria Results for the Colonel Blotto Game with Favoritism}
\label{sec:Corollary_results}

In the previous section, we successfully construct a set of OUDs of the \FCB game; we now show how one can use this result to deduce an equilibrium. As a preliminary result, we give a high-level condition under which the OUDs from \cref{def:OptDis_GRCBC} constitute an equilibrium of the $\FCB$ game; it is presented as a direct corollary from \cref{theo:OUDs} as follows: 

\begin{corollary} \label{corol:FCBEqui}
    For any game $\FCBn$ and any positive solution $\kappa=(\gA, \gB)$ of System~\eqref{eq:system_f}, if there exists a mixed-strategy of Player A (resp., Player B) whose univariate marginal distributions correspond to $\Ai$ (resp., $\Bi$) for all $i \in [n]$, then these mixed strategies constitute an equilibrium of~$\FCBn$.
\end{corollary}

\cref{corol:FCBEqui} is a standard statement in analyzing equilibria of Colonel Blotto games from their OUDs. In general, it is challenging to construct such mixed strategies as required in \cref{corol:FCBEqui}---this is, as discussed in \cref{sec:Intro}, a notorious difficulty in studying the Colonel Blotto game. In the literature, several alternative concepts of solutions are proposed based on the related OUDs, they are relaxed from the equilibrium of the Colonel Blotto game in one way or another. We show that these results can also be extended to the \FCB game thanks to \cref{theo:OUDs}: in \cref{sec:GL_game}, we analyze a trivial equilibrium of the \FGL game and in \cref{sec:IU_result}, we propose an approximate equilibrium of the \FCB game based on the rescaling technique of \cite{vu2019approximate}. Nevertheless, we start in \cref{sec:Equi_FCB} by listing special cases of \FCB where the exact equilibrium can be computed.

%
%
%
%
%
 
\subsection{Exact Equilibria of the $\FCB$ Game in Particular Cases}
\label{sec:Equi_FCB}

For several parameters configurations, we can leverage existing results in special cases of the Colonel Blotto game to solve special cases of the \FCB game. For instance, \cite{roberson2006} successfully constructs an equilibrium of the game with homogeneous battlefields---the key idea is that this game has a set of OUDs that are the same for all battlefields. Therefore, we can generalize this idea to construct an equilibrium from the set of OUDs $\left\{ \Ai,  \Bi: i \in [n] \right\}$ in any $\FCBn$ game whose parameters $w_i, p_i, q_i$ are such that $ \Ai(x) = F_{A^{\kappa }_j}(x), \forall x \in [0, \infty)$, $\forall i,j \in [n]$ (and $ \Bi(x) = F_{B^{\kappa }_j}(x), \forall x \in [0, \infty)$). A simple example where this condition holds is when $w_i = w_j$, $p_i = p_j$ and $q_i = q_j $, $\forall i, j \in [n]$ in which any solution of System~\eqref{eq:system_f} induces an indices set (as defined in \cref{tab:Opt_Uni_GRCBC}) that contains the whole set $\{1, \ldots, n\}$. It is also possible to extend this approach to \FCB games where the set of battlefields can be partitioned into groups with homogeneous OUDs and such that the cardinality of each group is sufficiently large (following an idea proposed by \cite{schwartz2014} in the case of classical Colonel Blotto~games).

%
%
 
\subsection{Equilibrium of the General Lotto Game with Favoritism}
\label{sec:GL_game}
In some applications of the \FCB game (and of the classical Colonel Blotto game), the budget constraints do not need to hold with probability 1; instead, they are only required to hold in expectation. In such cases, the General Lotto game with favoritism ($\FGL$) is relevant and applicable. 

Due to the relaxation in the budget constraints, any set of OUDs of a game instance $\FCBn$ can serve as a set of equilibrium marginals of the corresponding game $\FGLn$ (having the same parameters): it is trivial to deduce mixed strategies of $\FGLn$ from univariate distributions $\Ai, \Bi, i \in [n]$ (from \cref{def:OptDis_GRCBC}). Formally, from \cref{theo:OUDs}, we have the following corollary:
\begin{corollary} \label{corol:payoffGL}
   For any game $\FGLn$ and any positive solution $\kappa=(\gA, \gB)$ of System~\eqref{eq:system_f}, the strategy profile where Player A (resp., Player B) draws independently her allocation to battlefield $i \in [n]$ from $\Ai$ (resp., $\Bi$) is an equilibrium.
\end{corollary}
%
%
%
Naturally, in the \FGL game, when players follow the equilibrium described in \cref{corol:payoffGL}, they gain the same payoffs as in \eqref{eq:payoff_A_GL}-\eqref{eq:payoff_B_GL}. It is also trivial to check that \cref{corol:payoffGL} is consistent with previous results on the classical General Lotto game (\ie the \FGL game where $p_i=0$ and $q_i=1$ for any $i \in [n]$), \eg from \cite{myerson1993incentives,kovenock2020generalizations}.

\subsection{An Approximate Equilibrium of the \FCB Game}
\label{sec:IU_result}

In the game theory literature, approximate equilibria are often considered as alternative solution-concepts when it is impossible or inefficient to compute exact equilibria. Here, we focus on finding a good approximate equilibrium of the \FCB game that can be simply and efficiently constructed; this is relevant in applications where the budget constraints must hold precisely (\eg in security or telecommunication systems involving a fixed capacity of resources) but sub-optimality is acceptable if the error is negligible relative to the scale of the problem. We begin by recalling the definition of approximate equilibria \cite{myerson1991game,Nisan07} in the context of the \FCB game: 
\begin{definition}[$\eps$-equilibria]\label{def:appro_equi}
    For any $\varepsilon \ge 0$, an \emph{\mbox{$\varepsilon$-equilibrium}} of a game $\FCBn$ is any strategy profile $\left(s^{*},t^{*} \right)$ such that \mbox{$\Pi^A_{\FCBn}(s,t^*) \!\le\! \Pi^A_{\FCBn}(s^*,t^*)\!+\!\varepsilon$} and \mbox{$\Pi^B_{\FCBn}(s^*,t)\! \le\! \Pi^B_{\FCBn}(s^*,t^*)\! +\! \varepsilon$} for any strategy $s$ and $t$ of Players A and B.
\end{definition}
%
%

The set of OUDs constructed in \cref{sec:OptUni_GRCBC} allows us to apply directly a technique from the literature of classical Colonel Blotto game to look for an approximate equilibrium of the \FCB game: in recent work, \cite{vu2019approximate} propose an approximation scheme (called the IU strategy) for the Colonel Blotto game in which players independently draw their allocations from a set of OUDs, then rescale them to guarantee the budget constraint. The authors prove that IU strategies constitute an $\varepsilon W$-equilibrium of the CB game where $\varepsilon = \tilde{\mO}(n^{-1/2})$ and $W$ is the sum of battlefields' values. We extend this idea to the \FCB game and propose the following definition:
\begin{definition}[IU Strategies] \label{def:IU}
    For any game $\FCBn$ and any solution $\kappa = (\gA, \gB) \in \R^2_{>0}$ of System~\eqref{eq:system_f}, we define $\IU^{\play}_{\kappa}$ to be the mixed strategy of player $\play \in \braces*{A,B}$ such that her allocations, namely $\boldsymbol{x}^\play$, are randomly generated by the following simple procedure:\footnote{In fact, in this definition, when $\sum_{j \in [n]} a_j =0$ (resp. when $\sum_{j \in [n]} b_j =0$), we can assign any arbitrary $\boldsymbol{x}^A \in S^A$ (resp. any $\boldsymbol{x}^A \in S^B$). This choice will not affect the asymptotic results stated in this section (particularly, \cref{propo:IU}).} \emph{Player A} draws independently a real number $a_i$ from $\Ai, \forall i \in [n]$. If $\sum_{j=1}^n a_j = 0$, set $x^A_i = \frac{X^A}{n}$; otherwise, set $x^A_i = \frac{a_i}{\sum_{j=1}^n a_j} \cdot X^A$. \emph{Player B} draws independently a real number $b_i$ from $\Bi,\forall i \in [n]$. If $\sum_{j=1}^n b_j = 0$, set $x^B_i = \frac{X^B}{n}$; otherwise, set $x^B_i = \frac{b_i}{\sum_{j=1}^n b_j} \cdot X^B$.
%
\end{definition}
Intuitively, by playing IU strategies, players draw independently from the OUDs then normalize before making the actual allocations. It is trivial to check that the realizations from $\IU^{\play}_{\kappa}$ satisfies the corresponding budget constraint in the \FCB game. Now, we consider the following~assumption:
\begin{assumption}\label{assum:bound}
    $\exists \wmax, \wmin: 0<\wmin \le w_i \le \wmax < +\infty$, $\forall i \in [n]$. 
\end{assumption}
\noindent \cref{assum:bound} is a mild technical assumption that is satisfied by most (if not all) applications of the \FCB game. Intuitively, it says that the battlefields' values of the game $\FCB$ are bounded away from 0 and infinity. With a simple adaptation of the results of \cite{vu2019approximate} (for the classical Colonel Blotto game), we obtain the following proposition (we give its proof in \ref{appen:IU}):
\begin{proposition}[IU Strategies is an $\eps$-equilibrium]
\label{propo:IU}
   In any game $\FCBn$ satisfying \cref{assum:bound}, there exists a positive number \mbox{$\varepsilon = \tilde{\mO} \left(n^{-1/2}\right)$} such that for any solution $\kappa = (\gA, \gB) \in \R^2_{>0}$ of System~\eqref{eq:system_f}, the profile $\parens*{\IU^A_{\kappa}, \IU^B_{\kappa}}$ is an \mbox{$\varepsilon W^n$-equilibrium} where $W^n:= \sum_{i=1}^n w_i$.
\end{proposition}
%
%
Here, recall that the notation $\tilde{\mO}$ is a variant of the $\mO$-asymptotic notation where logarithmic terms are ignored. We can interpret \cref{propo:IU} as follows: consider a sequence of games $\FCBn$ in which $n$ increases (\ie games with larger and larger numbers of battlefields). Note that $W^n$ is an upper-bound of the players' payoffs in the game $\FCBn$, thus $W^n$ is relative to the scale of this game. To qualify the proposed approximate equilibrium based on the evolution of $n$, we consider the ratio between the involved approximation error $\eps W^n$ of $\parens*{\IU^A_{\kappa}, \IU^B_{\kappa}}$ and this relative-scaled quantity $W^n$; this tracks down the proportion of payoff that Player $\play$ might lose by following $\IU^{\play}_{\kappa}$ instead of the best-response against $\IU^{-\play}_{\kappa}$. As we consider $\FCBn$ games with larger and larger $n$, this ratio (which is exactly~$\varepsilon$) quickly tends to 0 with a speed in order $\tilde{\mO} (n^{-1/2})$. Therefore, in the \FCB games with large number of battlefields, the players can confidently play $\parens*{\IU^A_{\kappa}, \IU^B_{\kappa}}$ as an approximate equilibrium knowing that the involved level of error is negligible. Note also that the approximation error presented in \cref{propo:IU} also depend on other parameters of the game including $X^A, X^B$, $\wmin, \wmax$, $\max\{ \left|p_i\right|, i\in [n] \}, \min \{q_i, i \in [n]\}$ and $\alpha$ (these constants are hidden in the $\tilde{\bigoh}$ notation).

%
%
%
%

%
%
\section{Efficient Approximation of Optimal Univariate Distributions}
\label{sec:heuristic}

Our characterization of the (approximate) equilibria of the \FCB and \FGL games build upon System~\eqref{eq:system_f}. \cref{theo:OUDs} shows the existence of a solution of this system, but in practice it is also important to be able to compute such a solution. It is not clear how to do this efficiently: recall that \eqref{eq:system_f} is \emph{not} simply a system of quadratic equations in $\gA, \gB$ since as these variables change, the configuration of the indices sets (involved in the definitions of $g^A, g^B$) also changes. Given a game $\FCBn$, a naive way to solve System~\eqref{eq:system_f} would be to consider all possible partitions of \mbox{$[n]$} into $\Iplus$,  \mbox{$\Iminus$} and $[n] \backslash \bracks*{\Iplus \bigcup \Iminus}$, then solve the particular system of quadratic equations corresponding to each case. This approach, however, is inefficient as in the worst case, the number of partitions is exponential in $n$; as illustrated in the following toy example: 
%
 
\begin{figure}[h!]
\centering
\begin{tikzpicture}
\node (img){	\includegraphics[height = 0.15\textheight]{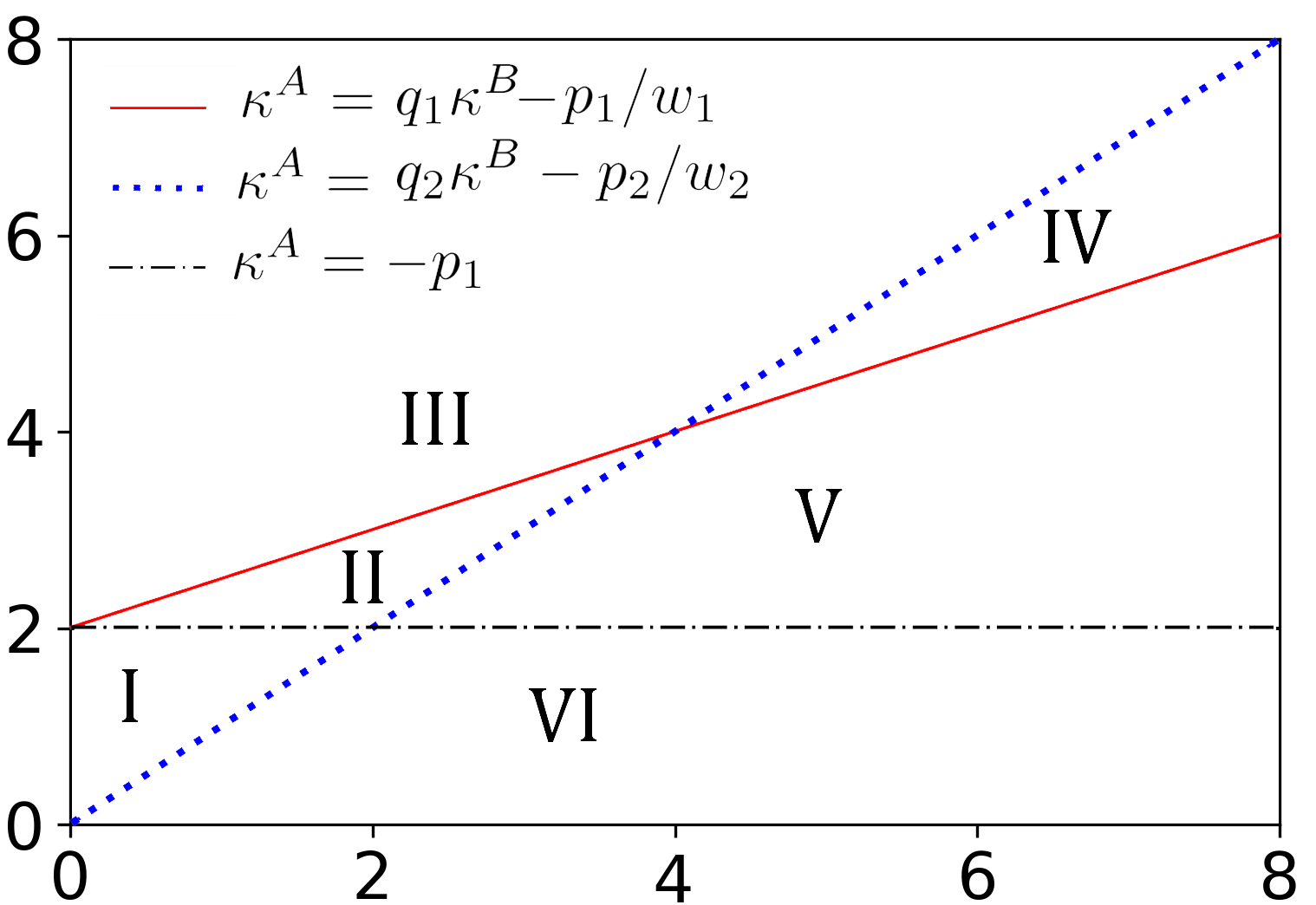}};
\node[below=of img, node distance=0cm, yshift=1.3cm, xshift = 0.3cm] {\scriptsize $\gB$};
\node[left=of img, node distance=-0.2cm, anchor=center,yshift=-0.0cm,xshift = 0.9cm] {\scriptsize $\gA$};
\end{tikzpicture}

\caption{Conditions to partition battlefields into the indices sets of the $\FCBn$ game in \cref{ex:example_GRCB_2}.} \label{fig:example}
\end{figure}
 
\begin{example}\label{ex:example_GRCB_2}
\emph{
Consider the game $\FCBn$ with $n =2$, $X^A = X^B= 2$, $w_1 = w_2 =1$, $p_1 = -2$, $p_2 = 0$, $q_1 = 1/2$ and $q_2 = 1$.  Even in this extremely simple game, there are 6 possible configurations of $\Iplus$ and $\Iminus$. In \cref{fig:example}, we illustrate these cases by 6 regions in the first quadrant of the $\gA$-$\gB$ plane separated by the axes and the polynomials involving in the conditions that determine the indices sets. Consider these cases, each inducing a system of quadratic equations, we
%
%
see that there is no positive solution of~\eqref{eq:system_f} in the cases corresponding to Regions I, II, III, IV and VI of \cref{fig:example}. Only when $2 < \gA < \min \{ \gB,  \gB /2 +2 \}$ (i.e., the point $(\gA, \gB)$ lies in Region V of \cref{fig:example}), we have $\Iminus =\{1\}$ and $\Iplus=\{2\}$ and thus, System~\eqref{eq:system_f}
has a unique positive solution in $\mathbb{R}^2_{>0}$ that is $\gA = 2 + \sqrt{4/3}$ and $\gB = 2 + \sqrt{12}$ (which satisfy the conditions of Region~V).
}
\end{example}

In large or even moderate instances, this naive approach is impossible, leading to the important question: can we (approximately) solve System~\eqref{eq:system_f} more efficiently? We answer this question positively: we propose in \cref{sec:approx_algo} an \emph{approximation algorithm} that computes an solution of System~\eqref{eq:system_f} with arbitrarily small error, and we analyze its running time in \cref{sec:efficiency}. Finally, in \cref{sec:approx_OUDs}, we analyze the impact of the approximation error on the OUDs from  \cref{sec:OptUni_GRCBC}. 


%
%
%
\subsection{An Approximation Algorithm Solving System~\eqref{eq:system_f}}
\label{sec:approx_algo}
Throughout this section, we focus on the following approximation concept: in any game $\FCBn$ (and $\FGLn$), for any $\delta>0$,  a point $(\tgA, \tgB) \in \mathbb{R}^2_{>0}$ is called a \emph{${\delta}$-approximate solution} of System~\eqref{eq:system_f} if there exists a solution $(\gA, \gB) \in \R^2_{>0}$ of System~\eqref{eq:system_f} satisfying the following conditions:
\begin{align}
    	& \left| \tgA - \gA \right| \le {\delta} \textrm{ and }  \left|\tgB - \gB \right| \le {\delta}, \label{eq:delta_close} \\
        \textrm{ and }&  g^A(\tgA, \tgB)\le 0 \textrm{ and } g^B(\tgA, \tgB)\le0. \label{eq:system_approx_f}
\end{align}
Intuitively, any $\tk = (\tgA, \tgB)$ satisfying~\eqref{eq:delta_close} is ${\delta}$-close to a solution of System~\eqref{eq:system_f} (in the metric induced by the $\| \cdot\|_{\infty}$ norm) and, by \eqref{eq:system_approx_f}, the distributions $\braces*{ \tAi, \tBi , i \in [n]}$ from \cref{def:OptDis_GRCBC} corresponding to $\tk$ satisfies Condition~\ref{condi:OUD2} of \cref{def:OUD} (i.e., budget constraints).\footnote{Note also that since $G$ is continuous (it is also Lipschitz-continuous \wrt $\| \cdot \|_{\infty}$ norm), the distance between $G(\tgA, \tgB)$ and $(0,0)$ also tends to 0 when $\delta \rightarrow 0$; to quantify this, one might look for the Lipschitz constants of $g^A, g^B$; but since this analysis is not relevant to results presented in this section, we omit the details.} In principle, we would like to find ${\delta}$-approximate solutions such that $\delta$ is as small as possible; naturally, this will come with a trade-off on the running time (we discuss this further in \cref{sec:efficiency}).

We propose an approximation algorithm, having a stopping-criterion parameter $\delta$, that quickly finds a $\delta$-approximate solution of~\eqref{eq:system_f} in any game $\FCBn$. A pseudo-code of this algorithm is given in \ref{appen:heuristic} along with many details; we discuss here only its main intuition. Recall that the function $G:\R^2 \rightarrow \R^2$ defined in~\eqref{eq:G_func} is a continuous mapping and that solving System~\eqref{eq:system_f} is equivalent to finding a zero of $G$. To do this, our approximation algorithm consists of a dichotomy procedure (i.e., a bisection method) such that at each loop-iteration, it considers a smaller subset of $\mathbb{R}^2$. It starts with an arbitrary rectangle $D \subset \mathbb{R}^2_{>0}$ (including its boundary and interior), then checks to see whether its image via $G$ contains the point $(0,0)$---this can be done by computing the winding number of the $G$-image of the boundary of $D$ (which is a closed parametric curve) around $(0,0)$: due to \cref{lem:wind}, if this winding number is non-zero, $G(D)$ contains $(0,0)$. If $G(D)$ does not contains $(0,0)$, we enlarge the rectangle $D$ (\eg by doubling its length and width) while maintaining that $D \subset \R^2_{>0}$. We repeat this enlargement step until we find a rectangle whose $G$-image contains $(0,0)$. Due to \cref{theo:OUDs} and \cref{lem:wind}, such a rectangle $D$ exists with a zero of $G$ inside. We then proceed by dividing the rectangle $D$ into smaller rectangles and checking which among them has a $G$-image containing $(0,0)$, then repeating this procedure on that smaller rectangle. The algorithm terminates as soon as it finds a rectangle, say $D^*$, such that $G(D^*)$ has a non-zero winding number around $(0,0)$ and $D^*$ has a diameter smaller than $\delta$ (thus any point in $D^*$ satisfies \eqref{eq:delta_close}); as a sub-routine of the computation of the involved winding number, our algorithm also determines a point $(\tgA, \tgB)$ in $D^*$ satisfying \eqref{eq:system_approx_f}. The output $(\tgA, \tgB)$ is a $\delta$-approximate solution of System~\eqref{eq:system_f}.

%
%
\subsection{Computational Time of the Approximation Algorithm}
\label{sec:efficiency}
In the approximation algorithm described above, the most complicated step is the computation of the winding number of the involved rectangles. To do this efficiently, we draw tools from the literature: for any parametric curve $\curve:[a,b] \rightarrow \R^2$ where $\min_{t \in [a,b]} \| \curve(t)\|_{\infty} = \delta$, the insertion procedure with control of singularity (IPS) algorithm proposed by \cite{zapata2012geometric} takes $\bigoh \parens*{(b-a) \delta^{-1}}$ time to output a special polygonal approximation of $\curve$---having a number $\bigoh \parens*{(b-a) \delta^{-1}}$ of vertices---such that the winding number of this polygonal approximation is precisely the winding number of $\curve$. To compute this winding number, we calculate the value of $\curve$ at all vertices of this polygon. Inserting IPS into our approximation algorithm running with the parameter $\delta$, in any game $\FCBn$, for any rectangle~$D$ in consideration, we can represent $G(\partial D)$ by a parametric curve $\curve:[a,b] \rightarrow \R^2$ and compute the winding number of $G(\partial D)$ in $\bigoh \parens*{n(b-a) \delta^{-1}}$ time (it takes $\bigoh(n)$ time to compute the $G$-value of a vertex of the polygonal approximation). Note that this computational time also depends on other parameters of the game $\FCBn$ (they are hidden in $\bigoh$-notation above); we discuss this point in details in \ref{appen:heuristic}. Based on this procedure, we have the following~proposition:
%
%

\begin{proposition}\label{propo:heuristic}
    For any game $\FCBn$ and $\delta <1$, the approximation algorithm described above finds a $\delta$-approximate solution of System~\eqref{eq:system_f} in $\tilde{\bigoh}(n \delta^{-1})$ time.
\end{proposition}
\cref{propo:heuristic} confirms the efficiency of our approximation algorithm, as the running time of our algorithm is only $\bigoh(n)$. The order $\tilde{\bigoh}(\delta^{-1})$ gives the trade-off between the running-time and the precision-level of solutions $\delta$. In fact, the running time of our algorithm also depends on the choice of the initial rectangle. More precisely, for a solution $(\gA, \gB)$ of System~\eqref{eq:system_f} such that \mbox{$\|(\gA, \gB)\|_{\infty} \!<\!R$}, our approximation algorithm, initialized with a rectangle whose center $(\gA_0,\gB_0)$ satisfying $\|(\gA_0,\gB_0) \|_{\infty} =  L_0$ and $\delta <1$, terminates after $\bigoh \parens*{\log \parens*{\frac{R}{\delta}} + \log \parens*{ \max \braces*{\frac{R}{L_0}, \frac{L_0}{R}}}}$ iterations and each iteration runs in $\bigoh \parens*{ {R n}{\delta}^{-1} }$ time. Intuitively, if the initialized rectangle is too small and/or the actual solution is too far away from this rectangle, the algorithm requires a longer time. We conduct several experiments to illustrate the computational time of our approximation algorithm. Due to space constraints, we place these results in~\ref{appen:heuristic}; globally, our proposed algorithm is very fast in comparison with the naive approach described~above.

%
%

%
%
\subsection{Approximations of Optimal Univariate Distributions of the \FCB Game}
\label{sec:approx_OUDs}
To conclude this section, we show that from a $\delta$-approximate solutions of System~\eqref{eq:system_f}, one gets an approximate equilibrium  for the \FCB and \FGL games. This is based on the following proposition:
  
\begin{proposition}\label{lem:approx_sol}
    In any game $\FCBn$ (and $\FGLn$), let $\kappa = (\gA,\gB)$ and $\tk = (\tgA, \tgB)$ be a positive solution and a $\delta$-approximate solution of System~\eqref{eq:system_f} respectively (such that $\kappa, \tk$ satisfy \eqref{eq:delta_close}-\eqref{eq:system_approx_f}). Then, the sets of distributions $\braces{\Ai,\Bi, i \in [n]}$ and $\braces{\tAi,\tBi, i \in [n]}$ from \cref{def:OptDis_GRCBC} corresponding to $\kappa$ and $\tk$~satisfy $\big| \Ai(x)\! -\! \tAi(x) \big| \le \bigoh \parens{\delta}$ and $\big| \Bi(x) - \tBi(x) \big| \le \bigoh\parens{\delta}$, for any $i \in [n]$ and~$x\in [0,\infty)$. 
\end{proposition}
A proof of \cref{lem:approx_sol} is given in \ref{appen:approx_OUD}. Intuitively, it shows that when $\tk = (\tgA, \tgB)$ is a $\delta$-approximate solution of \eqref{eq:system_f}, the distributions ${\tAi,\tBi}$ are approximations of the distributions ${\Ai,\Bi}$ with the approximation error in order $\bigoh(\delta)$ (note that it is also polynomial in terms of $1/\min\{\gA, \tgA, \gB, \tgB\}$, $\max\{\gA, \tgA, \gB, \tgB\}$, $1/\min_{i \in [n]}{q_i}$ and $\max_{i \in [n]}{|p_i|}$). From \cref{lem:approx_sol}, we can also deduce that the players' payoffs when they use strategies with marginals following ${\tAi,\tBi, i \in [n]}$ are $\bracks*{\bigoh ( \delta) W^n}$-close to the payoffs when players use strategies with marginal ${\Ai,\Bi, i \in [n]}$ (see the formulations given in \eqref{eq:payoff_A_GL}-\eqref{eq:payoff_B_GL}). As a consequence, by following the scheme leading to results in \cref{sec:Corollary_results}, for any games $\FCBn$ and $\FGLn$ (having the same parameters), for any $\delta$-approximate solution $\tk = (\tgA, \tgB)$ of System~\eqref{eq:system_f}, we have:
\begin{itemize}
    \item[$(i)$] In $\FGLn$, the strategies profile when Player A (resp. Player B) draws independently her allocation to battlefield $i$  from ${\tAi}$ (resp. $\tBi$) constitutes a $\bracks*{\bigoh(\delta) W^n}$-equilibrium.\footnote{These are indeed mixed strategies of $\FGLn$ since $\sum_{i \in [n]} \Ex_{x \sim \tAi} \left[ x  \right] < X^A$ and $\sum_{i \in [n]} \Ex_{x \sim \tBi} \left[ x  \right] < X^B$ due to~\eqref{eq:system_approx_f}.}
    \item[$(ii)$] In $\FCBn$, the strategies $(\IU^{A}_{\tk}, \IU^{B}_{\tk} )$ is a $\bracks*{\bigoh(\delta + \varepsilon ) W^n}$-equilibrium where $\varepsilon = \tilde{\bigoh}(n^{-1/2})$. In this case, we
    cannot obtain an approximate equilibrium with a level of error better than $\varepsilon$; to achieve this, we only need to run the approximation algorithm with $\delta = {\varepsilon}$.
\end{itemize}
	%


%
%
\section{Numerical Illustrations of the Effect of Favoritism in Colonel Blotto and General Lotto~Games}
\label{sec:NumExp}

In this section, we conduct numerical experiments illustrating the effect of favoritism in the \FCB and \FGL games. For each game instance, if its parameters satisfy \cref{assum:1} and~\cref{assum:2}, we run the approximation algorithm described in \cref{sec:heuristic} to find a ${\delta}$-approximate solution $\tk = (\tgA, \tgB)$ of the corresponding System~\eqref{eq:system_f} where we set ${\delta} = 10^{-6}$; then, we report the obtained results regarding the distributions $\tAi, \tBi, i \in [n]$ from \cref{def:OptDis_GRCBC} corresponding to~$\tk$. If \cref{assum:1} and~\cref{assum:2} are violated, we report the results corresponding to the trivial pure equilibria.

In the first experiment, we aim to illustrate the relation between parameters $p_i, q_i$ ($i \in [n]$) and the players' equilibrium payoffs in the \FGL game presented in \eqref{eq:payoff_A_GL}-\eqref{eq:payoff_B_GL} (which are also the payoffs of the corresponding \FCB game if the assumptions in \cref{corol:FCBEqui} hold). Since $\FGL$ and $\FCB$ are constant-sum games, we focus on the payoff of Player A. Although our results hold for \FCB and \FGL games in the general setting of parameters, we first focus on instances where the players' budgets are symmetric and all battlefields are homogeneous in order to single out the effect of favoritism. In particular, we consider a group of instances of the \FGL game with $n=4$ battlefields, $X^A = X^B =10$, $\alpha =1/2$, in which all battlefields have the same values ($w_i =1, \forall i$) and the same favoritism parameters $p_i = \bp, q_i = \bq, \forall i$ for given $\bp, \bq$. For comparison, recall that the instance where $\bp= 0, \bq =1$ corresponds to the classical Colonel Blotto/ General Lotto~game.
	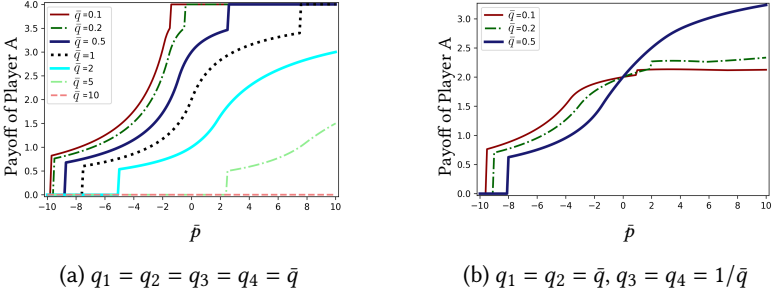
\begin{figure*}[htb!]
		\centering
		\subfloat[$q_1 = q_2 = q_3 = q_4 = \bar{q}$]{
			\begin{tikzpicture}
			\node (img){\includegraphics[height = 0.14\textheight]{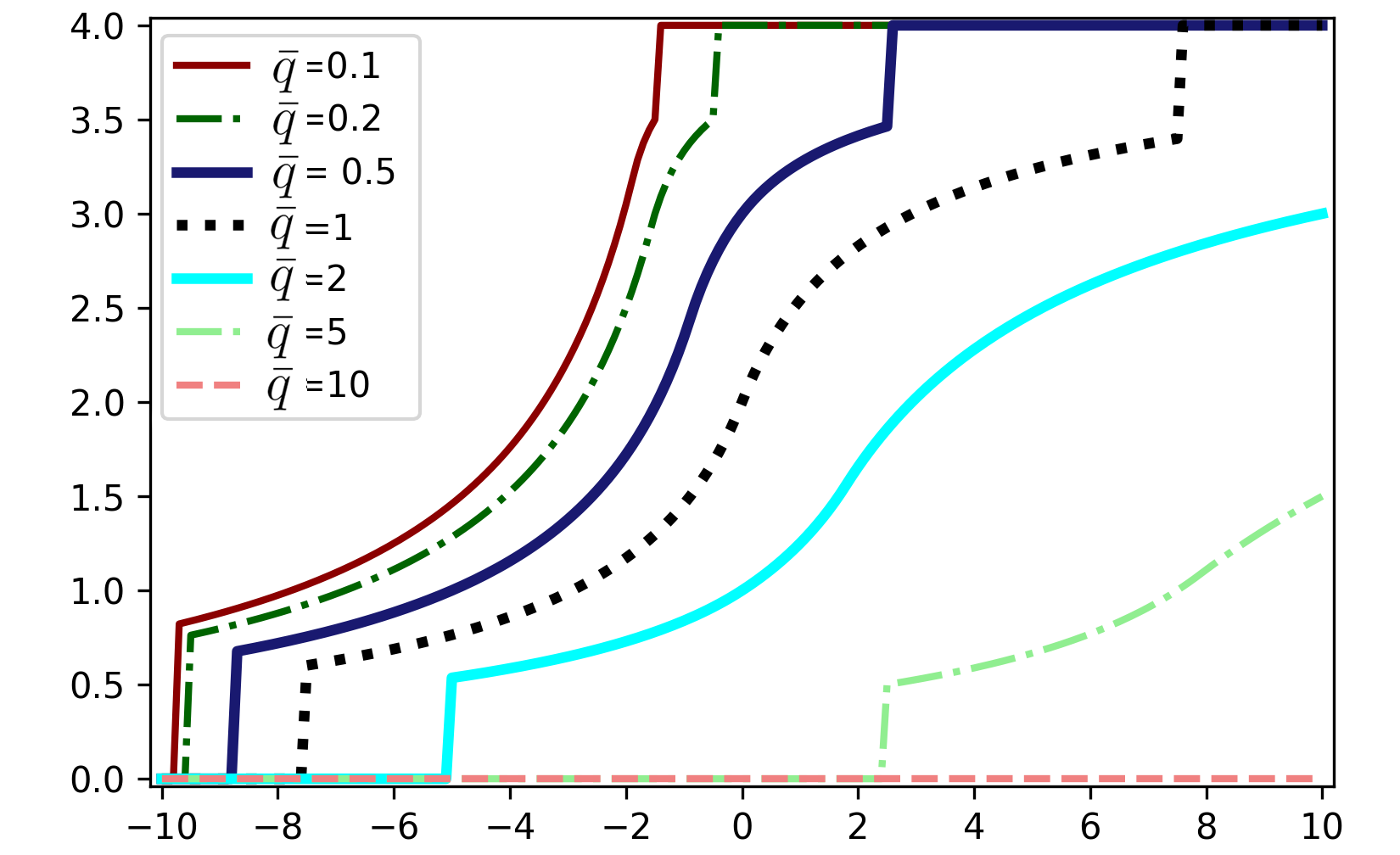}};
			\node[below=of img, node distance=0cm, yshift=1.1cm, xshift =0.2cm] {\scriptsize$\bp$};
			\node[left=of img, node distance=0cm, rotate=90, anchor=center,xshift=0cm, yshift=-1.2cm] {\scriptsize Payoff of Player A};
			\end{tikzpicture}
		}   
				\qquad
		%
		%
			\subfloat[$q_1 = q_2 = \bar{q}$, $q_3 = q_4 = 1/\bar{q}$]{
			\begin{tikzpicture}
			\node (img){\includegraphics[height = 0.14\textheight]{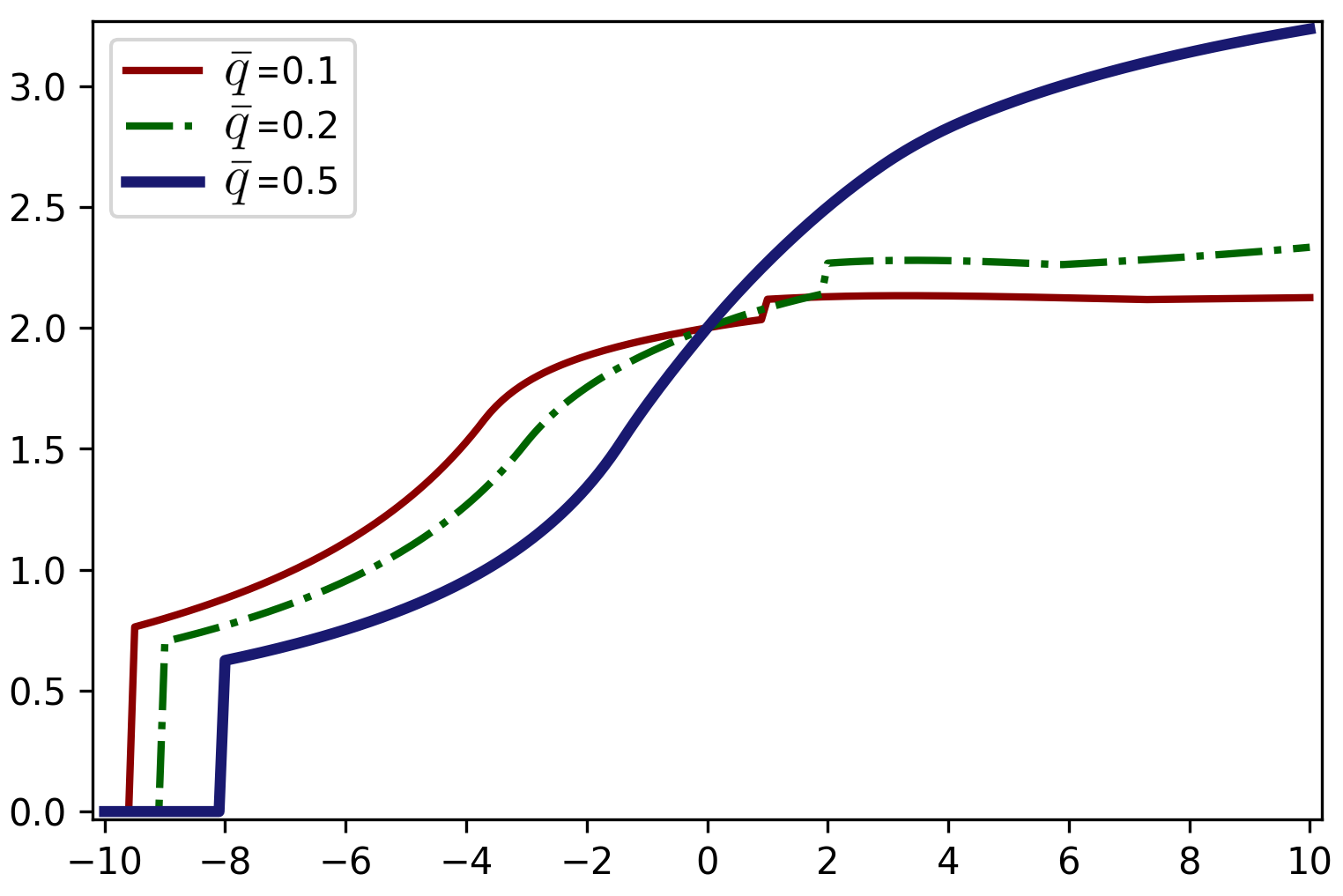}};
			\node[below=of img, node distance=0cm, yshift=1.1cm, xshift = 0.2cm] {\scriptsize$\bp$};
			\node[left=of img, node distance=0cm, rotate=90, anchor=center,xshift = 0cm,yshift = -1.0cm] {\scriptsize Payoff of Player A};
			\end{tikzpicture}
		}   
			\caption[]{Equilibrium payoffs of Player A in \FGL games where $n\!=\!4$, $X^A \!=\! X^B = 10$, $\alpha = \frac{1}{2}$; $w_i \!=\! 1$, $p_i\!=\!\bp$,~$\forall i \in [n]$.} \label{fig:Experiment1}
\end{figure*}

\cref{fig:Experiment1}(a) illustrates the (expected) equilibrium payoff of Player A in instances of the \FGL game where \mbox{$\bp \in \left\{- \! X^A,-\!0.99X^A,\ldots, 0, 0.01{X^A}, 0.02X^A, \ldots, X^A\right\}$} and \mbox{$\bq \in \{{1}/{10}, {1}/{5}, {1}/{2}, 1, 2, 5, 10 \}$}. First we observe that as $\bp$ increases and/or $\bq$ decreases, i.e., favoritism inclines towards Player~A, her expected payoff naturally increases (or at least does not decrease). Second,  most instances satisfy \cref{assum:1} and~\cref{assum:2}; in this case the curves representing Player A's payoffs are piecewise quadratic in $\bp$, which is consistent with its theoretical expression in \eqref{eq:payoff_A_GL}. Third, we observe that the equilibrium payoff of Player A, as a function of $\bp$, is discontinuous at several points. This is due to the fact that for instances where $\bp$ is very small or when $\bq$ is very large (i.e., Player B has strong favoritism), the game has trivial equilibria where Player B can guarantee to win all battlefields (and thus, Player A's payoff is 0). 
	 
Next, we consider game instances with the same $n, X^A, X^B, w_i$ and $p_i$, but we allow resource's effectiveness to vary across battlefields, in particular, $q_1 = q_2 = \bq$ and $q_3 = q_4 = 1/\bq$ where \mbox{$\bq \in \{0.1, 0.2,0.5 \}$}. \cref{fig:Experiment1}(b) reports Player A’s equilibrium payoff in these cases. First, we observe that when $\bp = 0$, the game is symmetric and each player's equilibrium payoff is precisely $ \sum_{i \in [n]} {w_i} /2 = 2$. Contrary to the case of \cref{fig:Experiment1}(a), however, in \cref{fig:Experiment1}(b) when $\bp$ is large, Player A can no longer guarantee to win all battlefields.
Moreover, in cases where $\bar{q_i} \in \{0.1, 0.2\}$, when $\bp \ge 2$ and it increases (i.e., Player A has strong pre-allocations), she cannot improve much her payoff. This is explained by the fact that although Player A can guarantee to win battlefields 1 and 2 (where $q_i$ is small), the effectiveness of her resources in battlefields 3 and 4 is too weak so she does not gain much in these battlefields. This illustrates the different effect on equilibrium of the favoritism in resources' effectiveness and in~pre-allocations.

In the next experiment, we consider the following situation: players compete on $n =4$ battlefields where $w_1 = w_2 = w_3 = 1$, $w_4 = 5$ and $q_i = 1, \forall i$ (i.e., resources have the same effectiveness). Player A has a total budget $X^* = 10$, but in this experiment a proportion $P < X^*$ taken out of this budget is pre-allocated. Then, Players A and B play an \FGL (or an \FCB) game where Player A's budget is $X^A = 10 - P$, Player B's budget is $X^B = 10$, and $p_i \ge 0$ such that $\sum_{i \in [n]} p_i = P$. We aim to analyze Player A's payoff when $P$ increases (i.e., when more and more of her budget is committed as pre-allocation).
While interesting, we leave the question of what is an optimal distribution of pre-allocation as future work; here we simply compare two simple distributions of Player A's pre-allocation: (i) in the \emph{spread strategy}, the pre-allocation is spread over all battlefields: $p_i = P/n, \forall i \in [n]$; (ii) in the \emph{focus strategy}, the pre-allocation is concentrated on battlefield 4 (the battlefield with a large value): $p_1 = p_2 = p_3 = 0$ and $p_4 = P$.
\begin{figure*}[h!]
		\centering
		\subfloat[]{
			\begin{tikzpicture}
			\node (img){\includegraphics[height = 0.20\textwidth]{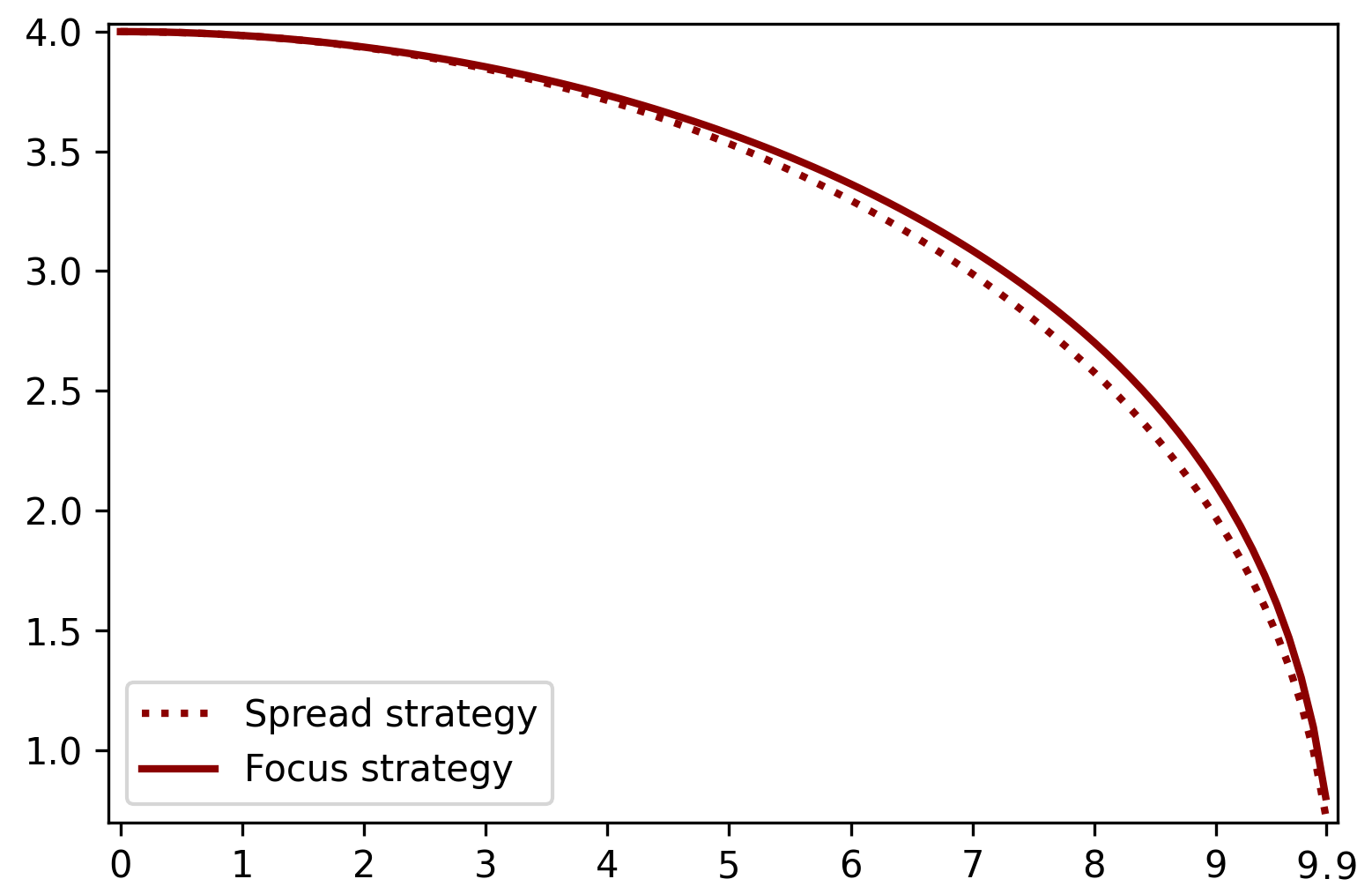}};
			\node[below=of img, node distance=0cm, yshift=1.1cm, xshift = 0.1cm] {\scriptsize $P$};
			\node[left=of img, node distance=0cm, rotate=90, anchor=center,xshift = 0cm,yshift = -1.0cm] {\scriptsize Payoff of Player A};
			\end{tikzpicture}
		}   
				\quad
		\subfloat[]{
			\begin{tikzpicture}
			\node (img){\includegraphics[height = 0.20\textwidth]{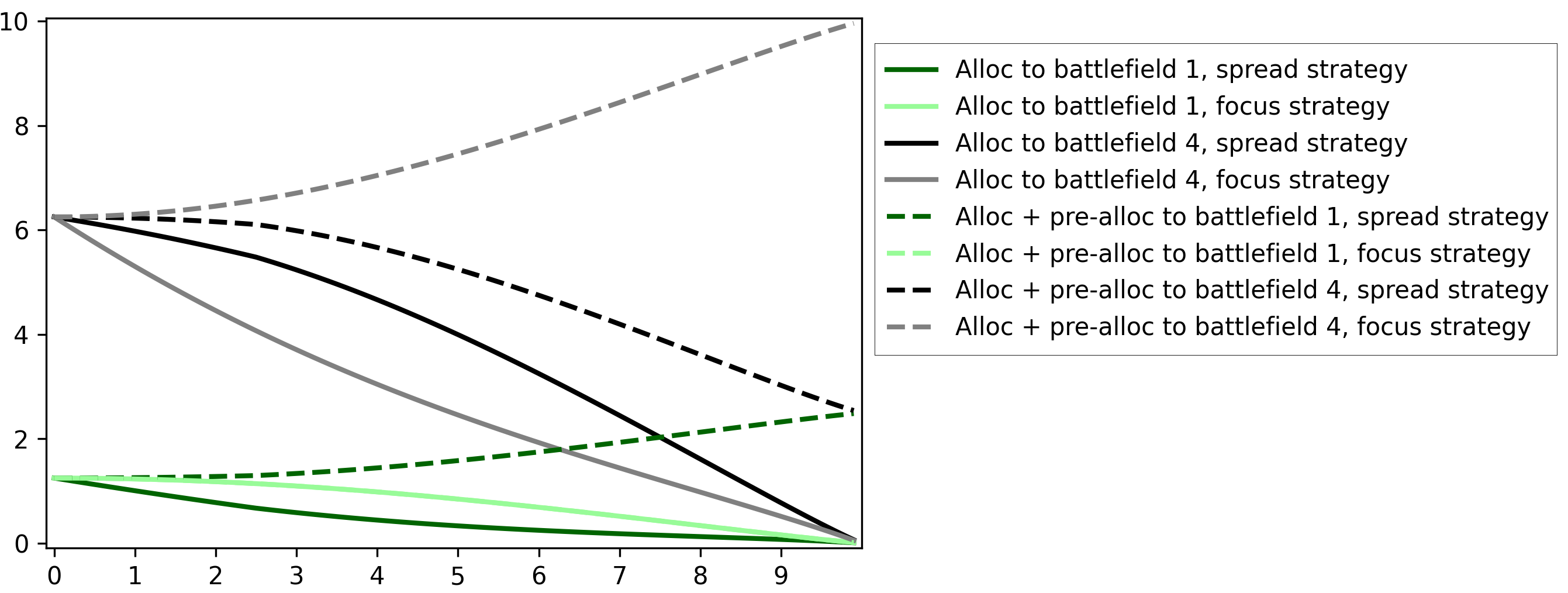}};
			\node[below=of img, node distance=0cm, yshift=1.1cm, xshift = -1.7cm] {\scriptsize $P$};
			\node[left=of img, node distance=0cm, rotate=90, anchor=center,xshift = 0cm,yshift = -1.0cm] {\scriptsize Allocations of Player A};
			\end{tikzpicture}
		}   
		%
%
		\caption[]{Player A's equilibrium payoffs and expected allocations in the \FGL game where $n=4$, \mbox{$w_1 = w_2 = w_3 = 1$}, $w_4 = 5$, $X^A = 10 - P$, $X^B = 10$, $q_i = 1, \forall i$ and $\sum_{i \in [n]}{p_i} =P$.} \label{fig:Experiment2}
\end{figure*}
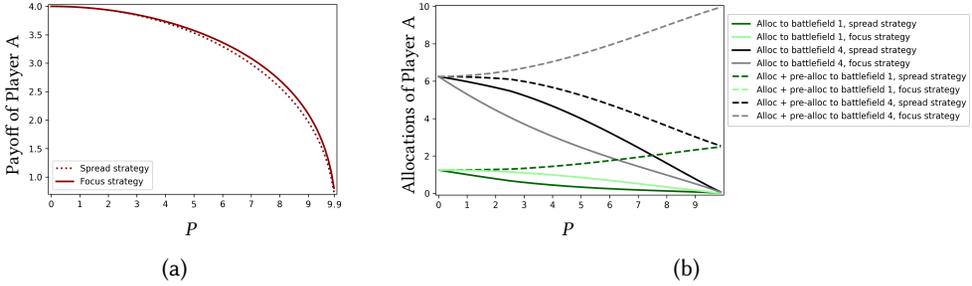

\cref{fig:Experiment2}(a) illustrates Player A's equilibrium payoff in the \FGL games as described above with $P \in \left\{0,0.1, \ldots, 9.9 \right\}$.  When $P=0$, it is the classical CB game and in this case, the game is symmetric, thus, each player has an equilibrium payoff $\sum_{i \in [n]}w_i /2 =4$. As $P$ increases, Player A's payoff decreases; intuitively, when the proportion of Player A's budget that is pre-allocated increases, she reveals more information about her (pre-)allocations and has less flexibility in play.
Interestingly, we observe that in instances where $P$ is relatively large, Player A gets a better payoff by distributing the pre-allocations using the focus strategy rather than by the spread strategy, \ie it is better for Player A to focus on ``securing'' the big battlefield. In \cref{fig:Experiment2}(b), we plot the expected allocations of Player A at equilibrium, alone and when added to her pre-allocations (note that since the parameters on battlefields 1, 2, 3 are identical, Player A's expected allocations are the same on these battlefields). As $P$ increases, Player A' expected allocations to the battlefields decrease since her budget \mbox{$X^A = X^* - P$} is reduced (although the aggregate of her allocation and pre-allocation increases in some cases). Interestingly, when $P$ is relatively small, Player A's expected allocations to battlefield 4 is much larger than that at battlefield 1 (this is because $w_4 > w_1$). However, when $P$ increases and under the focus strategy for pre-allocation, her allocation at battlefield 4 decreases quicker than that with the spread strategy. This is consistent with the intuition above that she already ``secures'' this battlefield by the focus strategy thus, she should not distribute a large allocation~there.
%
%

%
%
\section{Concluding Discussion}
\label{sec:conclu}

We introduced the \emph{Colonel Blotto game with favoritism} and analyzed its equilibria and approximate equilibria. We first characterized completely the equilibrium of all-pay auctions with favoritism. Using this, we then proved that there exists a set of optimal univariate distributions of the Colonel Blotto game with favoritism and gave a construction thereof. In several special cases of the Colonel Blotto game with favoritism, these univariate distributions give an exact equilibrium; in other cases, we derived an approximate equilibrium. We then proposed an algorithm that efficiently computes an approximation of the proposed optimal univariate distributions with arbitrarily small~error.

Our model of favoritism uses a linear-form of the winner-determination rule, defined in \cref{sec:FormCB}, similar to works on all-pay auctions with favoritism (see e.g., \cite{konrad2002investment,siegel2014asymmetric}). This is a natural formulation to capture the fundamental properties of favoritism due to its simplicity and to the natural interpretation of the parameters $p_i$ and $q_i$; and our Colonel Blotto game with favoritism---which is derived from this rule---provides a meaningful model for applications with favoritism (see \cref{sec:Intro} for several motivational examples). 
Nevertheless, an interesting direction for future investigations would be to consider more general winner-determination rules to model favoritism. A natural extension of our work is to consider polynomial winner-determination rules, that is, player A wins battlefield $i$ if $P^{m}_{A,i} (x^A_i) \ge P^{m}_{B,i} (x^B_i)$ and loses otherwise; where, $P^{m}_{A,i} (\cdot)$ and $P^{m}_{B,i} (\cdot)$ are some pre-determined polynomials of degree $m$ with coefficients dependent on the battlefield and the corresponding player (the first two coefficients are similar to $p_i$ and $q_i$). It would remain possible to map the corresponding Colonel Blotto game to a set of all-pay auctions; but these all-pay auctions would now have complex (polynomial) winner-determination rules. This raises two challenges: \emph{(i)} the equilibrium for such complex all-pay auctions (used to derive the optimal univariate distributions of the Colonel Blotto game) is not known and appears to be non-trivial to derive; and \emph{(ii)} proving the existence of the univariate distributions might require a fixed-point technique other than our solution in the linear-form case (cf. \cref{theo:OUDs})---or at least an adaptation of our proof technique.

%
%
\clearpage
\newpage
\bibliographystyle{plainnat}
\bibliography{mybibfile}

\begin{thebibliography}{49}
\providecommand{\natexlab}[1]{#1}
\providecommand{\url}[1]{\texttt{#1}}
\expandafter\ifx\csname urlstyle\endcsname\relax
  \providecommand{\doi}[1]{doi: #1}\else
  \providecommand{\doi}{doi: \begingroup \urlstyle{rm}\Url}\fi

\bibitem[Ahmadinejad et~al.(2016)Ahmadinejad, Dehghani, Hajiaghayi, Lucier,
  Mahini, and Seddighin]{Ahmadinejad16a}
Amir~Mahdi Ahmadinejad, Sina Dehghani, Mohammad~Taghi Hajiaghayi, Brendan
  Lucier, Hamid Mahini, and Saeed Seddighin.
\newblock {From Duels to Battlefields: Computing Equilibria of Blotto and Other
  Games}.
\newblock In \emph{Proceedings of the 13th AAAI Conference on Artificial
  Intelligence (AAAI)}, pages 369--375, 2016.

\bibitem[Baye et~al.(1994)Baye, Kovenock, and De~Vries]{baye1994}
M.~R. Baye, Dan Kovenock, and C.~G. De~Vries.
\newblock The solution to the tullock rent-seeking game when r> 2:
  Mixed-strategy equilibria and mean dissipation rates.
\newblock \emph{Public Choice}, 81\penalty0 (3-4):\penalty0 363--380, 1994.

\bibitem[Baye et~al.(1996)Baye, Kovenock, and de~Vries]{baye1996all}
M.~R. Baye, Dan Kovenock, and C.~G. de~Vries.
\newblock The all-pay auction with complete information.
\newblock \emph{Economic Theory}, 8\penalty0 (2):\penalty0 291--305, 6 1996.

\bibitem[Behnezhad et~al.(2017)Behnezhad, Dehghani, Derakhshan, Aghayi, and
  Seddighin]{Behnezhad17a}
Soheil Behnezhad, Sina Dehghani, Mahsa Derakhshan, Mohammad Taghi~Haji Aghayi,
  and Saeed Seddighin.
\newblock {Faster and Simpler Algorithm for Optimal Strategies of Blotto Game}.
\newblock In \emph{Proceedings of the 31st AAAI Conference on Artificial
  Intelligence (AAAI)}, pages 369--375, 2017.

\bibitem[Behnezhad et~al.(2018)Behnezhad, Blum, Derakhshan, HajiAghayi,
  Mahdian, Papadimitriou, Rivest, Seddighin, and
  Stark]{behnezhad2018battlefields}
Soheil Behnezhad, Avrim Blum, Mahsa Derakhshan, MohammadTaghi HajiAghayi,
  Mohammad Mahdian, Christos~H Papadimitriou, Ronald~L Rivest, Saeed Seddighin,
  and Philip~B Stark.
\newblock From battlefields to elections: Winning strategies of {B}lotto and
  auditing games.
\newblock In \emph{Proceedings of the 29th Annual ACM-SIAM Symposium on
  Discrete Algorithms}, pages 2291--2310. SIAM, 2018.

\bibitem[Behnezhad et~al.(2019)Behnezhad, Blum, Derakhshan, Hajiaghayi,
  Papadimitriou, and Seddighin]{Behnezhad19a}
Soheil Behnezhad, Avrim Blum, Mahsa Derakhshan, Mohammadtaghi Hajiaghayi,
  Christos~H. Papadimitriou, and Saeed Seddighin.
\newblock {Optimal Strategies of Blotto Games: Beyond Convexity}.
\newblock In \emph{Proceedings of the 2019 ACM Conference on Economics and
  Computation (EC)}, page 597–616, 2019.

\bibitem[Boix-Adser{\`a} et~al.(2020)Boix-Adser{\`a}, Edelman, and
  Jayanti]{Boix-Adsera20a}
Enric Boix-Adser{\`a}, Benjamin~L. Edelman, and Siddhartha Jayanti.
\newblock {The Multiplayer Colonel Blotto Game}.
\newblock In \emph{Proceedings of the 2020 ACM Conference on Economics and
  Computation (EC)}, 2020.

\bibitem[Borel(1921)]{borel1921}
E~Borel.
\newblock La th{\'e}orie du jeu et les {\'e}quations int{\'e}grales {\`a} noyau
  sym{\'e}trique.
\newblock \emph{Comptes rendus de l’Acad{\'e}mie des Sciences}, 173\penalty0
  (1304-1308):\penalty0 58, 1921.

\bibitem[Borel and Ville(1938)]{borel1938}
E~Borel and J~Ville.
\newblock \emph{Application de la th{\'e}orie des probabilit{\'e}s aux jeux de
  hasard}.
\newblock Gauthier-Villars, 1938.
\newblock original edition by Gauthier-Villars, Paris, 1938; reprinted at the
  end of Th{\'e}orie math{\'e}matique du bridge {\`a} la port{\'e}e de tous, by
  E. Borel \& A. Ch{\'e}ron, Editions Jacques Gabay, Paris.

\bibitem[Brouwer(1911)]{brouwer1911abbildung}
Luitzen Egbertus~Jan Brouwer.
\newblock {\"U}ber abbildung von mannigfaltigkeiten.
\newblock \emph{Mathematische annalen}, 71\penalty0 (1):\penalty0 97--115,
  1911.

\bibitem[Chandan et~al.(2020)Chandan, Paarporn, and Marden]{chandan2020showing}
Rahul Chandan, Keith Paarporn, and Jason~R Marden.
\newblock When showing your hand pays off: Announcing strategic intentions in
  {C}olonel {B}lotto games.
\newblock In \emph{Proceedings of the 2020 American Control Conference (ACC)},
  2020.

\bibitem[Chia(2012)]{chia2012}
Pern~Hui Chia.
\newblock {C}olonel {B}lotto in web security.
\newblock In \emph{The 11th Workshop on Economics and Information Security,
  WEIS Rump Session}, pages 141--150, 2012.

\bibitem[Chinn and Steenrod(1966)]{chinn1966first}
WG~Chinn and NE~Steenrod.
\newblock First concept of topology: The geometry of mapping of segments,
  curves, circles and disks.
\newblock \emph{6aed., AMS, New Mathematical Library}, 1966.

\bibitem[Corch{\'o}n(2007)]{corchon2007theory}
Luis~C Corch{\'o}n.
\newblock The theory of contests: a survey.
\newblock \emph{Review of Economic Design}, 11\penalty0 (2):\penalty0 69--100,
  9 2007.
\newblock ISSN 1434-4750.

\bibitem[Fu(2006)]{fu2006theory}
Qiang Fu.
\newblock A theory of affirmative action in college admissions.
\newblock \emph{Economic Inquiry}, 44\penalty0 (3):\penalty0 420--428, 2006.

\bibitem[Fu and Wu(2019)]{fu2019contests}
Qiang Fu and Zenan Wu.
\newblock Contests: Theory and topics.
\newblock In \emph{Oxford Research Encyclopedia of Economics and Finance}.
  Oxford University Press, 2019.

\bibitem[Gross(1950)]{gross1950}
Oliver Gross.
\newblock The symmetric {B}lotto game.
\newblock Technical report, US Air Force Project RAND Research Memorandum,
  1950.

\bibitem[Gross and Wagner(1950)]{grosswagner}
Oliver Gross and Robert Wagner.
\newblock A continuous {C}olonel {B}lotto game.
\newblock Technical report, RAND project air force Santa Monica CA, 1950.

\bibitem[{Hajimirsaadeghi} and {Mandayam}(2017)]{hajimirsaadeghi2017dynamic}
M.~{Hajimirsaadeghi} and N.~B. {Mandayam}.
\newblock A dynamic colonel {B}lotto game model for spectrum sharing in
  wireless networks.
\newblock In \emph{Proceedings of the 55th Annual Allerton Conference on
  Communication, Control, and Computing (Allerton)}, pages 287--294, 10 2017.

\bibitem[Hart(2008)]{hart2008}
Sergiu Hart.
\newblock Discrete {C}olonel {B}lotto and {G}eneral {L}otto games.
\newblock \emph{International Journal of Game Theory}, 36\penalty0
  (3):\penalty0 441--460, 2008.

\bibitem[Hillman and Riley(1989)]{hillman1989}
Arye~L. Hillman and John~G. Riley.
\newblock Politically contestable rents and transfers.
\newblock \emph{Economics \& Politics}, 1\penalty0 (1):\penalty0 17--39, 1989.

\bibitem[Hoeffding(1963)]{hoeffding1963probability}
W~Hoeffding.
\newblock Probability inequalities for sums of bounded random variables.
\newblock \emph{Journal of the American Statistical Association,}, 58:\penalty0
  13--30, 1963.

\bibitem[Hortala-Vallve and Llorente-Saguer(2012)]{hortala2012}
Rafael Hortala-Vallve and Aniol Llorente-Saguer.
\newblock {Pure strategy {N}ash equilibria in non-zero sum {C}olonel {B}lotto
  games}.
\newblock \emph{International Journal of Game Theory}, 41\penalty0
  (2):\penalty0 331--343, 5 2012.

\bibitem[Konrad(2002)]{konrad2002investment}
Kai~A Konrad.
\newblock Investment in the absence of property rights; the role of incumbency
  advantages.
\newblock \emph{European Economic Review}, 46\penalty0 (8):\penalty0
  1521--1537, 2002.

\bibitem[Konrad and Kovenock(2009)]{konrad2009}
Kai~A Konrad and Dan Kovenock.
\newblock Multi-battle contests.
\newblock \emph{Games and Economic Behavior}, 66\penalty0 (1):\penalty0
  256--274, 2009.

\bibitem[Kovenock and Roberson(2012{\natexlab{a}})]{kovenock2012}
Dan Kovenock and Brian Roberson.
\newblock Coalitional {C}olonel {B}lotto games with application to the
  economics of alliances.
\newblock \emph{Journal of Public Economic Theory}, 14\penalty0 (4):\penalty0
  653--676, 2012{\natexlab{a}}.

\bibitem[Kovenock and Roberson(2012{\natexlab{b}})]{kovenock2012conflicts}
Dan Kovenock and Brian Roberson.
\newblock Conflicts with multiple battlefields.
\newblock In \emph{The Oxford Handbook of the Economics of Peace and Conflict}.
  Oxford University Press, 2012{\natexlab{b}}.

\bibitem[Kovenock and Roberson(2020)]{kovenock2020generalizations}
Dan Kovenock and Brian Roberson.
\newblock {Generalizations of the General Lotto and Colonel Blotto games}.
\newblock \emph{Economic Theory}, pages 1--36, 2020.

\bibitem[Kruse and Deely(1969)]{kruse1969joint}
RL~Kruse and JJ~Deely.
\newblock Joint continuity of monotonic functions.
\newblock \emph{The American Mathematical Monthly}, 76\penalty0 (1):\penalty0
  74--76, 1969.

\bibitem[Kulpa(1997)]{kulpa1997poincare}
Wladyslaw Kulpa.
\newblock The poincar{\'e}-miranda theorem.
\newblock \emph{The American Mathematical Monthly}, 104\penalty0 (6):\penalty0
  545--550, 1997.

\bibitem[Laslier(2002)]{laslier2002}
J.~F. Laslier.
\newblock How two-party competition treats minorities.
\newblock \emph{Review of Economic Design}, 7\penalty0 (3):\penalty0 297--307,
  2002.

\bibitem[Li and Yu(2012)]{li2012contests}
Sanxi Li and Jun Yu.
\newblock Contests with endogenous discrimination.
\newblock \emph{Economics Letters}, 117\penalty0 (3):\penalty0 834--836, 2012.

\bibitem[Macdonell and Mastronardi(2015)]{macdonell2015}
Scott~T. Macdonell and Nick Mastronardi.
\newblock Waging simple wars: a complete characterization of two-battlefield
  {B}lotto equilibria.
\newblock \emph{Economic Theory}, 58\penalty0 (1):\penalty0 183--216, 1 2015.

\bibitem[Masucci and Silva(2014)]{masucci2014}
Antonia~Maria Masucci and Alonso Silva.
\newblock Strategic resource allocation for competitive influence in social
  networks.
\newblock In \emph{Proceedings of the 52nd Annual Allerton Conference on
  Communication, Control, and Computing (Allerton)}, pages 951--958, 9 2014.

\bibitem[Masucci and Silva(2015)]{masucci2015}
Antonia~Maria Masucci and Alonso Silva.
\newblock Defensive resource allocation in social networks.
\newblock In \emph{Proceedings of the 54th IEEE Conference on Decision and
  Control (CDC)}, pages 2927--2932, 12 2015.

\bibitem[Myerson(1991)]{myerson1991game}
Roger~B Myerson.
\newblock \emph{Game Theory: Analysis of Conflict}.
\newblock Harvard University Press, 1991.
\newblock ISBN 9780674341159.

\bibitem[Myerson(1993)]{myerson1993incentives}
Roger~B Myerson.
\newblock Incentives to cultivate favored minorities under alternative
  electoral systems.
\newblock \emph{American Political Science Review}, 87\penalty0 (4):\penalty0
  856--869, 1993.

\bibitem[Nisan et~al.(2007)Nisan, Roughgarden, Tardos, and Vazirani]{Nisan07}
Noam Nisan, Tim Roughgarden, Eva Tardos, and Vijay~V. Vazirani.
\newblock \emph{Algorithmic Game Theory}.
\newblock Cambridge University Press, 2007.
\newblock ISBN 0521872820.

\bibitem[Pastine and Pastine(2012)]{pastine2012incumbency}
Ivan Pastine and Tuvana Pastine.
\newblock Incumbency advantage and political campaign spending limits.
\newblock \emph{Journal of Public Economics}, 96\penalty0 (1-2):\penalty0
  20--32, 2012.

\bibitem[Roberson(2006)]{roberson2006}
Brian Roberson.
\newblock The {C}olonel {B}lotto game.
\newblock \emph{Economic Theory}, 29\penalty0 (1):\penalty0 1--24, 2006.
\newblock ISSN 09382259, 14320479.

\bibitem[Schwartz et~al.(2014)Schwartz, Loiseau, and Sastry]{schwartz2014}
Galina Schwartz, Patrick Loiseau, and Shankar~S Sastry.
\newblock The heterogeneous {C}olonel {B}lotto game.
\newblock In \emph{Proceedings of the 7th International Conference on Network
  Games, Control and Optimization (NetGCoop)}, pages 232--238, 2014.

\bibitem[Siegel(2009)]{siegel2009all}
Ron Siegel.
\newblock All-pay contests.
\newblock \emph{Econometrica}, 77\penalty0 (1):\penalty0 71--92, 2009.

\bibitem[Siegel(2014)]{siegel2014asymmetric}
Ron Siegel.
\newblock Asymmetric contests with head starts and nonmonotonic costs.
\newblock \emph{American Economic Journal: Microeconomics}, 6\penalty0
  (3):\penalty0 59--105, 2014.

\bibitem[Thomas(2017)]{thomas2017}
Caroline Thomas.
\newblock N-dimensional {B}lotto game with heterogeneous battlefield values.
\newblock \emph{Economic Theory}, pages 1--36, 2017.

\bibitem[Viro et~al.(2008)Viro, Ivanov, Netsvetaev, and
  Kharlamov]{viro2008elementary}
O~Ya Viro, OA~Ivanov, N~Yu Netsvetaev, and VM~Kharlamov.
\newblock \emph{Elementary topology}.
\newblock American Mathematical Soc., 2008.

\bibitem[Vu et~al.(2018)Vu, Loiseau, and Silva]{vu18a}
Dong~Quan Vu, Patrick Loiseau, and Alonso Silva.
\newblock {Efficient Computation of Approximate Equilibria in Discrete Colonel
  Blotto Games}.
\newblock In \emph{Proceedings of the 27th International Joint Conference on
  Artificial Intelligence and the 23rd European Conference on Artificial
  Intelligence (IJCAI-ECAI)}, pages 519--526, 2018.

\bibitem[Vu et~al.(2020{\natexlab{a}})Vu, Loiseau, and
  Silva]{vu2019approximate}
Dong~Quan Vu, Patrick Loiseau, and Alonso Silva.
\newblock {Approximate Equilibria in Generalized Colonel Blotto and Generalized
  Lottery Blotto Games}, 2020{\natexlab{a}}.
\newblock arXiv:1910.06559v2.

\bibitem[Vu et~al.(2020{\natexlab{b}})Vu, Loiseau, Silva, and
  Tran-Thanh]{vu2020}
Dong~Quan Vu, Patrick Loiseau, Alonso Silva, and Long Tran-Thanh.
\newblock Path planning problems with side observations—when colonels play
  hide-and-seek.
\newblock In \emph{Proceedings of the 34th AAAI Conference on Artificial
  Intelligence (AAAI)}, 2020{\natexlab{b}}.

\bibitem[Zapata and Mart{\'\i}n(2012)]{zapata2012geometric}
Juan-Luis~Garc{\'\i}a Zapata and Juan Carlos~D{\'\i}az Mart{\'\i}n.
\newblock A geometric algorithm for winding number computation with complexity
  analysis.
\newblock \emph{Journal of Complexity}, 28\penalty0 (3):\penalty0 320--345,
  2012.

\end{thebibliography}

%
%
\newpage
\appendix
\renewcommand{\thesection}{Appendix~\Alph{section}}%
\setcounter{lemma}{0}
\setcounter{definition}{0}
\renewcommand{\thelemma}{\Alph{section}-\arabic{lemma}}
\renewcommand{\thedefinition}{\Alph{section}-\arabic{definition}}
\renewcommand{\theproposition}{\Alph{section}-\arabic{proposition}}
\setcounter{equation}{0}
\renewcommand\theequation{\Alph{section}.\arabic{equation}}
\renewcommand{\theexample}{\Alph{section}-\arabic{lemma}}


%
\section{Supplementary Materials for Results in Section~\ref{sec:EquiAPA}}

\subsection{Known results on equilibria of all-pay auctions}
		\label{appen:preli}
		
		To ease the comparison between the state-of-the-art and our main results on the \FAPA game (Section~\ref{sec:EquiAPA}), we review here several results in the literature. The results stated in this section are extracted from previous works and rewritten in our notations. 
		

		\begin{theorem}[extracted from \cite{baye1994,hillman1989}]
			\label{theo:clasAPA}
			In the classical two-player all-pay auction (i.e., an $\FAPA$ with $p = 0$, $q =1$ and $\alpha = 1/2$), if $u^A \ge u^B$, there exists a unique mixed equilibrium where Players A and B bid according to the following distributions:
			\begin{align}
			& A(x) =   \left\{ \begin{array}{l}
			\frac{x}{u^B}, \forall x \in \left[ 0, u^B \right], \\
			1  \quad  , \forall x > u^B, 
			\end{array} \right.
			\textrm{ and } 
			& B(x) =   \left\{ \begin{array}{l}
			\frac{u^A-u^B}{u^A} + \frac{x}{u^A}, \forall x \in \left[ 0, u^B \right] \\
			1 \qquad \qquad , \forall x > u^B.
			\end{array} \right. 
			\end{align}
			In this equilibrium, Player A's payoff is $\Pi^A = u^A-u^B$, Player B's payoff is $\Pi^B = 0$. 
		\end{theorem}

		In intuition, $A(x)$ is the uniform distribution on $[0, u^B]$ and $B(x)$ is the distribution with a (strictly positive) probability mass at $0$ and the remaining mass is distributed uniformly in $(0, u^B]$. In the case where $u^B > u^A$, players exchange their roles and a similar statement to Theorem~\ref{theo:clasAPA} can be easily deduced.

		\begin{theorem}[extracted from \cite{konrad2002investment}]
			\label{theo:incumAPA}
			In the $\FAPA$ where $u^A = u^B = u$, $ q>0$, $0< q <1$ and $\alpha = 1/2$, 
			\begin{itemize}
				\item[$(i)$] If $q u - p \le 0$, there exists a unique pure equilibrium where players' bids are $x^A = x^B = 0$ and their equilibrium payoffs are $\Pi^A = u$ and $\Pi^B = 0$.
				\item[($ii$)] If $0 < q u - p$, there exists no pure equilibrium; the unique mixed equilibrium is where Players A and B draw their bids from the following distributions:
				\begin{align}
				& \bar{A}(x) =   \left\{ \begin{array}{l}
				\frac{p}{q u} + \frac{x}{q u}, \forall x \in \left[ 0, q u - p \right], \\
				1                  \qquad \qquad, \forall x > q u -p, 
				\end{array} \right.
				\textrm{ and } 
				& \bar{B}(x) =   \left\{ \begin{array}{l}
				1 - q + \frac{p}{u}, \forall x \in \left[ 0, \frac{p}{q} \right) \\
				1 - q + \frac{q \cdot x}{u}, \forall x \in \left[\frac{p}{q}, u \right], \\
				1 \qquad \qquad , \forall x > u.
				\end{array} \right. 
				\end{align}
				In this mixed equilibrium, players' payoffs are $\Pi^A = u(1-q) + p$ and $\Pi^B =0$.
			\end{itemize}
		\end{theorem}
		
		Intuitively, $\bar{A}(x)$ is the distribution placing a positive mass at 0 and distributing the remaining mass uniformly on $(0, qu - p]$ and $\bar{B}(x)$ is the distribution placing a mass at 0 and distributing the remaining mass uniformly on $\left( p/q, u \right)$. It is possible to deduce similar results for the case where $p < 0 $ and $ q > 1$ (it is not stated explicitly in \cite{konrad2002investment}). However, \cite{konrad2002investment} does not consider the cases where the additive asymmetric parameter $p$ is in favor of one player while the multiplicative asymmetric parameter $q$ is in favor of the~other.

%
%
\subsection{Proof of Theorem~\ref{theo:positive}}\label{appen:proofAPAPos}
	\begin{proof}
	
	\textbf{Proof of Result $(i)$:} For any $x^B \ge u^B$ and any $x^A$, we have \mbox{$\Pi_{\FAPA}^B\left( x^A, x^B \right) < 0$}. Moreover, due to the condition $q u^B - p \le 0$, we have $x^A > q x^B - p$ for any $x^A \ge 0$ and $0 \le x^B < u^B$; that is, player B always loses if she bids strictly lower than $u^B$. Trivially, $x^B = 0$ is the unique dominant strategy of player B. Player A's best response against $x^B =0$ is $x^A=0$. In conclusion, we have: 
	
	\begin{align*}
	& \Pi_{\FAPA}^A\left (0, 0\right) = u^A \textrm{ and } \Pi_{\FAPA}^A \left(x^A,0\right) = u^A - x^A < u^A, \forall x^A >0,\\
	& \Pi_{\FAPA}^B\left (0, 0\right) = 0 \textrm{ and }  \Pi_{\FAPA}^B \left(0, x^B\right) <0, \forall x^B >0,\\
	\end{align*}

	\textbf{Proof of Result $(ii)$} First, from $0 < q u^B - p \le u^A$, we have $0 \le p/q < u^B$. We prove (by contradiction) that there exists no pure equilibrium under this condition. Assume that the profile $x^A, x^B$ is a pure equilibrium of the $\FAPA$ game. We consider two cases:
	
	\begin{itemize}
		\item Case 1: If $x^A = 0$, then player B's best response is to choose $ x^B = p/q + \varepsilon$ with an infinitesimal $\varepsilon >0$ since by doing it, she can guarantee to win (since $q(p/q+\varepsilon) -p =q \varepsilon > 0$) and gets the payoff $u^B - p/q - \varepsilon > 0$.\footnote{Note that if player B choose $x^B = 0$, she loses and hes payoff is only $0$.} However, player A's best response against $ x^B = p/q + \varepsilon$ \emph{is not} $x^A= 0$.\footnote{Player A's best response against $ x^B = p/q + \varepsilon$ is $x^A= q \varepsilon + \delta$ (with an infinitesimal $\delta >0$ such that $u^A - q \varepsilon - \delta > 0$).}
		\item Case 2: If $x^A > 0$, then player B's best response is either $x^B = (x^A + p)/q + \varepsilon$ if there exists $\varepsilon > 0$ small enough such that $q u^B - p -x^A - \varepsilon > 0$ or $x^B = 0$ if there is no such $\varepsilon$. However, $x^A > 0$ is not the best response of player A against neither $x^B = (x^A + p)/q + \varepsilon$ nor against $x^B = 0$.\footnote{The best response of player A against $x^B = (x^A + p)/q + \varepsilon$ is $x^A + q\varepsilon + \delta$ where $0< \delta < u^A -x^A - q \varepsilon$ ($\delta$ exists thanks to the condition on $\varepsilon$ and that $q u^B -p \le u^A $) and her best response against $x^B =0 $ is $x^A =0$.}
	\end{itemize}
	We conclude that $x^A, x^B$ cannot the best response against each other; thus, there exists no pure equilibrium in this case.
	
	Now, we prove that if player B plays according to $\Biip$, player A has no incentive to deviate from playing according to $\Aiip$. Denote by $A^+_2$ and $B^+_2$ the random variables that correspond to $\Aiip$ and $\Biip$, since $\Aiip$ is a continuous distribution on $\left( 0, q u^B - p \right]$, we have:
	\begin{align}
	\Pi_{\FAPA}^A \left(\Aiip, \Biip \right) & = \left[u^A \prob\left(B^+_2 < \frac{p}{q} \right) - 0 \right] \prob\left (A^+_2 = 0 \right)     
	+ \left[\alpha u^A \prob\left(B^+_2 = \frac{p}{q} \right) - 0 \right] \prob\left (A^+_2 = 0 \right) \nonumber \\
	& \qquad \qquad + \int_{0}^{q u^B - p} \left[ u^A \prob \left( B^+_2 < \frac{x + p}{q} \right) - x \right]  \de \Aiip(x) \nonumber \\
	& = u^A \Biip\left( \frac{p}{q}\right) \frac{p}{q u^B} + 0 + \int_{0}^{q u^B - p} \left[ u^A \Biip \left(\frac{x + p}{q} \right) - x \right]  \de \Aiip(x) \label{eq:case1.2} \\
	& = \left( u^A - qu^B + p \right) \frac{p}{q u^B} + \int_{0}^{q u^B - p} \left( u^A - qu^B + p \right)\frac{1}{q u^B} \de x \nonumber\\
	& = u^A - q u^B +p. \nonumber
	\end{align} 
	Here, \eqref{eq:case1.2} comes from the fact that $\prob \left(B^+_2 = p/q \right) =0,$ due to definition. Now, if player A plays a pure strategy $x^A > q u^B -p$ while player B plays $\Biip$, her payoff~is:
	\begin{equation*}
	\Pi_{\FAPA}^A \left(x^A, \Biip \right) \le  u^A - x^A < u^A - q u^B +p = \Pi_{\FAPA}^A \left(\Aiip, \Biip \right).
	\end{equation*}
	Moreover, for any pure strategy $x^A \in [0, q u^B - p]$, we have:
	\begin{align*}
	\Pi_{\FAPA}^A \left(x^A, \Biip \right) &=  u^A  \prob\left(B^+_2 < \frac{x^A \!+\! p}{q} \right) +  \alpha u^A  \prob\left(B^+_2 = \frac{x^A \!+\! p}{q} \right) \!-\! x^A\\
	& \le  u^A \Biip \left( \frac{x^A \!+\! p}{q} \right) \!-\! x^A = u^A \left[1 \!- \!\frac{q u^B}{u^A} \!+\! \frac{q}{u^A}  \frac{(x^A \!+\! p)}{q}\right] \!-\! x^A \\
	& \!=\! u^A \!-\! q u^B \!+\! p\\
	& \!=\! \Pi_{\FAPA}^A \left(\Aiip, \Biip \right).
	\end{align*}
	In conclusion, $\Pi^A\left(\Aiip, \Biip \right) \ge \Pi^A \left(x^A, \Biip \right)$ for any $x^A \ge 0$.
	
	Similarly, we prove that when player A plays $\Aiip$, player B has no incentive to deviate from $\Biip$. Indeed, since $\Biip$ is a continuous distribution on $[p/q, u^B]$, we~have
	\begin{align}
	\Pi_{\FAPA}^B \left(\Aiip, \Biip \right) & = \left[u^B \prob\left(A^+_2 < 0 \right) - \frac{p}{q} \right] \prob\left (B^+_2 = \frac{p}{q} \right) \nonumber\\
	& \qquad \qquad + \left[(1-\alpha) u^B \prob\left(A^+_2 = 0 \right) - \frac{p}{q} \right] \prob\left (B^+_2 = \frac{p}{q} \right) \nonumber \\
	& \qquad \qquad + \int_{p/q}^{u^B} \left[ u^B \prob \left( A^+_2 < qx - p \right) - x \right]  \de \Biip(x) \nonumber \\
	& = 0 + 0 + \int_{p/q}^{u^B} \left[ u^B \Aiip \left( qx - p \right) - x \right]  \de \Biip(x) \label{eq:explain} \\
	& = \int_{p/q}^{u^B} \left[ u^B \left(\frac{p}{ q u^B} + \frac{qx - p}{q u^B} \right) - x \right]\frac{q}{u^A} \de x  \nonumber \\
	& = 0. \nonumber
	\end{align} 
	Here, \eqref{eq:explain} comes from the fact $\prob \left(B^+_2 = p/q  \right) = 0$ and that $\prob (A^+_2 = z) =0$ for any \mbox{$z \in (0, q u^B -p]$} due to definition. Now, as stated above, for any pure strategy \mbox{$x^B > u^B$}, trivially $\Pi^B_{\FAPA}(\Aiip, x^B)$ < 0. Moreover, \mbox{$\Pi_{\FAPA}^B \left( \Aiip, x^B \right)  \le u^B \Aiip \left( q x^B \!-\! p\right) \! -\! x^B \!= \!0$} for any \mbox{$x^B \in [0, u^B]$}. Therefore, we conclude that \mbox{$\Pi_{\FAPA}^B\left(\Aiip, \Biip \right) \ge \Pi_{\FAPA}^B \left(\Aiip,x^B \right)$} for any $x^B \ge 0$.
	
	\paragraph{Proof of Result~$(iii)$} 
	Similarly to the proof of Result $(ii)$, we can prove that there exists no pure equilibrium if $ q u^B - p > u^A >0$. Now, let us denote by $A^+_3$ and $B^+_3$ the random variables that correspond to $\Aiiip$ and~$\Biiip$; we prove that if player B plays according to $\Biiip$, player A has no incentive to deviate from playing according to $\Aiiip$.
	\begin{align*}
	\Pi_{\FAPA}^A \left(\Aiiip, \Biiip \right) & = \left[u^A \prob\left(B^+_3 < \frac{p}{q} \right) - 0 \right] \prob\left (A^+_3 = 0 \right)     
	+ \left[\alpha u^A \prob\left(B^+_3 = \frac{p}{q} \right) - 0 \right] \prob\left (A^+_3 = 0 \right) \nonumber \\
	& \qquad \qquad + \int_{0}^{u^A} \left[ u^A \prob \left( B^+_3 < \frac{x + p}{q} \right) - x \right]  \de \Aiiip(x) \nonumber \\
	& = 0 + 0 + \int_{0}^{u^A} \left[ u^A \Biip \left(\frac{x + p}{q} \right) - x \right]  \de \Aiiip(x) \\
	& =  \int_{0}^{u^A} \left[u^A \left(\frac{-p}{u^A} + \frac{q}{u^A} \frac{(x + p)}{q} \right) - x \right]  \de \Aiiip(x)\\
	& = 0. \nonumber
	\end{align*}
	Moreover, trivially, for any $x^A > u^B$, we have $\Pi_{\FAPA}^A \left(x^A, \Biiip \right) < 0$ and for any \mbox{$x^A \in [0, u^B]$}, we have
	\begin{align*}
	\Pi_{\FAPA}^A \left(x^A, \Biiip \right) \le  & u^A \Biiip \left( \frac{x^A + p}{q} \right) - x^A \\
	= & u^A \left[\frac{-p}{u^A} + \frac{q}{u^A} \frac{(x^A + p)}{q} \right] - x^A \\
	= & 0 = \Pi_{\FAPA}^A \left(\Aiiip, \Biiip \right).
	\end{align*}
	Therefore, $\Pi_{\FAPA}^A\left(\Aiiip, \Biiip \right) \ge \Pi_{\FAPA}^A \left(x^A, \Biiip \right)$ for any $x^A \ge 0$.
	
	On the other hand, since $\Biiip$ is a continuous distribution on $\left[\frac{p}{q}, \frac{u^A+p}{q} \right]$, we do not need to consider the tie cases and we can deduce that: 
	\begin{align}
	\Pi_{\FAPA}^B \left(\Aiiip, \Biiip \right) & = \int_{p/q}^{\frac{u^A + p}{q}} \left[ u^B \Aiiip \left(qx - p \right) - x \right]  \de \Biiip(x) \nonumber \\
	& = \int_{p/q}^{\frac{u^A+p}{q}} \left[ u^B \left( 1- \frac{u^A}{q u^B} + \frac{qx- p}{q u^B} \right) - x \right]\frac{q}{u^A} \de x  \nonumber \\
	& = u^B - \frac{u^A + p}{q}. \nonumber
	\end{align} 
	Moreover, trivially, for any $x^B > u^B$, we have $\Pi^B\left( \Aiiip, x^B \right) < 0 < u^B - \frac{u^A + p}{q}$; and for any \mbox{$x^B \in [0, u^B]$}, we have:
	\begin{equation*}
		\Pi_{\FAPA}^B \left( \Aiiip, x^B \right)  \le u^B \Aiiip \left( q x^B - p\right) - x^B =  u^B \left( 1- \frac{u^A}{q u^B} + \frac{qx^B- p}{q u^B} \right) - x^B =  u^B - \frac{u^A + p}{q}.
	\end{equation*}
	
    Therefore, we can conclude that \mbox{$\Pi_{\FAPA}^B\left(\Aiiip, \Biiip \right) \ge \Pi^B \left(\Aiiip,x^B \right)$} for any $x^B \ge 0$.

    Finally, for a proof of uniqueness of the mixed equilibrium in Result~$(ii)$ and~$(iii)$, we can follow the scheme presented by~\cite{baye1996all} and check through a series of lemmas. This is a standard approach in the literature of all-pay auction and we omit the detailed proof here. 	
\end{proof}

%
%
%
\subsection{Equilibrium of \FAPA when $p <0$, $q > 0$} \label{sec:appen:APAneg}
		We now consider the $\FAPA$ game in the case $p < 0$. We first define $p^{\prime} = -p/q$ and $q^{\prime} = 1/q$. Since $p<0$, we have $p^{\prime} >0$. Moreover, for any $x^A, x^B$, we have:
\begin{equation*}
\be \left(x^A, q x^B - p \right) = \be \left( (x^A +p)/q, x^B\right) = \be \left(  q^{\prime} x^A - p^{\prime}, x^B \right).
\end{equation*}
Therefore, the $\FAPA$ game with $p < 0$ (and $q>0$) is equivalent to an \FAPA with $p^{\prime} >0 $ (and $q^{\prime} >0$) in which the roles of players are exchanged. Applying Theorem~\ref{theo:positive} to this \emph{new} game, we can deduce the following theorem:

		\begin{theorem}
			\label{theo:negative}
			In the $\FAPA$ game where $p < 0$, we have the following results:
			
			\begin{itemize}
				\item[$(i)$] If $(u^A + p)/q \le 0$,\footnote{That is $q^{\prime} u^B - p^{\prime} \le 0$.}
				there exists a unique pure equilibrium where players' bids are $x^A = x^B = 0$ and their equilibrium payoffs are $\Pi^A = 0$ and $\Pi^B = u^B$ respectively.
				\item[($ii$)] If $0 < (u^A + p)/q \le  u^B$,\footnote{That is $0 \le q^{\prime} u^A - p^{\prime} \le u^B$.}
				there exists no pure equilibrium; there is a mixed equilibrium where Player A (resp. Player B) draws her bid from the distribution $\Aiip$ (resp. $\Biip$) defined as follows.
				\begin{align}
				& \Aiim(x) =   \left\{ \begin{array}{l}
				1-\frac{u^A}{q u^B} - \frac{p}{q u^B}, \forall x \in \left[ 0, - p \right), \\
				1-\frac{u^A}{q u^B} + \frac{x}{q u^B}, \forall x \in \left[ -p, u^A \right],\\
				1             \qquad     \qquad \qquad, \forall x > u^A, 
				\end{array} \right.
				\textrm{ and } 
				& \Biim(x) =   \left\{ \begin{array}{l}
				- \frac{p}{u^A} + \frac{qx}{u^A}, \forall x \in \left[ 0, \frac{u^A+p}{q} \right] \\
				1 \qquad \qquad, \forall x > \frac{u^A+p}{q}.
				\end{array} \right.  \label{eq:FA-Def}
				\end{align}
				In this mixed equilibrium, players' payoffs are $\Pi^A =0$ and $\Pi^B =u^B - (u^A+p)/q$.
				\item[$(iii)$]  If $ (u^A + p)/q > u^B$,\footnote{That is $q^{\prime} u^A - p^{\prime} > u^B$.}
				there exists no pure equilibrium; there is a mixed equilibrium where Player A (resp. Player B) draws her bid from the distribution $\Aiiim$ (resp. $\Biiim$) defined as follows.
				\begin{align}
				& \Aiiim(x) =   \left\{ \begin{array}{l}
				0 \qquad \qquad, \forall x \in \left[ 0, -p \right), \\
				\frac{p}{q u^B} + \frac{x}{q u^B}, \forall x \in \left[ -p, q u^B -p\right],\\
				1 \qquad \qquad, \forall x > q u^B -p,
				\end{array} \right.
				\textrm{ and } 
				& \Biiim(x) =   \left\{ \begin{array}{l}
				1- \frac{q u^B}{u^A} + \frac{q \cdot x}{u^A}, \forall x \in \left[0, u^B \right], \\
				1 \qquad \qquad, \forall x > u^B.
				\end{array} \right. \label{eq:bFA-Def}
				\end{align}
				In this mixed equilibrium, players' payoffs are $\Pi^A = u^A -q u^B + p$ and $\Pi^B = 0$.
				
			\end{itemize}
		\end{theorem}
		
\begin{figure*}[htb]
			\centering
			\subfloat[$\FAPA$ instance with $u^A = 3$, $u^B = 4$, \mbox{$p = -1$}, $q =1$ (i.e., \mbox{$0 \le (u^A+p)/q < u^B$}). ]{
				\begin{tikzpicture}
				\node (img){\includegraphics[width=0.42\textwidth]{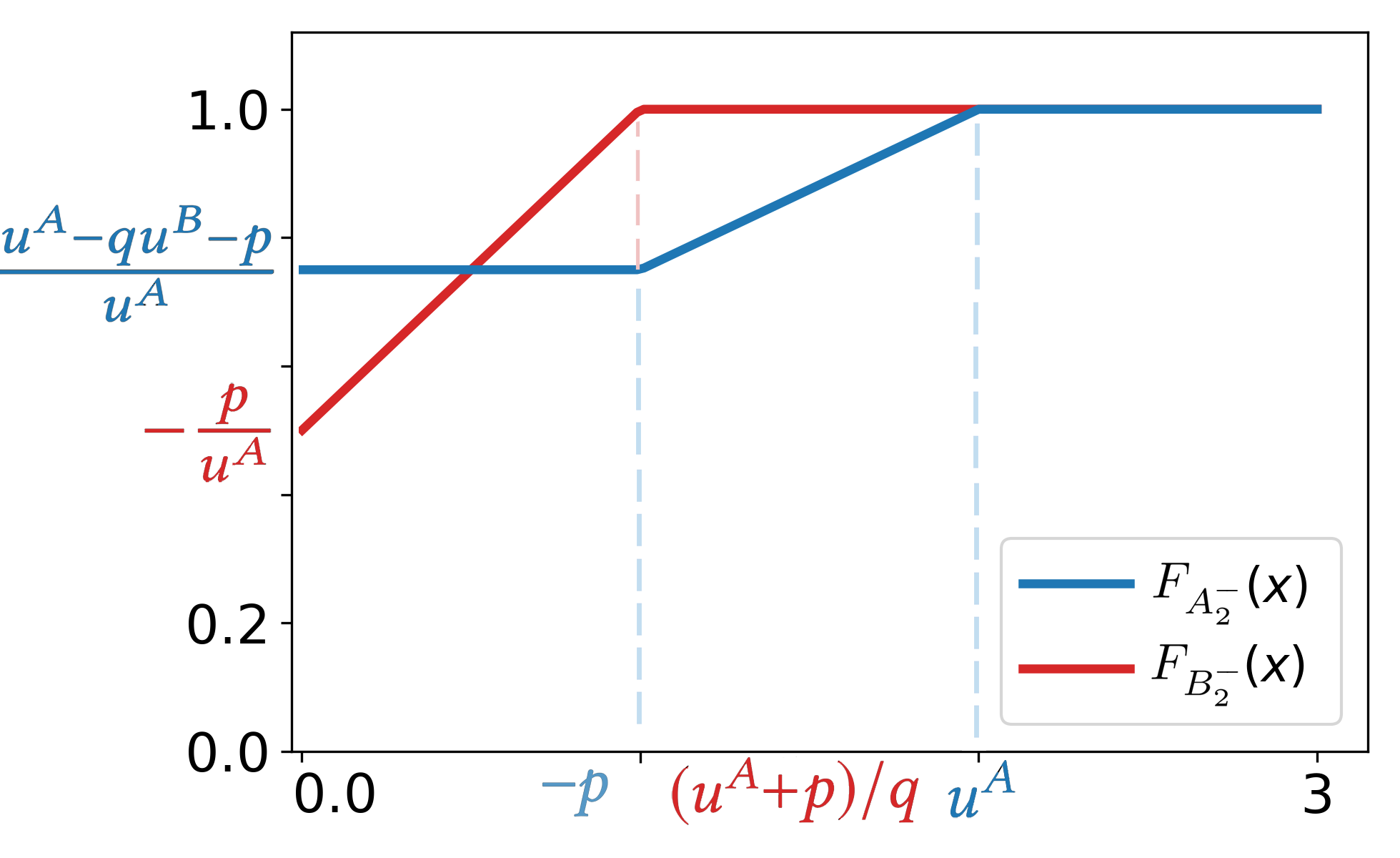}};
				\node[below=of img, node distance=0cm, yshift=1.1cm, xshift = 0.5cm] {$x$};
				\end{tikzpicture}
			}   
			\qquad
			\subfloat[$\FAPA$ instance with $u^A = 4$, $u^B = 2$, \mbox{$p = -1$}, $q =1$ (i.e., \mbox{$ (u^A+p)/q > u^B$}). ]{
				\begin{tikzpicture}
				\node (img){\includegraphics[width=0.40\textwidth]{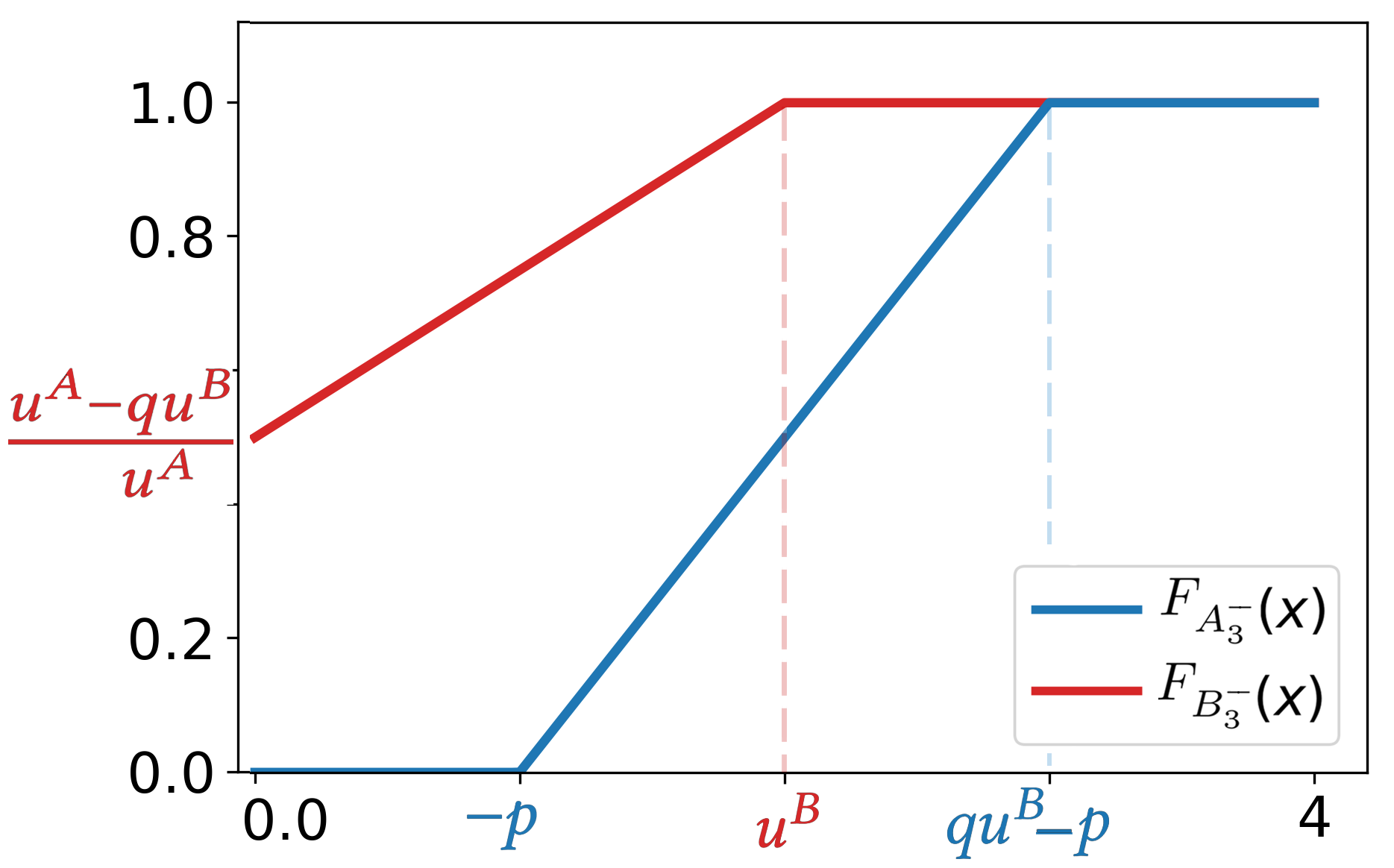}};
				\node[below=of img, node distance=0cm, yshift=1.1cm,xshift = 0.5cm] {$x$};
				\end{tikzpicture}
			}%
			\caption{The mixed equilibrium of the $\FAPA$ with $p <  0$.}
			\label{fig2}
		\end{figure*}

		Similarly to the case where $p \ge 0$, we can verify that in Theorem~\ref{theo:negative}, all the functions $\Aiim, \Biim,\Aiiim$ and $\Biiim$ satisfy the conditions of a distribution and they are continuous on $[0, \infty)$. These distributions are also in the class of uniform-type distributions. The interpretation of these functions are very similar to the analysis for $\Aiip, \Biip,\Aiiip$ and $\Biiip$ and their illustration are given in Figure~\ref{fig2}.
		
	We compare the results in Theorem~\ref{theo:negative} with the results in the classical all-pay auction (Theorem~\ref{theo:clasAPA}). 
	

%
\section{Supplementary Materials for Results in Section~\ref{sec:OptUni_GRCBC}}
    We will provide the proof of Theorem~\ref{theo:OUDs} in \ref{sec:appen_proof_OUD} and the proof of Proposition~\ref{propo:payoffEQ} in \ref{sec:appen_proof_payoffEQ}. The main tool used in these proofs is the notion of the winding number of parametric curves---an important notion in topology; for the sake of completeness, we first revisit the definitions related to this concept in \ref{sec:appen_winding}.

%
%
\subsection{Preliminaries on Winding Number of Parametric Curves}
\label{sec:appen_winding}
Intuitively, the winding number of a 2-dimensional curve relative to a given (2-dimensional) point is the number of rotations the curve goes around the point. A 2-dimensional curve can either be defined as a continuous function from $\mathbb{R}$ to $\mathbb{R}^2$ (often used in topology) or either as a complex-valued function of a real variable (often used in complex analysis); as a consequence, the winding number can also be defined by using either topological terminology or Cauchy integral formula (\ie contour integral) of complex curves. In this work, we choose the former approach, \ie defining the winding number in topological sense. The related definitions presented below are mostly extracted from \cite{chinn1966first} and rewritten in our notation.

We begin with the basic concepts of parametric curves in~topology. Given a range $[a,b] \subset \mathbb{R}$, any continuous mapping $\curve: [a,b] \rightarrow \mathbb{R}^2$ is called a \emph{parametric curve}. Henceforth, we refer to a parametric curve simply as a \emph{curve}. For a curve $\curve$, we abuse the notation and denote its image (\ie $\curve([a,b]) \subset \R^2$) also by $\curve$. There are two type of curves that are of special interest for us in this work as follows:

\begin{definition}[Closed curve and short curve] 
{$~$}
    \begin{itemize}
        \item $ \curve: [a,b] \rightarrow \mathbb{R}^2$ is a closed curve if and only if $\curve(a) = \curve(b)$.
        \item $\curve: [a,b] \rightarrow \mathbb{R}^2$ is a short curve relative to to a point $y \in \mathbb{R}^2$ if there exists a ray, say $R$, coming from $y$ which does not intersect $\curve$.
    \end{itemize}
\end{definition}
Now, in the $\R^2$ plan, for a given ray $R$ starting from $y \in \mathbb{R}^2$, we call $\pole (R,y)$ the polar coordinate system whose pole (\ie the reference point) is $y$ and the polar axis (\ie the reference direction) is~$R$. For any point $x \in \R^2$, we denote its angular coordinate in $\pole(R,y)$ by $\mathbb{A}^{(R,y)}_x$; WLOG, we assume that $\mathbb{A}^{(R,y)}_x \in [0,2\pi], \forall x \in \R^2$ (the counterclockwise rotation is positive).

Given a \emph{short} curve $\curve:[a,b] \rightarrow \R^2$, a point $y \notin \curve$ (\ie $\nexists t \in [a,b]: \curve(t) =y$) and a ray $R$ coming from $y$ which does not intersect $\curve$, the value of the \emph{angle swept} by $\curve$ relative to the point $y$ is defined as~follows:
\begin{equation*}
    \A(\curve, y ) = \mathbb{A}^{(R,y)}_{\curve(b)}  - \mathbb{A}^{(R,y)}_{\curve(a)}. 
\end{equation*}
In other words, the angle swept by a short curve relative to a point is the difference between the angular coordinates of its two ending-points in the corresponding polar coordinate system.
\begin{lemma}[extracted from \cite{chinn1966first}]
\label{lem:angle_swept} 
 The angle swept of short curves is additive; more formally, let $a<b<c$ be real numbers and let $\curve: [a,c] \rightarrow \R^2$ be a short curve relative to a point $y$. Denote $\curve_1= \restr{\curve}{[a,b]}$ and $\curve_2= \restr{\curve}{[b,c]}$, then $\curve_1, \curve_2$ are also short curves relative to $y$ and we have $\A(\curve,y) =\A(\curve_1,y) + \A(\curve_2,y) $.
\end{lemma}

Next, we define the notion of sufficiently-fine partitions of a curve. For any curve $\curve:[a,b] \rightarrow \R^2$, let $t_0,t_1, \ldots, t_m$ be a sequence of real numbers such that $a=t_0 < t_1<\ldots< t_m =b$. A \emph{sufficiently-fine partition} of the curve $\curve$ relative to a point $y$ is a sequence of curves $\{\curve_1, \curve_2, \ldots, \curve_m\}$ such that $\curve_i = \restr{\curve}{[t_{i-1}.t_i]}$ is a \emph{short} curve relative to $y$ for any $i =1, 2,\ldots, m$. Importantly, for any curve $\curve$ and a point $y \notin \curve$, there always exists a sufficient partition of $\curve$ relative to $y$. Moreover, if $\{\curve_1, \curve_2, \ldots, \curve_m\}$
and $\{\curve^{\prime}_1, \curve^{\prime}_2, \ldots, \curve^{\prime}_k\}$ are two  sufficiently-fine partitions of the curve $\curve$ relative to $y$, then \mbox{$\sum_{j=1}^m \A(\curve_j,y) = \sum_{j=1}^k \A(\curve^{\prime}_j,y)$} (see \cite{chinn1966first} for proofs of these~statements).

Based on Lemma~\ref{lem:angle_swept} concerning the angle swept by short curves and the notion of sufficiently fine partition, we can define the angle swept by any generic parametric curve (not necessarily short) which induces the definition of the winding number.

\begin{definition}[Angle swept by a curve]
    For any curve $\curve$ and a point $y \notin \curve$, the angle swept by $\curve$ relative to $y$ is defined as $\A(\curve,y) = \sum_{i=1}^m {\A(\curve_i,y)}$ where $\{\curve_1, \curve_2, \ldots, \curve_m\}$ is any sufficiently-fine partition of $\curve$ relative to $y$ (note that for any $i=1,2,\ldots, m$, $\A(\curve_i,y)$ is well-defined since $\curve_i$ is a short curve relative to $y$).
\end{definition}
\begin{definition}[Winding number]
 Given a \emph{closed} curve $\curve$ and a point $y \notin \curve$, the winding number of $\curve$ around $y$ is defined~as:
 \begin{equation*}
     \W(\curve,y) = \frac{\A(\curve,y)}{2\pi}.
 \end{equation*}
\end{definition}
Trivially, since $\curve$ is a closed curve, its winding number $\W(\curve,y)$ is an integer number. Finally, we present the following fixed-point theorem based on the notion of winding number. 
\begin{lemma}[Fixed-point theorem (extracted from \cite{chinn1966first})] \label{lem:winding}
    Let $D$ be a disk in $\R^2$ (or any topologically equivalence of a disk) and $\partial D$ be its boundary. Let $G: D \rightarrow \R^2$ be a continuous mapping and $y\in \R^2$ such that $y \notin G(\partial D)$. If the closed curve $\curve:= G \parens*{\partial D}$ has a non-zero winding number around $y$, then $y \in G(D)$; in other words, there exists $x \in D$ such that $G(x) = y$. 
\end{lemma}
A proof of this theorem is given in \cite{chinn1966first}. We note that this theorem can be considered as a generalization of the intermediate value theorem (for 1-dimensional functions).

%
%
%
\subsection{Proof of Theorem~\ref{theo:OUDs} (OUDs Construction of the $\FCB$ Game)}
\label{sec:appen_proof_OUD}

While the proof of Result~$(ii)$ of Theorem \ref{theo:OUDs} has been presented completely in Section~\ref{sec:OptUni_GRCBC}; we will prove Result~$(i)$ of this theorem in this section. In other words, given a game instance $\FCBn$, we look for a set $D \in \R^2$, topologically equivalent to a $\R^2$-disk, such that when combined with the function $G$ corresponding to $\FCBn$, this set satisfies sufficient conditions of Lemma~\ref{lem:wind} (which is an adaptation of Lemma~\ref{lem:winding} into the problem in consideration. Particularly, \emph{we want to find $D$ such that $G(\partial D)$ (where $G$ is defined as in \eqref{eq:G_func}) is a closed curve and it has a non-zero winding number around $(0,0)$}.

As discussed in Section~\ref{sec:OptUni_GRCBC}, if $p_i=0, \forall i \in [n]$, we can follow the approach of \cite{kovenock2020generalizations} to convert System~\ref{eq:system_f} into a real-valued 1-dimensional function and proceed to prove the existence of its solution via the intermediate value theorem. In the remainder of this section, we assume that in the game $\FCBn$, there exists $i \in [n]$ such that $p_i \neq 0$. For the sake of brevity, we use the notations $I_{\ge0}:= \braces*{i \in [n]: p_i \ge 0}$ and $I_{<0}:= \braces*{i \in [n]: p_i < 0}$. For any given $(\gA, \gB) \in \R^2$, we also define $\Iplus:= \braces*{j \in I_{\ge0}: {\gB > \frac{p_j}{q_j w_j }} }$ and  \mbox{$\Iminus := \braces*{j \in I_{<0}: {\gA > \frac{-p_j}{w_j }} }$}.

Now, let us choose $D$ to be a rectangle whose vertices are $(\delta,\delta)$, $(L, \delta)$, $(L,L)$ and $(L,\delta)$ where $\delta, L$ will be defined later such that $0< \delta < L$. More formally, we define the following closed curve:
\begin{align*}
    \curve \colon& [\delta, 4L-3\delta] \longrightarrow\R^2 \nonumber \\
    & \phantom{+++}  t \phantom{+++} \longmapsto \left\{ \begin{array}{l}
			 (t, \delta), \forall t \in [\delta, L], \\
			(L, \delta-L+t), \forall t \in [L,2L-\delta],\\
			 (3L - t -\delta, L), \forall t \in [2L-\delta, 3L-2\delta],\\
			 (\delta, 4L - 2 \delta - t), \forall t \in [3L-2\delta, 4L-3\delta].
			\end{array} \right.
\end{align*}
In other words, $\curve$ is a parametric curve form of $\partial D$ that starts at $(\delta, \delta)$, goes along the sides of the rectangle $D$ in the counter-clockwise direction and stops when it reaches $(\delta, \delta)$ again. for the sake of notation, we will denote by $G(\curve)$ the $G$-image of $\curve$; importantly, by the definition of $G$ and the fact that $G$ is a continuous mapping, $G(\curve)$ is also a closed curve in $\R^2$. If $(0,0) \in G(\curve)$ (\ie $(0,0)$ lies on the $G$-image of $\partial D$, then we have proved that there exists a positive zero of $G$; thus, we can conclude the proof in this case. In the remainder of the proof, we assume that  $(0,0) \notin G(\curve)$.

Now, we also define
\begin{itemize}
    \item  $\curve_1:= \restr{\curve}{[\delta, L]}$ representing the curve going along the side of the rectangle $D$ from $(\delta, \delta)$ to~$(L,\delta)$,
    \item  $\curve_2:= \restr{\curve}{[L,2L-\delta]}$ representing the curve going along the side of the rectangle $D$ from $(L,\delta)$ to~$(L,L)$,
    \item  $\curve_3:= \restr{\curve}{[2L-\delta, 3L-2\delta]}$ representing the curve going along the side of the rectangle $D$ from $(L,L)$ to~$(\delta, L)$,
    \item  $\curve_4:= \restr{\curve}{[3L-2\delta, 4L-3\delta]}$ representing the curve going along the side of the rectangle $D$ from $(L,L)$ to~$(\delta,\delta)$.
\end{itemize}
For any $i=1,2,3,4$, we also denote $G(\curve_i)$ the $G$-image of $\curve_i$ and note that $G(\curve_i)$ is also a parametric curve in $\R^2$. We will proceed the proof by showing that $\braces*{G(\curve_1),  G(\curve_2), G(\curve_3), G(\curve_4})$ is a sufficiently-fine partition of $G(\curve)$ (\ie of $\partial D$) relative to $(0,0)$; moreover, $\sum_{j=1}^4 \A(\curve_j) \neq 0$. The curves $G(\curve_1),  G(\curve_2), G(\curve_3), G(\curve_4)$ have different forms and features; each requires a different analysis. We do this in six steps as~follows.

%
%
\textit{\underline{Step 0:} Prove that $G(\delta,\delta) \in \R^2_{<0}$ and  $G(L,L) \in \R^2_{>0}$ when $\delta$ is small enough and $L$ is large enough.}

Let us choose $\delta$ such that
\begin{align}
    & \delta < \bar{\delta}  =\left\{ \begin{array}{l}
			 \min \braces*{ \min \limits_{i \in I_{\ge 0}} \braces*{\frac{p_i}{q_i w_i}}, \min\limits_{i \in I_{<0}} \braces*{\frac{-p_i}{w_i}} }, \textrm{ if } I_{\ge0} \neq \emptyset \textrm{ and }  I_{p<0} \neq \emptyset , \\
			 \min\limits_{i \in I_{\ge0}} \braces*{\frac{p_i}{q_i w_i}}, \textrm{ if }  I_{<0} = \emptyset,\\
			 \min \limits_{i \in  I_{<0}} \braces*{\frac{-p_i}{w_i}}, \textrm{ if } I_{\ge0} = \emptyset.
			\end{array} \right. \label{eq:delta_choice}
\end{align}
Trivially, under the assumptions mentioned above, $\delta >0$; moreover, $\Ip(\delta, \delta) = \In(\delta, \delta) = \emptyset$; thus, 
\begin{equation*}
    g^A(\delta, \delta) = -X^B \delta <0 \quad \textrm{ and } \quad  g^B(\delta, \delta) = -X^A \delta <0 .
\end{equation*}

On the other hand, for any $L > L_0:= \max \braces*{ \max \limits_{I_{\ge0}} \braces*{\frac{p_i}{q_i w_i}}, \max \limits_{i \in I_{<0}} \braces*{\frac{-p_i}{w_i}}}$, we have $I^+(L,L) = I_{\ge0}$ and $I^-(L,L) = I_{<0}$. Therefore, we have
\begin{align}
    	g^A (L,L )   &\!=\!   \sum_{i  \in  I_{\ge0}} \frac{ \bracks*{h_i(L,L)}^2 \!-\!  {p_i}^2 }{2 q_i w_i } + \sum_{i  \in  I_{<0}} \frac{ \bracks*{h_i(L,L)}^2 }{2 q_i w_i} - X^B L \label{eq:LLA} \\
	g^B (\gA,\gB)   &\!=\!   \sum_{i  \in  I_{\ge0}} \frac{ \left[h_i(L,L) \!-\! p_i\right]^2}{2 q_i w_i } \!+\!  \sum_{i  \in   I_{<0}} \frac{ \left[ h_i(L,L) \!-\!p_i \right]^2 \!-\! p_i^2 }{2 q_i w_i}  \!-\! X^A L. \label{eq:LLB}
\end{align}
By definition, $h_i(L,L) := \min \{ q_i w_i L, w_i L + p_i \}$; therefore, the right-hand-sides of~\eqref{eq:LLA} and~\eqref{eq:LLB} are quadratic expressions in terms of $L$ with a strictly-positive second-degree coefficient; in other words, they are convex functions in terms of $L$. Therefore, there exists a constant $L_1>0$ such that for any $L \ge \max \braces*{L_0,L_1}$, we have $g^A (L,L ) >0$ and $g^B (L,L ) >0$.

%
%
\textit{\underline{Step 1:} Prove that $G(\curve_1)$ is a short curve and  $\A(G(\curve_1), (0,0)) <0$ when $\delta$ is small enough and $L$ is large enough.}

We recall that $\curve_1:[\delta, L] \rightarrow \R^2$ and $\curve_1(t) = (t, \delta)$. Now, from the choice of $\delta$ as in \eqref{eq:delta_choice}, for any $t \in [\delta, L]$, we have $\delta < \frac{p_i}{q_i w_i}$ for any $i \in I_{\ge0}$; therefore, $I^+(t,\delta) = \emptyset$. Thus, 
\begin{equation*}
    g^A (t,\delta) = \sum_{i \in I^-(t, \delta)} \frac{h_i(t, \delta)^2}{2q_i w_i} - X^B t.
\end{equation*}
If $I_{<0} = \emptyset$, trivially $g^A (t,\delta) <0$. Now, if $I_{<0} = \emptyset$, we have:
\begin{align*}
    g^A (t,\delta) & \le \sum_{i \in I^-(t, \delta)} \frac{(q_i w_i \delta)^2}{2q_i w_i} - X^B t \\
        & = \sum_{i \in \braces*{ j \in I_{<0}: t >\frac{-p_j}{w_j}}
        (t, \delta)} \bracks*{ \frac{q_i w_i \delta^2}{2} - \frac{X^B t}{|  I^-(t, \delta) |}} \\
        & \le \sum_{i \in \braces*{ j \in I_{<0}: t >\frac{-p_j}{w_j}}
        (t, \delta)} \bracks*{ \frac{q_i w_i \delta^2}{2} - \frac{X^B \delta}{|  I^-(t, \delta) |}} \quad \textrm{(since } \delta < \frac{-p_i}{ w_i} < t) \\
\end{align*}
Therefore, for any $\delta < \frac{X^B}{|  I^-(t, \delta) | } \frac{2}{\max{q_i w_i}}$, we always have $ g^A (t,\delta)<0$. In other words, for any $t \in [\delta, L]$, the curve $G(t, \delta)$ lies on the left of the Oy axis in the the $\R^2$-plane. We conclude that $G(\curve_1)$ is a short curve relative to $(0,0)$ (we can use $R_1: [0, + \infty) \rightarrow \R^2$ such that $R_1(t) = (t,0)$ as the reference~ray). On the other hand, 
\begin{equation*}
    g^B (t,\delta) = \sum_{i \in I^-(t, \delta)} \frac{ \bracks*{h_i(t, \delta) - p_i}^2 - p_i^2}{2q_i w_i} - X^A \delta.
\end{equation*}
For any $i \in I^-(t, \delta)$, $\bracks*{h_i(t, \delta) - p_i}^2 - p_i^2$ is an increasing function of $t$. Intuitively, this means that as $t$ increases from $\delta$ to $L$, the curve $G(\curve_1)$ lies on the left of the Oy-axis and only goes upward. Therefore, the angle swept by $G(\curve_1)$ relative to $(0,0)$ is negative, \ie  $\A(G(\curve_1),(0,0)) < 0$.

%
%
\textit{\underline{Step 2:} Prove that $G(\curve_2)$ is a short curve and  $\A(G(\curve_2), (0,0)) <0$ when $\delta$ is small enough and $L$ is large enough.}

We recall that $\curve_2:[L, 2L-\delta] \rightarrow \R^2$ and $\curve_2(t) = (L, t-L + \delta)$. In this step, for the sake of brevity, let us denote $t^{\prime} = t- L +\delta$; as $t$ increases from $L$ to $2L - \delta$, we have $t^{\prime}$ increases from $\delta$ to $L$. In other words, we can rewrite $\curve_2(t) = (L,\tp)$.

First, to prove that $G(\curve_2)$ is a short curve, we will prove that it does not intersect the ray $R_4: [0, + \infty) \rightarrow \R^2$ such that $R_4(t) = (0,-t)$; intuitively, this means that all intersection points of $G(\curve_2)$ and the Oy-axis have positive x-coordinates. Indeed, fix a number $\tp \in [\delta, L]$ such that \mbox{$g^A(L, \tp) = 0$},\footnote{There exists such $\tp$ since $g^A(L,\delta)<0$ and $g^A(L,L)>0$ (see Step 1 and 2).} we have $h_i(L, \tp):= \min \braces*{q_i w_i \tp, w_i L + p_i  } = q_i w_i \tp$ for any $i \in I^+(L,\tp)$. We can see this by using proof by contradiction: assume otherwise, \ie assume there exists $j \in  I^+(L,\tp)$ such that $q_j w_j \tp > w_j L + p_j$, then $g^A(L, \tp) > \frac{(w_j L + p_j)^2}{2 q_j w_j} - X^B L >0 $ when we choose $L$ large enough;\footnote{The right-hand-side of this inequality is a quadratic expression in terms of $L$ with a strictly-positive second-degree coefficient} this contradicts with the assumption that $g^A(L,\tp) = 0$. As a~consequence,
\begin{align}
    & g^A(L,\tp) = 0  \nonumber\\
    \Leftrightarrow &  \sum_{i \in I^+(L, \tp)} \frac{h_i(L,\tp)^2 - p_i^2 }{2q_i w_i} + \sum_{i \in I^-(L, \tp)}\frac{h_i(L,\tp)^2}{2q_i w_i} - X^B L = 0  \label{eq:curve2_mid}\\
    \Leftrightarrow &  \sum_{i \in I^+(L, \tp)} \frac{(q_i w_i \tp)^2 - p_i^2 }{2q_i w_i} + \sum_{i \in I^-(L, \tp)}\frac{(q_i w_i \tp)^2}{2q_i w_i} - X^B L = 0, \nonumber
\end{align}
which implies that
\begin{equation}
    \sum_{i \in I^+(L, \tp)} \frac{(q_i w_i \tp)^2 - p_i^2 }{2q_i w_i} - X^B L \le 0. \label{eq:curve2_A}
\end{equation}
On the other hand, we have:
\begin{align}
    g^B (L,\tp) & = \sum_{i \in I^+(L, \tp)} \frac{ \bracks*{h_i(t, \delta) - p_i}^2}{2q_i w_i} +  \sum_{i \in I^-(L, \tp)} \frac{ \bracks*{h_i(t, \delta) - p_i}^2 - p_i^2}{2q_i w_i} - X^A \tp \nonumber\\
     & =  \sum_{i \in I^+(L, \tp)} \frac{-2 h_i(L,\tp) p_i + p_i^2}{2q_i w_i} + \sum_{i \in I^-(L, \tp)}\frac{-2 h_i(L,\tp) p_i}{2q_i w_i} + X^B L - X^A \tp      \quad \textrm{(from \eqref{eq:curve2_mid})} \nonumber\\
     & \ge -\sum_{i \in I^+(L, \tp)} \frac{2 q_i w_i \tp p_i + p_i^2}{2q_i w_i} + X^B L - X^A \tp     \quad \textrm{(since } p_i <0, \forall i \in I^-(L, \tp)). \label{eq:curve2_end}
\end{align}

Now, if $I^+(L,\tp) = \emptyset$, from \eqref{eq:curve2_end}, trivially $g^B (L,\tp) \ge 0$ (since $X^A\le X^B$ and $\tp \le L $). In cases where $I^+(L,\tp) \neq \emptyset$, we have $\sum_{i \in I^+(L,\tp)} \frac{q_i w_i}{2} >0$ and we define
\begin{equation*}
    C_2(L): = \sqrt{\frac{\sum_{i \in I^+(L,\tp)} \frac{p_i^2}{2 q_i w_i} + X^B L}{\sum_{i \in I^+(L,\tp)} \frac{q_i w_i}{2}}}.
\end{equation*}
From \eqref{eq:curve2_A}, we have $\tp \le C_2(L)$. Combining this to \eqref{eq:curve2_end}, we have 

\begin{equation}
      g^B (L,\tp)  \ge  -\sum_{i \in I^+(L, \tp)} \frac{2 q_i w_i \tp p_i + p_i^2}{2q_i w_i} + X^B L - X^A \cdot C_2(L).\label{eq:curve2_stop}
\end{equation}
Now, since $C_2(L) = \bigoh(\sqrt{L})$, the right-hand-side of \eqref{eq:curve2_stop} is a quadratic expression in terms of $\sqrt{L}$ with a strictly-positive second-degree coefficient; therefore, there exists $L_2>0$ large enough such that for any $L\ge \max\braces*{L_0, L_1, L_2}$, if $g^A(L,\tp) =0$, we always have $ g^B (L,\tp) \ge 0 $.

Finally, we conclude that $G(\curve_2)$ is a short curve relative to $(0,0)$ (we can choose $R_4$ defined above as the reference ray). Note that $(L,\delta)$ and $(L,L)$ are respectively the starting point and ending point of $G(\curve_2)$; moreover, from Steps 1 and 2, we have proved $g^A(L,\delta) <0 $ and $g^A(L,L)>0$, therefore, the angle swept by $G(\curve_2)$ relative to $(0,0)$ is also negative, \ie $\A(G(\curve_2),(0,0)) <0$.

%
%
\textit{\underline{Step 3:} Prove that $G(\curve_3)$ is a short curve and  $\A(G(\curve_3), (0,0)) <0$ when $\delta$ is small enough and $L$ is large enough.}

We recall that $\curve_3:[2L-\delta, 3L-2\delta] \rightarrow \R^2$ and $\curve_3(t) = (3L-t-\delta, L)$. In this step, for the sake of brevity, let us denote $t^{\prime} = t- 2L +2\delta$; as $t$ increases from $2L-\delta$ to $3L- 2\delta$, we have $t^{\prime}$ increases from $\delta$ to $L$. In other words, we can rewrite $\curve_3(t) = (L + \delta - \tp, L)$.

First, we aim to prove that $G(\curve_3)$ does not intersect the ray $R_3:[0,\infty) \rightarrow \R^2$ where \mbox{$R_3(t) = (-t,0)$}, \ie we prove that if $g^B (L+\delta - \tp,L) =0$ then $g^A (L+\delta - \tp,L) >0$. To do this, we notice that for any $\tp \in \bracks{\delta,L}$ such that $ g^B(L+\delta - \tp,L) =0$, we have
\begin{equation}
    h_i(L+ \delta - \tp,L) = \min\braces*{q_i w_i L, w_i (L+\delta-\tp) + p_i} = w_i (L+\delta-\tp) + p_i, \forall i \in I^-(L+\delta-\tp,L).\label{eq:curve_3_contr}
\end{equation}
Indeed, we can prove \eqref{eq:curve_3_contr} by proof of contradiction: assume otherwise, \ie assume that there exists $j \in I^-(L+\delta-\tp,L)$ such that $q_j w_j L < w_j (L+\delta-\tp) + p_j$, then \mbox{$g^B(L+\delta - \tp,L) > \frac{(q_j w_j L - p_j)^2 - p_j^2}{2q_j w_j} - X^A L >0$} when we choose $L$ large enough; this contradicts with the assumption that $g^B(L+\delta - \tp,L) =0$.

As a consequence of \eqref{eq:curve_3_contr}, we have:
\begin{align}
                &  g^B(L+\delta - \tp,L)  = 0 \\
 \Leftrightarrow &  \sum_{i \in I^+(L+\delta - \tp,L) } \frac{ \bracks*{h_i(L+\delta - \tp,L)  - p_i}^2}{2q_i w_i} +  \sum_{i \in I^-(L+\delta - \tp,L) } \frac{ \bracks*{h_i(L+\delta - \tp,L)  - p_i}^2 - p_i^2}{2q_i w_i} - X^A L = 0 \label{eq:curve_3:gB} \\
 \Rightarrow  &  \sum_{i \in I^-(L+\delta - \tp,L) } \frac{ \bracks*{h_i(L+\delta - \tp,L)  - p_i}^2 - p_i^2}{2q_i w_i} - X^A L \le 0 \nonumber \\
  \Rightarrow  &  \sum_{i \in I^-(L+\delta - \tp,L) } \frac{ \bracks*{w_i(L+\delta - \tp,L)}^2 - p_i^2}{2q_i w_i} - X^A L \le 0
 \label{eq:curve_3:mid}
\end{align}

On the other hand, 

\begin{align}
    & g^A(L+\delta - \tp,L)  \\
    =&  \sum_{i \in I^+(L+\delta - \tp,L)} \frac{h_i(L+\delta - \tp,L)^2 - p_i^2 }{2q_i w_i} + \sum_{i \in I^-(L+\delta - \tp,L)}\frac{h_i(L+\delta - \tp,L)^2}{2q_i w_i} - X^B L \nonumber \\
    =& \sum_{i \in I^+(L+\delta - \tp,L)} \frac{2 h_i(L+\delta - \tp,L) p_i }{2q_i w_i} + \sum_{i \in I^-(L+\delta - \tp,L)} \frac{w_i(L+\delta - \tp) p_i }{2 q_i w_i} + X^B L - X^A(L+\delta - \tp) \quad \textrm{(due to \eqref{eq:curve_3:gB}}) \nonumber\\
    \ge &   \sum_{i \in I^-(L+\delta - \tp,L)} \frac{w_i(L+\delta - \tp) p_i }{2 q_i w_i} + X^B L - X^A(L+\delta - \tp) \textrm{(since } p_i >0, \forall i \in I^+(L+\delta - \tp,L). \label{eq:curve_3:almost}
\end{align}
Now, if $I^-(L+\delta - \tp,L) = \emptyset$, from \eqref{eq:curve_3:almost}, trivially $g^A(L+\delta - \tp,L)>0$ since $X^B\ge X^A$ and $ L \ge L+\delta-\tp , \forall \tp \in [\delta, L]$. Reversely, if $I^-(L+\delta - \tp,L) \neq \emptyset$, we have $\sum_{i \in I^-(L+\delta - \tp,L)} \frac{w_i^2}{2 q_i w_i} >0$, and we can define:

\begin{equation*}
    C_3(L): = \sqrt{\frac{\sum \limits_{i \in I^-(L+\delta - \tp,L)} \frac{p_i^2}{2q_i w_i} + X^A L  }{\sum \limits_{i \in I^-(L+\delta - \tp,L)} \frac{w_i^2}{2 q_i w_i} }}.
\end{equation*}

From \eqref{eq:curve_3:mid}, we have $ (L+\delta - \tp) \le C_3(L)$; therefore, combine with \eqref{eq:curve_3:almost}, we have
\begin{equation}
    g^A(L+\delta - \tp,L) \ge   \sum_{i \in I^-(L+\delta - \tp,L)} \frac{w_i(L+\delta - \tp) p_i }{2 q_i w_i} + X^B L - X^A \cdot C_3(L).\label{eq:curve3_end}
\end{equation}
Now, since $C_3(L) = \bigoh(\sqrt{L})$, the right-hand-side of \eqref{eq:curve3_end} is a quadratic expression in terms of $\sqrt{L}$ with a strictly-positive second-degree coefficient; therefore, with $L$ large enough, we have $ g^A(L+\delta - \tp,L) \ge 0$.

In conclusion, we have prove that if $g^B(L+\delta - \tp,L)=0$, we always have $g^A(L+\delta - \tp,L)\ge 0$; thus, $G(\curve_3)$ does not intersect the ray $R_3$. Therefore, $G(\curve_3)$ is a short curve relative to $(0,0)$. Note that $(L,L)$ and $(\delta,L)$ are respectively the starting point and ending point of $G(\curve_3)$; moreover, $g^B(L,L)>0$ and $g^B(\delta,L)>0$, we conclude that the angle swept by $G(\curve_3)$ relative to $(0,0)$ is negative, \ie $\A(G(\curve_3),(0,0)) <0$.

%
%
\textit{\underline{Step 4:} Prove that $G(\curve_4)$ is a short curve and  $\A(G(\curve_4), (0,0)) <0$ when $\delta$ is small enough and $L$ is large enough.}

We recall that $\curve_4:[3L-2\delta, 4L-3\delta] \rightarrow \R^2$ and $\curve_4(t) = (\delta, 4L-2\delta-t)$. In this step, for the sake of brevity, let us denote $\tp = 3L -3\delta - t$; as $t$ increases from $3L-2\delta$ to $4L- 3\delta$, we have $t^{\prime}$ increases from $\delta$ to $L$. In other words, we can rewrite $\curve_4(t) = (\delta, L+\delta-\tp)$.

As $\delta$ is chosen as in~\eqref{eq:delta_choice}, we have $I^-(\delta, L+\delta-\tp)= \emptyset$; therefore, 

\begin{align}
    g^B (\delta, L+\delta-\tp) = \sum_{i \in I^+(\delta, L+\delta-\tp)} \frac{ \bracks*{ h_i(\delta, L+\delta-\tp) - p_i }^2 }{2q_i w_i} - X^A (L+\delta-\tp) \label{eq:curve_4_gb}
\end{align}

If $I^+(\delta, L+\delta-\tp)= \emptyset$, then from \eqref{eq:curve_4_gb}, trivially $g^B (\delta, L+\delta-\tp)<0$. Reversely, if $I^+(\delta, L+\delta-\tp) \neq \emptyset$, we can rewrite \eqref{eq:curve_4_gb} as follows:

\begin{align}
    g^B (\delta, L+\delta-\tp) 
        = & \sum_{i \in I^+(\delta, L+\delta-\tp)} \bracks*{\frac{ \bracks*{ h_i(\delta, L+\delta-\tp) - p_i }^2 }{2q_i w_i} \frac{X^A (L+\delta-\tp)}{ | I^+(\delta, L+\delta-\tp)|}} \nonumber\\
        \le & \sum_{i \in I^+(\delta, L+\delta-\tp)} \bracks*{ \frac{ \bracks*{ w_i \delta}^2 }{2q_i w_i} \frac{X^A \delta}{ | I^+(\delta, L+\delta-\tp)|} } \nonumber\\
\end{align}
Therefore, if $\delta < \frac{X^A}{| I^+(\delta, L+\delta-\tp)| } \min_{I_{\ge0} \braces*{ \frac{2q_i}{w_i}}}$, we have $ g^B (\delta, L+\delta-\tp) < 0$. Therefore, the curve $G(\curve_4)$ does not intersect the ray $R_2: [0,\infty) \rightarrow \R^2$ such that $R_2(t) = (0,t)$; thus, $G(\curve_4)$ is a short curve. Note that $(\delta, L)$ and $(\delta,\delta)$ are respectively the starting point and ending point of $G(\curve_4)$; moreover, for $L$ large enough, we have
\begin{equation*}
    g^B(\delta, L) = \sum_{i \in I^+(\delta, L)} \frac{ \bracks*{ h_i(\delta, L) - p_i }^2 }{2q_i w_i} - X^A L \le \sum_{i \in I^+(\delta, L)} \frac{ \bracks*{ w_i \delta }^2 }{2q_i w_i} - X^A L \le  - X^A \delta = g^B(\delta, \delta).
\end{equation*}
Therefore, the angle swept by $G(\curve_4)$ relative to $(0,0)$ is negative, \ie $\A(G(\curve_4) ,(0,0)) <0$.

\textit{\underline{Step 5:} Conclusion}
By choosing $\delta>0$ small enough and $L \gg \delta$ large enough such that all conditions mentioned in Steps~0-4 hold, we conclude that $\braces*{G(\curve_1),G(\curve_2),G(\curve_3),G(\curve_4)}$ is a sufficiently fine partition of the curve $G(\curve)$; therefore, 
\begin{equation*}
    \W(G(\curve),(0,0)) = \frac{\sum_{j=1}^4} \A(G(\curve_j) ,(0,0)) {2\Pi} <0.
\end{equation*}
Recall that $\curve$ is the boundary of the rectangle $D$ whose vertices are $(\delta,\delta)$, $(L, \delta)$, $(L,L)$ and $(L,\delta)$. Apply Lemma~\ref{lem:winding}, we conclude that $(0,0) \in G(D)$. This concludes the proof.

%
%
%
\subsection{Proof of Proposition~\ref{propo:payoffEQ}}
\label{sec:appen_proof_payoffEQ}

In this section, we focus on Proposition~\ref{propo:payoffEQ}. We aim to compute the payoffs of the players in a game $\FCBn$ when they play the strategies having marginals $\left\{\Ai, \Bi \right\}_{i \in [n]}$ where $(\gA, \gB)$ is a solution of System~\eqref{eq:system_f}; more precisely, it is when the allocation of Player A (resp. Player B) to battlefield $i$ follows $\Ai$ (resp. $\Bi$). 

We denote $A_i$ (respectively, $B_i$) the random variable corresponding to $\Ai$ (respectively,~$\Bi$). In this case, the expected payoff that Player A gains in battlefield $i \in [n]$ is defined as follows $\Pi^A_i :=  \alpha w_i \prob(A_i = q_i B_i - p_i) + w_i \prob \left( B_i < \frac{A_i + p_i}{q_i} \right)$. We have the following remark:
		
		\begin{remark}\label{remark:tie}
			If tie allocations happen with probability zero, i.e., $\prob \left(  B_i = \frac{A_i + p_i}{q_i}  \right) = 0$, 
			\begin{equation}
			\prob \left( B_i < \frac{x + p_i}{q_i}   \right) =  \Bi\left(  \frac{x+ p_i}{q_i}\right) , \forall x \sim A_i.
			\end{equation}
			
		\end{remark}

Now, $\Ai$ and $\Bi$ are define in Table~\ref{tab:Opt_Uni_GRCBC} which involve 6 cases of parameter configurations which corresponds to 6 indices sets $I^+_1 (\gA, \gB)$, $I^+_2 (\gA, \gB)$, $I^+_3 (\gA, \gB)$, $I^-_1 (\gA, \gB)$, $I^-_2 (\gA, \gB)$ and $I^-_3 (\gA, \gB)$. In the following, we consider these 6 cases:

\underline{Case 1:} $i \in I^+_1$. In this case, both players allocate zero with probability 1. Therefore, if $p_i >0$, Player A wins this battlefield with probability 1. On the other hand, if $p_i =0$, a tie situation happens. Therefore, we~have

\begin{align*}
\Pi^A_i = w_i \mathbb{I}_{\{p_i > 0\}} + \alpha w_i \mathbb{I}_{\{p_i =0\}} .
\end{align*}

\underline{Case 2:} $i \in I^+_2$. First, we show that tie situations only happen with probability 0 in this case. Indeed, we have:
\begin{itemize}
	\item If $p_i =0$, trivially we see that $  \Ai$ is a uniform distribution (on $[0, q_i w_i \gB]$), thus, $\prob(x^A = q_i x^B -p_i ) =0$ for any $x^A \sim A_i, x^B \sim B_i$.
	\item If $p_i >0$, note that $  \Ai$ and $  \Bi$ are distributions with an atom at 0; therefore, $A_i = 0$ and $B_i =0$ might happen with a positive probability. However, if that is the case, Player A wins this battlefield (since $0 < 0 - p_i$). On the other hand, due to the continuity of $  \Ai$ on $(0, q_i w_i \gB -p_i)$, $\prob(x^A = q_i x^B -p_i ) =0$ for any $x^A \sum A_i$ and $x^A \neq 0$ and $x^B \sim B_i$.
\end{itemize}
		
Using Remark~\ref{remark:tie} and the fact that the derivative of $  \Ai$ equals zero on $(q_i w_i \gB -p_i, \infty)$, we the following:

\begin{align}
\Pi^A_i &:=   w_i \frac{p_i}{q_i w_i \gB } \left( 1 - \frac{q_i \gB}{\gA}  + \frac{p_i}{w_i \gA} \right)  + w_i\int \limits_{0}^{\infty}   \Bi\left( \frac{x + p_i}{q_i}\right) \de   \Ai(x)   \nonumber \\ 
& =  w_i \frac{p_i}{q_i w_i \gB } \left( 1 - \frac{q_i \gB}{\gA}  + \frac{p_i}{w_i \gA} \right) +  w_i \int \limits_{0}^{q_i w_i \gB - p_i } \left(1 - \frac{q_i \gB}{\gA} + \frac{x + p_i}{w_i \gA} \right) \frac{1}{q_i \gB w_i} \de x	 \nonumber \\ 
& =  w_i \left(1 - \frac{q_i \gB}{\gA} + \frac{p_i}{w_i \gA} \right)  + \frac{(q_i w_i \gB - p_i )^2}{2 w_i \gA q_i \gB}.
\end{align}

\underline{Case 3:} $i \in I^+_3$. In this case, since $  \Bi$ is the uniform distribution on $\left[\frac{p_i}{q_i}, \frac{w_i \gA + p_i}{q_i } \right]$, ties happen with probability zero. Therefore,

\begin{align*}
\Pi^A_i &:=  w_i\int \limits_{0}^{\infty}   \Bi\left( \frac{x + p_i}{q_i}\right) \de   \Ai(x)   \\ 
& =    w_i\int \limits_{0}^{w_i \gA} \left(\frac{-p_i}{w_i\gA} + \frac{x+p_i}{w_i \gA} \right) \frac{1}{q_i w_i \gB} \de x\\
& = \frac{w_i \gA}{2 q_i \gB}.
\end{align*}

\underline{Case 4:}	  $i \in I^-_1$. In this case, both players allocate zero with probability 1. Since conditions of the indices set $I^-_1$ require that $p_i < 0$, Player B wins with probability 1; therefore, Player A's payoff is:
\begin{align*}
\Pi^A_i &:= 0.
\end{align*}

\underline{Case 5:}	  $i \in I^-_2$. In this case, ties happens with probability zero (note that although \mbox{$\prob(A_i =0) >0$} and $\prob(B_i =0) >0$, if both players allocate zero, Player B wins since $p_i <0$). Therefore,

\begin{align*}
\Pi^A_i &:=  w_i\int \limits_{0}^{\infty}   \Bi\left( \frac{x + p_i}{q_i}\right) \de   \Ai(x)   \\ 
& =    w_i\int \limits_{-p_i}^{w_i \gA} \left(\frac{-p_i}{w_i\gA} + \frac{x+p_i}{w_i \gA} \right) \frac{1}{q_i w_i \gB} \de x\\
& = \frac{w_i \gA}{2 q_i \gB} - \frac{p_i ^2}{2 w_i \gA q_i \gB}.
\end{align*}

\underline{Case 6:}	  $i \in I^-_3$. In this case, since $  \Ai$ is the uniform distribution on $\left[{-p_i}, q_i w_i \gB - p_i \right]$, ties happen with probability zero. Therefore,

\begin{align*}
\Pi^A_i &:=  w_i\int \limits_{0}^{\infty}   \Bi\left( \frac{x + p_i}{q_i}\right) \de   \Ai(x)   \\ 
& =    w_i\int \limits_{-p_i}^{q_i w_i \gB - p_i} \left(1 - \frac{q_i \gB}{\gA} + \frac{x+ p_i}{w_i \gA} \right) \frac{1}{q_i w_i \gB} \de x\\
& = w_i \left( 1 - \frac{q_i \gB}{\gA} + \frac{p_i}{w_i \gA} \right) + \frac{(q_i w_i \gB - p_i)^2 - p_i^2}{2 w_i \gA q_i \gB} \\
& = w_i - \frac{q_i \gB w_i }{2\gA}. 
\end{align*}

In conclusion, the total expected payoff of each player is simply the aggregate of her payoffs in each battlefields; therefore, given a pair of $\gA, \gB$, the total payoff of Player A is:

\begin{align}
\Pi^A &:= \sum \limits_{i \in I^+_1} {\left[w_i \mathbb{I}_{\{p_i > 0\}} + \alpha w_i \mathbb{I}_{\{p_i =0\}} \right]}  + \sum \limits_{i \in I^+_2} {\left[ w_i \left(1 - \frac{q_i \gB}{\gA} + \frac{p_i}{w_i \gA} \right)  + \frac{(q_i w_i \gB - p_i )^2}{2 w_i \gA q_i \gB} \right]} \nonumber \\
& \hspace{5mm} +  \sum \limits_{i \in I^+_3} \left[ \frac{w_i \gA}{2 q_i \gB} \right] + \sum \limits_{i \in I^-_2} \left[ \frac{w_i \gA}{2 q_i \gB} - \frac{p_i ^2}{2 w_i \gA q_i \gB} \right] +  \sum \limits_{i \in I^-_3} \left[w_i - \frac{q_i \gB w_i }{2\gA} \right] \label{payoff_A_GL}.
\end{align}	

Player B's expected payoff is simply 

\begin{equation}
\Pi^B = \sum_{i \in [n]} w_i - \Pi^A. \label{payoff_B_GL}
\end{equation}

\section{Supplementary Materials for Results in Section~\ref{sec:Corollary_results}}

In Section~\ref{sec:IU_result}, we have presented Proposition~\ref{propo:IU} concerning the IU strategies in the \FCB game. As mentioned above, the result of this proposition can be obtained by following the proof of \cite{vu2019approximate} for the case of classical Colonel Blotto game. For the sake of completeness, in this section, we present the main ideas of the proof of Proposition~\ref{propo:IU}. 

To prove that $(\IU^A_\kappa,\IU^B_\kappa)$ constitutes an $\eps W^n$-equilibrium of a game $\FCBn$ where $\eps = \bigoh(n^{-1/2})$, we need to prove the following inequalities hold for any pure strategies $\boldsymbol{x}^A$ and $\boldsymbol{x}^B$ of Players A and B:

\begin{align}
    & \Pi^A_{\FCBn} (\boldsymbol{x}^A, \IU^B_\kappa) \le \Pi^A_{\FCBn} (\IU^A_\kappa, \IU^B_\kappa)  + \eps W^n \label{eq:IU_A}\\ 
    & \Pi^B_{\FCBn} (\IU^A_\kappa, \boldsymbol{x}^B) \le \Pi^B_{\FCBn} (\IU^A_\kappa, \IU^B_\kappa)  + \eps W^n.\label{eq:IU_B}
\end{align}
We focus on \eqref{eq:IU_A} (the proof for \eqref{eq:IU_B} can be done similarly). For the sake of brevity, we only present the proof where the tie-breaking rule parameter $\alpha$ is set to 1 (\ie if a tie happens at a battlefield $i$, Player A gains the value $w_i$). The case with a general value of $\alpha$ can be done similarly by noticing that the distributions $\Ai, \Bi$ are continuous at almost all points of their supports except at a single point (either $0$ or $p_i/q_i$---depending on the index set to which $i$ belongs); thus, the probability of a ties happens are 0 almost everywhere; even at the points where $\Ai, \Bi$ are discontinuous, the probability that a tie happens also goes quickly to 0 when $n$ increases with a speed much faster than the approximation error that we consider; therefore, one can also ignore these tie cases (see \cite{vu2019approximate} for a detailed discussions on similar phenomenon in the classical Colonel Blotto~game).

Now, let us denote $A_i$ and $B_i$ the random variables corresponding to distributions $\Ai, \Bi$. For any $i \in [n]$, we also define the random~variables:
\begin{equation*}
    A^n_i = \frac{A_i \cdot X^A}{\sum_{j \in [n] }  A_j}  \textrm{ and }  B^n_i = \frac{B_i \cdot X^B}{\sum_{j \in [n] }  B_j},
\end{equation*}
and call the corresponding distributions by $\Ani$ and $\Bni$. By definition, $\Ani, i \in [n]$ are the marginals of the $\IU^A_{\kappa}$ strategy and $\Bni, i \in [n]$ are the marginals of the $\IU^B_{\kappa}$ strategy. Therefore, we can rewrite the involved payoffs as follows:

\begin{align*}
    \Pi^A_{\FCBn} (\boldsymbol{x}^A, \IU^B_\kappa) & = \sum_{i \in [n]}  {    w_i \prob \parens*{B^n_i \le \frac{x^A_i + p_i}{q_i}  }} = \sum_{i \in [n]} w_i \Bni \parens* {\frac{ x^A_i + p_i}{q_i} } ,\\
    \Pi^A_{\FCBn} (\IU^A_\kappa, \IU^B_\kappa)  &= \sum_{i \in [n]} w_i \int_{0}^{\infty}   \Bni \parens* {\frac{ x + p_i}{q_i} } \textrm{d} \Ani (x) .
\end{align*}

Now, to make connection between these payoffs and the players' payoffs when they have marginals $\Ai, \Bi, i\in [n]$ (which are OUDs of the game), we need the following important lemma:

\begin{lemma}\label{lem:damn}
In any game $\FCBn$,  we have 
\begin{equation*}
    \sup_{x \in [0,\infty)} \abs*{ \Ai(x) - \Ani(x) } < \bigoh(n^{-1/2})  \textrm{ and }  \sup_{x \in [0,\infty)} \abs*{ \Bi(x) - \Bni(x) } < \bigoh(n^{-1/2}) .
\end{equation*}
\end{lemma}
A proof of this lemma can be obtained by following Lemma~B4 of \cite{vu2019approximate}. At a high-level, Lemma~\ref{lem:damn} comes from applying Hoeffding’s inequality \cite{hoeffding1963probability} and the fact that $\kappa=(\gA, \gB)$ is a solution of System~\ref{eq:system_f} (thus $ \mathbb{E} \bracks*{ \sum_{j \in [n]} A_j } = X^A $) to show that as $n \rightarrow \infty$,
\begin{equation*}
   A^n_i = \frac{A_i \cdot X^A}{\sum_{j \in [n] }  A_j}  \rightarrow \frac{A_i \cdot X^A}{\mathbb{E} \bracks*{ \sum_{j \in [n]} A_j }} = \frac{A_i X^A}{X^A} = A_i.
\end{equation*}
We also have a similar result for $B^n_i$ and $B_i$. Based on this, we can prove that $\Ani$ and $\Bni$ uniformly converge toward $A_i$ and $B_i$ as $n$ increases. Importantly, by working out the details, we can also determine the rate of this convergence (which gives the upper-bounds presented in Lemma~\ref{lem:damn}).

Now, based on Lemma~\ref{lem:damn}, we can show that as 
\begin{equation}
     \Pi^A_{\FCBn} (\boldsymbol{x}^A, \IU^B_\kappa) \le  \sum_{i \in [n]} w_i \Bi \parens* {\frac{ x^A_i + p_i}{q_i} } + \sum_{i=1}^n w_i \bigoh(n^{-1/2}).\label{eq:IU_a}
\end{equation}
On the other hand, we can also combine Lemma~\ref{lem:damn} with a variant of the portmanteau theorem (see \eg Lemma B5 of \cite{vu2019approximate}) to obtain that:
\begin{equation}
    \abs*{\int_{0}^{\infty}   \Bni \parens* {\frac{ x + p_i}{q_i} } \textrm{d} \Ani (x) -  \int_{0}^{\infty}   \Bi \parens* {\frac{ x + p_i}{q_i} } \textrm{d} \Ai (x) } < \bigoh(n^{-1/2})\label{eq:IU_b}
\end{equation}

By definition, $\Ai$ and $\Bi$ are equilibrium of the corresponding \FAPA game (see Definition~\ref{def:OptDis_GRCBC}); therefore, they are best-response against one another. In other words, for any pure strategy $\boldsymbol{x}^A$ of Player A, we have:
\begin{equation*}
    \sum_{i \in [n]} w_i \Bi \parens* {\frac{ x^A_i + p_i}{q_i} } \le \sum_{i \in [n]} w_i \int_{0}^{\infty}   \Bi \parens* {\frac{ x + p_i}{q_i} } \textrm{d} \Ai (x) .
\end{equation*}
Combining this inequality with~\eqref{eq:IU_a} and \eqref{eq:IU_b}, we have:

\begin{align*}
    \Pi^A_{\FCBn} (\boldsymbol{x}^A, \IU^B_\kappa) & = \sum_{i \in [n]} w_i \Bni \parens* {\frac{ x^A_i + p_i}{q_i} } ,\\
    & \le \sum_{i \in [n]} w_i \Bi \parens* {\frac{ x^A_i + p_i}{q_i} } +  \sum_{i=1}^n w_i \bigoh(n^{-1/2}) \\
    & \le \sum_{i \in [n]} w_i \int_{0}^{\infty}   \Bi \parens* {\frac{ x + p_i}{q_i} } \textrm{d} \Ai (x) +  \sum_{i=1}^n w_i \bigoh(n^{-1/2}) \\
    & \le \sum_{i \in [n]} w_i \int_{0}^{\infty}   \Bni \parens* {\frac{ x + p_i}{q_i} } \textrm{d} \Ani (x) +  \sum_{i=1}^n w_i \bigoh(n^{-1/2}) \\
    & \le  \Pi^A_{\FCBn} (\IU^A_\kappa, \IU^B_\kappa)  +  W^n\bigoh(n^{-1/2}) .
\end{align*}
Similarly, we can prove that $\Pi^B_{\FCBn} (\IU^A_\kappa, \boldsymbol{x}^B) \le \Pi^B_{\FCBn} (\IU^A_\kappa, \IU^B_\kappa)  +  W^n \bigoh(n^{-1/2}) $. This concludes the~proof.
\label{appen:IU}

\section{Supplementary Materials for Results in Section~\ref{sec:heuristic}}

In Section~\ref{sec:heuristic}, we presented the main ideas of the approximation algorithm that we propose in order to efficiently compute a $\delta$-approximate solution of System~\ref{eq:system_f} in any given \FCB game instance. In this section, we give re-discuss this algorithm in more details. First, we discuss a pseudo-code of this algorithm in~\ref{appen:pseudo}. We then give more details on the computational time results of this algorithm in \ref{appen:compute}

%
%
\subsection{A Pseudo-code of the Approximation Algorithm}
\label{appen:pseudo}

\begin{algorithm}[h!]
			\KwIn{$\FCBn$ game.}
			\textbf{Parameters}:  ${\delta}>0$, $m>0$ and $M \gg m$\;
			\KwOut{A $\delta$-approximate solution $(\tgA, \tgB) \in \mathbb{R}^2_{>0}$ of System~\ref{eq:system_f} of $\FCBn$ (satisfying~\eqref{eq:delta_close} and \eqref{eq:system_approx_f})}
			Initialize $D$ to be the rectangle with four vertices $(m,m), (m,M), (M,M), (M, m) $ \;
			Compute $\omega_D = $ the winding number of $G(D)$ around $(0,0)$ via Algorithm~\ref{algo:aux}\;
			\uIf{$\omega_D = 0$ }{
				M:= 2M and ${m} := {m}/2$ , then repeat line 1
			}
			\ElseIf{$\omega_D \neq 0$ }{
				Divide $D$ into two rectangles, $D_1$ and $D_2$, with equal areas\;
				Compute $\omega_{D_1} = $ the winding number of $G(D_1)$ around $(0,0)$ via Algorithm~\ref{algo:aux}\;
				\uIf{$\omega_{D_1} \neq 0$}{
					\uIf{diameter of $D_1$ is less than ${\delta}$}{Stop and return $(\tgA, \tgB) \in D_1$ satisfying ~\eqref{eq:system_approx_f} computed by Algorithm~\ref{algo:aux}\;}
					\lElse{Set $D:= D_1$ and repeat line 6}
				}
				\Else{
			    	Compute $\omega_{D_2} = $ the winding number of $G(D_2)$ around $(0,0)$ via Algorithm~\ref{algo:aux}\;
					\uIf{diameter of $D_2$ is less than ${\delta}$}{Stop and return $(\tgA, \tgB) \in D_2$ satisfying ~\eqref{eq:system_approx_f} computed by Algorithm~\ref{algo:aux}}
					\lElse{Set $D:= D_2$ and repeat line 6}	
				} 
			}
			\caption{Approximation algorithm finding a ${\delta}$-approximate solution of System~\eqref{eq:system_f}} \label{algo:heuristic_GRCB}
\end{algorithm}

Algorithm~\ref{algo:heuristic_GRCB} follows precisely the template described in Section~\ref{sec:heuristic}. Note that Algorithm~\ref{algo:heuristic_GRCB} takes 3 parameters as inputs: $\delta$ controls the preciseness level of the output approximation solutions, $m$ and $M$ controls the initialized rectangle. Moreover, it also uses a sub-routine to compute the winding number of the $G$-image of the rectangles involved in the dichotomy procedure. We present a pseudo-code of this sub-routine procedure as Algorithm~\ref{algo:aux}. Intuitively, to compute a winding number of $G(D)$ where $D$ is a rectangle having the parametric (closed) curve form as $\curve:[a,b] \rightarrow \R^2$, we compute a polygonal approximation of $G(D)$ via IPS algorithm proposed by \cite{zapata2012geometric}; we then calculate the winding number of this polygon by checking how many time one cross the Ox-axis in the $\R^2$-plane when tracking the curve $G(\curve)$by following the sides of this polygon (if it crosses in counterclockwise direction, we increase the counting by 1 unit and it crosses in clockwise direction, we decrease it by 1 unit). Moreover, while doing this, Algorithm~\ref{algo:aux} also computes the $G$-value of all vertices of the involved polygon; Algorithm~\ref{algo:aux} will record a any point $(x,y) \in D$ whose $G$-image is one of the vertex of the polygon and that $(x,y)$ satisfies \eqref{eq:system_approx_f}. If the winding number of $G(D)$ is non-zero, such $(x,y)$ is guaranteed to exist due to Lemma~\ref{lem:winding}.

\begin{algorithm}
    \KwIn{$\FCBn$ game, a rectangle $D$ presented as a parametric (closed) curve $\curve:[a,b] \rightarrow \R^2$}
    \KwOut{$\omega_{D} = $ the winding number of $G(D)$ around $(0,0)$ and a point in $D$ satisfying~\eqref{eq:system_approx_f}}
    Initialize $\omega_{D}=0$\;
    Use IPS Algorithm from \cite{zapata2012geometric} to find an array $(t_0 = a, t_1, \ldots, t_k = b)$ satisfying properties of connection relative to $G(\curve)$ (\ie $\{G(\curve(t_i)), i = 0,\ldots,k\}$ forms a polygonal approximation of~$G(D)$)\;
    \For{$i =0, \ldots, k$}{
        Compute $G(\curve(t_i))$, $G(\curve(t_{i+1}))$\;
        \lIf{$G(\curve(t_i))$ satisfy \eqref{eq:system_approx_f}}{Return the point $\curve(t_i)$}
        \lIf{Segment from $G(\curve(t_i))$ to $G(\curve(t_{i+1}))$ crosses from $\{(x,y) \in R^2: x>0, y<0 \}$ to $\{(x,y) \in R^2: x>0, y>0 \}$ } {$\omega_{D} = \omega_{D}+1$}
        \lElseIf{Segment from $G(\curve(t_i))$ to $G(\curve(t_{i+1}))$ crosses from $\{(x,y) \in R^2: x>0, y>0 \}$ to $\{(x,y) \in R^2: x>0, y<0 \}$ }{$\omega_{D} = \omega_{D}-1$}
    }
\caption{Winding number computation} \label{algo:aux}
\end{algorithm}

%
%
\subsection{Computational Time of the Approximation Algorithm}
\label{appen:compute}

Proposition~\ref{propo:heuristic} states that by running our approximation algorithm in $\tilde{\bigoh}(n \delta^{-1})$ time, we will find a $\delta$-approximate solution of System~\eqref{eq:system_f}. In this section, we first give a proof of this proposition.

\paragraph{Proof of Proposition~\ref{propo:heuristic}}{
Recall the notation $R$ and $L_0$ denoting the max-norm of an actual solution $(\gA,\gB)$ of System~\eqref{eq:system_f} and that of the center of the initialized rectangle. In Algorithm~\ref{algo:heuristic_GRCB}, we observe that after each enlargement step (Lines~5), we end up with a rectangle that is double in size; therefore, the loop in Lines 4-5 will terminate after $\bigoh(\log \parens*{ \max \braces*{ {R}/{L_0}, {L_0}/{R}} })$ iterations; when this loops end, we guarantee to find a rectangle containing $(\gA, \gB)$ (thus, the $G$-image of the boundary of this rectangle has non-zero winding number around $(0,0)$). 

Now, each time the loop in Lines 6-17 of Algorithm~\ref{algo:heuristic_GRCB} repeats, the size of the rectangle in consideration is reduced by one half. Therefore, after at most $\bigoh\parens*{\log \parens{R/\delta }}$ iterations (we assume that $\delta<1$), we end up with a rectangle whose diameter is smaller than $\delta$. The fact that there is always a sub-rectangle (obtained by dividing the rectangle considered in the previous loop run) such that the winding number of its $G$-image is non-zero is guaranteed by Lemma~\ref{lem:winding}; this guarantees that this loop cycle terminates after $\bigoh\parens*{\log \parens{R/\delta }}$ iterations.

Finally, we see that each time we need to compute a winding number in Algorithm~\ref{algo:heuristic_GRCB}, we call for a run of Algorithm~\ref{algo:aux}. From Theorem 4 of \cite{zapata2012geometric}, it takes IPS Algorithm $\bigoh \parens*{(b-a) \delta^{-1}}$ time to output the array $(t_0 = a, t_1, \ldots, t_k = b)$ as described in Line~2 of Algorithm~\ref{algo:aux} where $k= \bigoh \parens*{(b-a) \delta^{-1}}$ (thus, it induces a polygon with $k$ vertices). Note that since $G$ is Lipschitz-continuous, $G(D)$ is also a Lipschitz curve; thus the sufficient conditions of this theorem holds. Finally, each time Algorithm~\ref{algo:aux} computes a value $G(\curve(t_i))$, it takes $\bigoh(n)$ time; this is due the the definition of $G$ in~\eqref{eq:G_func}. In conclusion, each call of Algorithm~\ref{algo:aux} takes $\bigoh \parens*{(b-a) \delta^{-1}} n $ time and the result follows.
\qed
}

Now, to illustrate the efficiency of our approximation algorithm, we conduct several experiments. First, re-visit the toy-example (Example~\ref{ex:example_1234}) considered in Section~\ref{sec:heuristic} where we showed that a naive approach for computing its solution is very inefficient. The application of our approximate algorithm to solve this problem is given as Example~\ref{ex:damn}.
\begin{example}\label{ex:damn}
    \emph{Recall the game instance $\FCBn$ (with $n=2$) considered in Example~\ref{ex:example_GRCB_2} where System~\eqref{eq:system_f} has one positive (exact) solution $(\gA,\gB) := (2 + \sqrt{4/3}, 2+\sqrt{12}) \approx (3.1547005,5.4641016) $. With the parameter \mbox{$\delta=10^{-6}$}, our approximation algorithm outputs the solution $(\tgA, \tgB) =( 3.1547010, 5.4641018) $. The computation time is $\sim 2.78$~seconds when initializing with the rectangle whose vertices are $(\delta,\delta)$, $(\delta, 10X^A)$, $(10X^A, 10X^A)$ and~$(10X^A,\delta)$.}
\end{example}

\begin{figure}[h!]
	\centering
	\begin{tikzpicture}
	\node (img){\includegraphics[height = 0.3\textwidth]{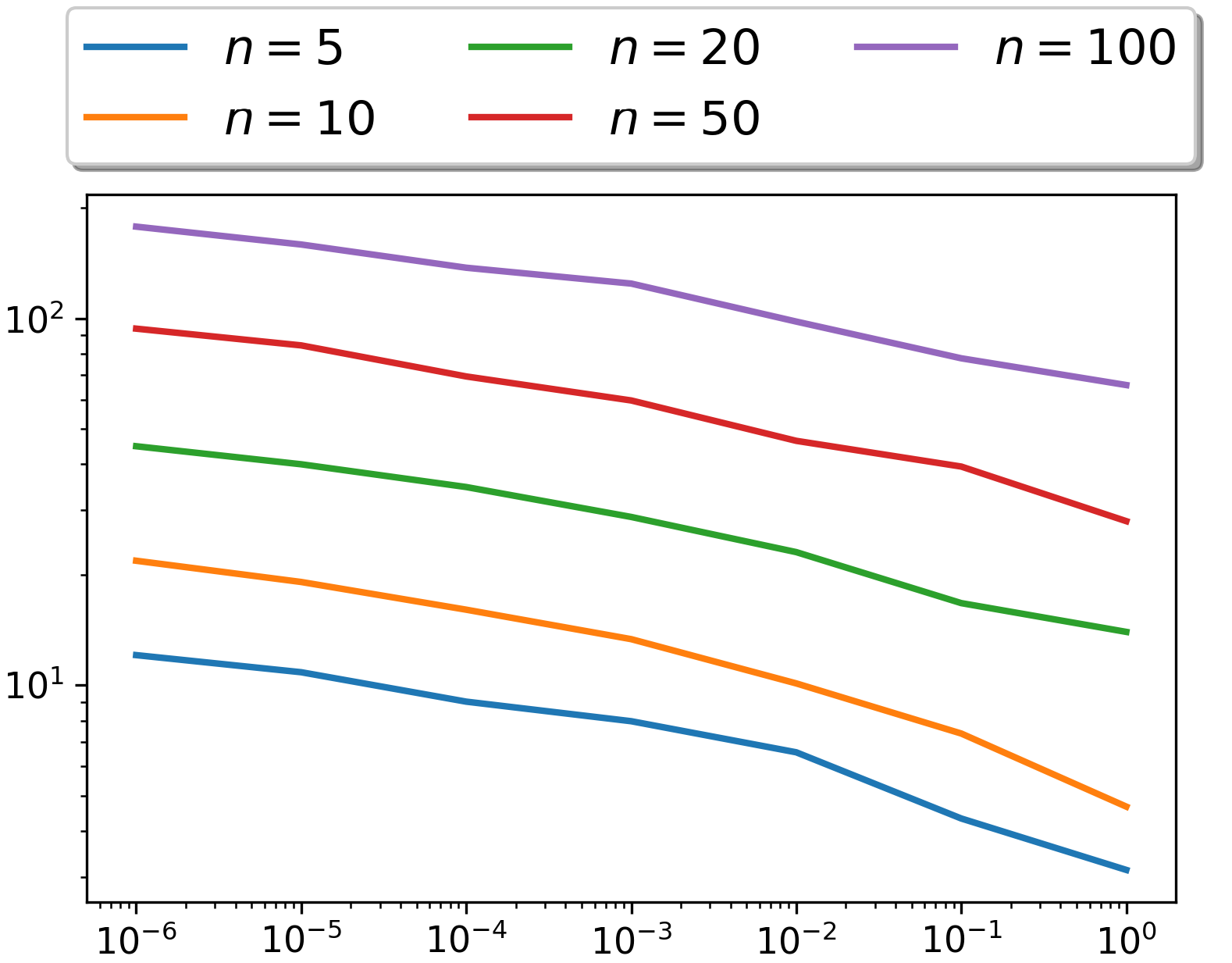}};
	\node[below=of img, node distance=0cm, yshift=1.1cm, xshift = 0.3cm] {$  {\delta}$};
	\node[left=of img, node distance=0cm, rotate=90, anchor=center,yshift=-0.7cm] {seconds};
	\end{tikzpicture}
	\caption[Trade-off between running time and accuracy of~\ref{algo:heuristic_GRCB}]{The trade-off between the running time of~\ref{algo:heuristic_GRCB} and $  {\delta}$. Both axes are drawn with log-scale.} \label{fig:example_time_GRCB}
\end{figure}

Next, we conduct the following experiment (running with a machine with an Intel Xeon CPU \@2.20GHz and 12Gb RAM). For each $n \in \{5,10,20,50,100\}$, we randomly generate 10 instances\footnote{We choose $X^A, X^B \in \{1,2,\ldots, 100\}$ randomly at uniform ($X^A \le X^B$); then, for each $i \in [n]$, we randomly generate a battlefield value $w_i \sim \mathcal{U}(0, X^A]$ and with equal probability, we choose either $p_i>0$ or $p_i=<0$ or $p_i = 0$; then draw $p_i$ from $\mathcal{U}(0, X^A)$ or $\mathcal{U}(-X^A, 0)$ or set it equal $0$ respectively; then, with equal probability, we choose either $q_i>1$ or $q_i \in (0,1)$ or $q_i = 1$; then draw $q_i$ from $\mathcal{U}(1, X^A)$ or $\mathcal{U}(1/X^A, 1)$ or set it equal $1$ respectively.} of $\FCBn$. We then run~\ref{algo:heuristic_GRCB} on each game instance with the input $  {\delta} \in \{10^{-1}, 10^{-2}, \ldots, 10^{-6}\}$ and $M:= 10 \cdot \min\{X^A,X^B\}$; we then measure the time it takes to output the $  {\delta}$-approximate solution of the corresponding System~\eqref{eq:system_f}. Figure~\ref{fig:example_time_GRCB} shows the average running time of~\ref{algo:heuristic_GRCB} taken from the 10 instances for each $n$ and ${\delta}$.
\subsection{Approximations of Optimal Univariate Distributions}
\label{appen:approx_OUD}
In this section, we give the proof of Proposition~\ref{lem:approx_sol} which shows the relation between the distributions $\tAi \tBi, i \in [n]$ from Definition~\ref{def:OptDis_GRCBC} that corresponds with any $\delta$-approximate solution of System~\eqref{eq:system_f} to and the OUDs $\Ai, \Bi, i \in [n]$ (based on the solution $(\gA, \gB)$ of System~\ref{eq:system_f}.

\paragraph{Proof of Proposition~\ref{lem:approx_sol}}{

Fix an $i \in [n]$, we look for upper-bounds of $ |\Ai(x) - \tAi(x) |$ and  $ |\Bi(x) - \tBi(x) |$. To do this, we consider two main cases: where $p_i \ge 0$ and where $p_i <0$.

We start with the case where $p_i \ge 0$. WLOG, let us assume $\gA \le \tgA$ and $\gB \le \tgB$ (the case where either $\gA > \tgA$ or $\gB > \tgB$ can be done similarly by switching the roles of $\gA, \gB$ and $\tgA, \tgB$). Given the value $(\gA, \gB)$, battlefield $i$ belongs to one of the indices sets $I^+_1(\gA, \gB), I^+_2(\gA, \gB)$ or $I^+_3(\gA, \gB)$. Similarly, we know that $i$ also belongs to one of the indices sets $I^+_1(\tgA, \tgB), I^+_2(\tgA, \tgB)$ or $I^+_3(\tgA, \tgB)$.}

\textit{Case 1.1:} If $i$ belongs to $I^+_1(\gA, \gB) \cap  I^+_1(\tgA, \tgB)$. Trivially, $\Ai(x) = \tAi(x)=1, \forall x$ and $\Bi(x) = \tBi(x)=1, \forall x$. Trivially, $ |\Ai(x) - \tAi(x) |=0 <\delta$ and  $ |\Bi(x) - \tBi(x) |= 0<\delta$ for any $x$.

\textit{Case 1.2:} If $i$ belongs to $I^+_2(\gA, \gB) \cap  I^+_2(\tgA, \tgB)$. We have
\begin{align*}
    & \Ai(x)  =  \left\{ \begin{array}{l}
			\frac{p_i}{q_i w_i \gB} +  \frac{x}{q_i w_i \gB}, \forall x  \in  \left[ 0, q_i w_i \gB  -  p_i \right], \\
			1              \qquad  \quad  \qquad \qquad, \forall x > q_i w_i \gB -p_i, 
			\end{array} \right.  \\
			\textrm{ and } &
			 \tAi(x)  =  \left\{ \begin{array}{l}
			\frac{p_i}{q_i w_i \tgB} +  \frac{x}{q_i w_i \tgB}, \forall x  \in  \left[ 0, q_i w_i \tgB  -  p_i \right], \\
			1              \qquad  \quad  \qquad \qquad, \forall x > q_i w_i \tgB -p_i, 
			\end{array} \right.
\end{align*}

For any $x \in [0,  q_i w_i \gB  -  p_i]$, we also have $x \in  [0,q_i w_i \tgB  -  p_i]$ and thus
\begin{align*}
 |\Ai(x) - \tAi(x)  | & = \abs*{  	\frac{p_i}{q_i w_i \gB} +  \frac{x}{q_i w_i \gB}  - \frac{p_i}{q_i w_i \tgB} - \frac{x}{q_i w_i \tgB} }\\
 & = \abs*{ \frac{p_i + x}{q_i w_i} \parens*{ \frac{\tgB - \gB}{\gB \tgB }}      }\\
 & \le \abs*{ \frac{q_i w_i \tgB }{q_i w_i} \parens*{ \frac{\delta}{\gB \tgB }}      }\\
 &= \frac{\delta}{\gB}
\end{align*}

Now, for any $ x$ such that $q_i w_i \gB  -  p_i < x \le q_i w_i \tgB  -  p_i $, we have:

\begin{align*}
 |\Ai(x) - \tAi(x)  | & = \abs*{  1  - \frac{p_i}{q_i w_i \tgB} - \frac{x}{q_i w_i \tgB} }\\
 & = \abs*{ \frac{q_i w_i \tgB - p_i -x}{q_i w_i \tgB} }\\
 & <  \abs*{ \frac{q_i w_i \tgB - p_i - q_i w_i \gB + p_i}{q_i w_i \tgB} }\\
 &= \frac{\delta}{\tgB}
\end{align*}
Finally, for any $x > q_i w_i \tgB  -  p_i>  q_i w_i \gB  -  p_i$, trivially $|\Ai(x) - \tAi(x)| = |1-1|= 0 < \delta$.

Therefore, we conclude that in this case, for any $x$, $|\Ai(x) - \tAi(x)| \le  \delta/ \min\braces*{\gB,\tgB}$. A similar proof can be done to show that $|\Bi(x) - \tBi(x)| \le  \delta/ \min\braces*{\gA,\tgA}$.

\textit{Case 1.3:} If $i$ belongs to $I^+_3(\gA, \gB) \cap  I^+_3(\tgA, \tgB)$. We have:

\begin{align*}
        & \Ai(x)  =    \left\{ \begin{array}{l}
			1- \frac{\gA}{q_i \gB} +  \frac{x}{q_i w_i \gB}, \forall x  \in  \left[ 0, w_i \gA \right], \\
			1         \qquad \qquad  \qquad \qquad, \forall x > w_i \gA, \end{array} \right. \\
\textrm{ and } & \tAi(x)  =    \left\{ \begin{array}{l}
			1- \frac{\tgA}{q_i \tgB} +  \frac{x}{q_i w_i \tgB}, \forall x  \in  \left[ 0, w_i \tgA \right], \\
			1         \qquad \qquad  \qquad \qquad, \forall x > w_i \tgA, \end{array} \right.
\end{align*}

For any $x \in [0, w_i \gA]$, we also have $x \in [0, w_i \tgA]$, therefore, 

\begin{align*}
    \abs*{\Ai(x) - \tAi(x) } 
        = & \abs*{  1- \frac{\gA}{q_i \gB} +  \frac{x}{q_i w_i \gB}   -1+ \frac{\tgA}{q_i \tgB} -  \frac{x}{q_i w_i \tgB}    }\\
        = & \abs*{ \frac{1}{q_i} \parens*{ \frac{\tgA}{\tgB} - \frac{\gA}{\gB}} + \frac{x}{q_i w_i} \frac{\tgB - \gB}{ \tgB \gB} }\\
        \le & \abs*{\frac{2}{q_i} { \frac{\tgA}{\tgB} }} +\abs*{ \frac{w_i \tgA}{q_i w_i} \frac{\tgB - \gB}{ \tgB \gB} }\\
        \le & \abs*{\frac{2}{q_i} { \frac{\tgA}{\tgB} }} +\abs*{ \frac{\tgA}{q_i } \frac{\delta}{ \tgB \gB} }\\
        = & \bigoh(\delta)
\end{align*}

For any $x$ such that $ w_i \gA < x \le  w_i \tgA$, we have 
\begin{align*} 
 \abs*{\Ai(x) - \tAi(x) } 
        = \abs*{ 1-1+ \frac{\tgA}{q_i \tgB} -  \frac{x}{q_i w_i \tgB}    } = \abs*{\frac{w_i \tgA- x}{q_i w_i \tgB} } \le  { \frac{\tgA - \gA}{q_i \tgB} } \le \frac{\delta}{q_i \tgB}.
\end{align*}
Finally, for any $ x > w_i \tgA \ge  w_i \gA$, trivially, we have $\abs*{\Ai(x) - \tAi(x) } =0$. We conclude that in this case, for any $x$, we also obtain $\abs*{\Ai(x) - \tAi(x) } < \bigoh (\delta)$. In a similar manner, we have $\abs*{\Bi(x) - \tBi(x) } < \bigoh (\delta)$.

\textit{Case 1.4:} We consider the case where $i \in I^+_1(\gA, \gB) \cap I^+_2(\tgA, \tgB)$, \ie when $ q_i w_i \gB - p_i \le 0 < q_i w_i \tgB - p_i \le w_i \tgA$ (this might happen since $\gA \le \tgA$). In this case, if $x \in q_i w_i \tgB - p_i$, we have:
\begin{align*}
    \abs*{ \Ai(x) - \tAi(x)} = \abs*{  1  - \frac{p_i}{q_i w_i \tgB} - \frac{x}{q_i w_i \tgB} } \le \frac{\delta}{\tgB}.
\end{align*}
Moreover, when $x >  q_i w_i \tgB - p_i$, we have $\abs*{ \Ai(x) - \tAi(x)} = 0 $. Therefore, we conclude that $\abs*{ \Ai(x) - \tAi(x)} \le \bigoh(\delta)$ for any $x$. A similar proof can be done for $\abs*{ \Bi(x) - \tBi(x)} \le \bigoh{\delta}$.

\textit{Case 1.5:} The case where $i \in I^+_1(\gA, \gB) \cap I^+_3(\tgA, \tgB)$ can be done similar to Case 1.4.

\textit{Case 1.6:} The case where $i \in I^+_2(\gA, \gB) \cap I^+_3(\tgA, \tgB)$, we have:

\begin{align*}
    & \Ai(x)  =  \left\{ \begin{array}{l}
			\frac{p_i}{q_i w_i \gB} +  \frac{x}{q_i w_i \gB}, \forall x  \in  \left[ 0, q_i w_i \gB  -  p_i \right], \\
			1              \qquad  \quad  \qquad \qquad, \forall x > q_i w_i \gB -p_i, 
			\end{array} \right.  \\
			\textrm{ and } 
			& \tAi(x)  =    \left\{ \begin{array}{l}
			1- \frac{\tgA}{q_i \tgB} +  \frac{x}{q_i w_i \tgB}, \forall x  \in  \left[ 0, w_i \tgA \right], \\
			1         \qquad \qquad  \qquad \qquad, \forall x > w_i \tgA, \end{array} \right.
\end{align*}

First, if $x \le \min \braces*{q_i w_i \gB -p_i, w_i \tgA }$, we have:
\begin{align*}
    \abs*{ \Ai(x) - \tAi(x)} & = \abs*{\frac{p_i}{q_i w_i \gB} +  \frac{x}{q_i w_i \gB} -  	1 + \frac{\tgA}{q_i \tgB} -  \frac{x}{q_i w_i \tgB}}\\
    & = \abs*{ \frac{p_i + x - q_i w_i \gB }{q_i w_i \gB} + \frac{w_i \tgA - x}{q_i w_i \tgB}  }\\
    & = \abs*{ \frac{1}{q_i w_i} \parens*{ \frac{p_i - q_i w_i \gB}{ \gB} + \frac{w_i \tgA}{\tgB} + \frac{x (\tgB -\gB)}{\tgB \gB} }    }\\
    & \le \bigoh(\delta).
\end{align*}

Therefore, we conclude that when $p_i \ge 0$, for any $x$, we can always prove that \mbox{$\abs*{\Ai(x) -\tAi(x)} < \delta $} and $\abs*{\Bi(x) -\tBi(x)} < \delta $.

Now, for the case where $p_i <0$, we can do similarly to the analysis when $p_i \ge 0$ by simply exchanging the roles of A and B, then replace $q_i = 1/q_i$, $p_i = -\frac{p_i}{q_i}$. We conclude the proof.

\qed
\label{appen:heuristic}

\end{document}